\documentclass[prb,twocolumn,floatfix]{revtex4-2}
\usepackage{latexsym}
\usepackage{graphicx}

\newcommand{\Tr}{{\rm Tr}}
  
\usepackage{pict2e}

\usepackage{subfigure}
\usepackage{array}
\usepackage{verbatim}
\usepackage{amsmath}
\usepackage{color}
\usepackage{xcolor}
\usepackage{soul}
\usepackage{tabu}
\usepackage{multirow}
\usepackage{lipsum}
\usepackage[normalem]{ulem}
\begin{document}
\title{Accurate electron correlation-energy functional:\\ 
  Expansion in
  an interaction renormalized by the random-phase approximation}
  \author{Mario Benites, Angel Rosado and Efstratios Manousakis}
  \address{
    Department  of  Physics, Florida State University, Tallahassee, FL 32306-4350}
  \begin{abstract}
    We present an accurate local density-functional for electronic-structure calculations
    within the density functional theory (DFT). The functional is
    derived by analyzing the structure of the standard perturbative expansion
    of the correlation energy of the interacting uniform electron gas.
    Then, the expansion is partially re-summed  and reorganized  as a
    self-consistent series in powers
    of a renormalized  electron-electron interaction vertex based on the screened frequency-momentum dependent dielectric matrix given by
    the well-known random-phase approximation.
    First, we demonstrate that
    the range of $r_s$, where this reorganized and renormalized series converges, contains and is significantly larger than
    the average range realized in real crystalline materials.
    Using a combination
    of analytical, numerical, and stochastic integration techniques we are
    able to calculate all the diagrams which have contribution up to the same
    leading order. 
     We benchmarked the functional
    using the Quantum ESPRESSO implementation of the DFT
    applied to the same list of materials, selected previously by other authors, in its entirety without any modification of the list. We find that for ground-state properties in general, such as, equilibrium atomic distances and bulk moduli, the functional presented here is more accurate than the currently available most popular one.
  \end{abstract}
  \maketitle
\section{Introduction} 

The density functional theory (DFT) as introduced by Hohenberg Kohn and Sham\cite{PhysRev.136.B864,PhysRev.140.A1133} (HKS) is broadly used to provide insight into the electronic structure of materials.
In addition to the assumed-known interaction potential
formed by the contribution to the total one-body pseudopotential by all
the screened-ionic cores, DFT requires an accurate 
  ``exchange-correlation''
potential $V_{xc}$. The $V_{xc}$ is a functional of the spatially varying local
electronic density-field
  $n(\vec r)$ produced by the collective presence of all of the electrons
  in the interacting ground-state. In fact, provided that the ground-state
  is not degenerate, one can show the existence of 
  a one-to-one correspondence between the interacting ground-state
  and the diagonal part of the interacting one-body density matrix.

  The important feature of this functional is its universal nature, namely,
  it is the same for all materials. Therefore, one can conceptualize the
  question of the electronic structure of any given
  material, as
  a problem where the response of a system of interacting electrons, in the presence of an external
  one-body potential presented by the ions, is sought. Its universal
  character, allows us to determine its form, in principle and at least its local part,
  by calculating it for the pure system of interacting electrons in the presence of
  a uniform background of positive-charge of density equal to
  the spatial average of the electronic density (the so-called Jellium model).

  While this approach of determining the universal density-functional
  by focusing our effort on the Jellium model started more than half
  a century ago, the ground-state properties of many materials
  still cannot be calculated with the desired accuracy.
  
  Most functionals, especially the one used by most authors\cite{PhysRevLett.77.3865,PhysRevB.44.13298,PhysRevB.45.13244},
  are based on an ad hoc functional form which is general enough
  to reduce itself and reproduce the leading terms of 
  extreme $r_s \to 0$ limit of the random phase approximation (RPA) (a limit which is
  expected to be captured correctly by the RPA), while the low-density limit (large $r_s$) and the
intermediate regime is fitted to the quantum Monte Carlo results\cite{PhysRevLett.45.566} of
the Jellium model.

However, quantum Monte Carlo (QMC) simulation of fermions is hindered by the infamous sign-problem
which forces us to limit our calculations to small-size systems
of the order of a few hundred\cite{PhysRevLett.45.566} or at most several hundred up to one thousand\cite{PhysRevB.50.1391,PhysRevLett.82.5317} electrons.
These
system sizes are too small to allow extrapolation to the infinite-size
limit to remove the finite-size effects
to a satisfactory level of accuracy. To give an example of the severity of
the finite-size effects, for 368 electrons
the occupied states of the non-interacting un-polarized determinant used as part of the initial and guiding trial-state in diffusion Monte Carlo, does not have the form of
a sphere at all and the kinetic energy of an occupied state on the $k_z=0$ plane and
that of the nearest unoccupied
state differ by a 4/9 fraction of the Fermi energy. Therefore, the simulated
system
is far away from the actual uniform Jellium metallic-system and flipping a spin in this simulated-system case
costs such a large amount of energy. What is worse is that there is no
known scaling
function to use in order to extrapolate in the infinite-size limit.

In this work, we illustrate that we should expect a
reorganized perturbative expansion,
based on using a renormalized RPA-screened interaction vertex,
  to have a wider range of validity in the parameter  $r_s$ than previously
  thought\cite{PhysRev.46.1002,PhysRev.106.364}. 
  In fact, we show that the range of $r_s$, where
  one should expect such a reorganized perturbative scheme to be
  convergent, contains the range of $r_s$ realized in most materials.
  It is worth noting that, as is well-known, using
  an effective-Hamiltonian approach which separates
  the degrees of freedom into collective (charge-density fluctuations)
  and elementary excitations above the ground-state (quasiparticle/quasihole)
  one could achieve a very different picture of the
  rate of convergence\cite{PhysRev.109.741,Nozieres}.

  We, therefore, undertake a systematically organized perturbation expansion
  in orders classified by the number of renormalized RPA-screened interaction-lines. The correlation energy for any value of the spin-polarization
  and any value of $r_s$ begins with two-leading families of diagrams in the
  RPA-renormalized interaction. The first is the well-known series of rings-of-bubbles-like diagrams
  and a series which we call the ``kite'' diagram series.
  To determine our functional, first,
  we very accurately calculate the sum of the ring diagrams for a  very
  wide range of $r_s$ and as a function of spin-polarization. We also
  determine analytically the small $r_s$ and the large $r_s$ singular
  behaviors of the sum of rings-of-bubbles diagrams.
  The kite diagram using the bare-Coulomb interaction
  was first estimated by
Gellmann and Brueckner\cite{PhysRev.106.364} by means of Monte Carlo integration
and was later calculated exactly by Onsager\cite{Onsager1966IntegralsIT}.
We calculate the correction to the kite-diagram family when a fully
RPA-renormalized interaction is used by first carrying out
imaginary frequency integrals analytically by appropriately choosing
the integration contour in the complex-frequency plane to avoid the
brunch-cut of the dielectric frequency-momentum-dependent function. This allows the remaining 11-dimensional integral to be well-behaved
for a stochastic integration method to be effective.
We find that, in the range of
$r_s$ accessible in most materials, this correction previously ignored
is as significant as the other contributions to the correlation energy
of the interacting electron fluid. The reason for this is that this
series of diagrams consists of a contribution to
the exchange-correlation energy where the effects of exchange and correlation
are both profound and cannot be disentangled.

We then use the numerical results of this calculation and our analytical
knowledge of the behavior of the various different terms in the
small and large $r_s$ limits to determine a functional form that fits
very accurately our numerical results as a function of $r_s$ and
spin-polarization. 

In order to assess the accuracy of our functional relative to the
currently popular local-density Perdew-Wang (PW) functional\cite{PhysRevB.45.13244}, we modified
the Quantum ESPRESSO package\cite{QE-2017} to include our functional and we carried out DFT calculations within the local density approximation (LDA)
for the same entire list of crystalline
materials given in Refs.~\onlinecite{PhysRevB.79.085104,PhysRevB.75.115131,10.1063/1.5116025} (where the performance of various other functionals
was evaluated). We find that our functional overall outperforms the
PW functional\cite{PhysRevB.45.13244} in the calculation of the equilibrium lattice
constants and bulk moduli. These observables are the
main indicators of the accuracy of a functional for ground-state
properties. We have avoided comparing gaps, as
the Kohn-Sham orbitals are only {\it auxiliary entities} needed in the
HKS theory, which is purely a ground-state theory and is not meant to describe {\it real} single-particle excitations.

The paper is organized as follows.
In Sec.\ref{calculation} the calculation of the electron gas is
described. In Sec.~\ref{our-functional} our functional is presented, while in Sec.~\ref{comparison-with-other} is compared with other functionals. In Sec.~\ref{performance} its performance is benchmarked. Last, in Sec.~\ref{conclusions} we give our discussion and our
concluding remarks.

    \section{Calculation of the correlation energy}
    \label{calculation}
In this section we describe how we obtain the
ground-state (GS) correlation energy of the homogeneous electron gas for any value of
$r_s$ and spin-polarization $\zeta$. We will assume a density $n_{\uparrow}$
of electrons with spin-up and $n_{\downarrow}$
of electrons with spin-down with total density $n = n_{\uparrow} + n_{\downarrow}$
and
\begin{eqnarray}
  \zeta &=& {{n_{\uparrow}-n_{\downarrow}} \over {n_{\uparrow}+n_{\downarrow}}}\\
  r_s &=& \Bigl ({{3 } \over {4 \pi n}} \Bigr )^{1 \over 3},
\end{eqnarray}
with corresponding Fermi wavevector:
\begin{equation}
  k_{F \sigma} = k_F x_{\sigma}, \hspace{1cm} k_F \equiv \left({3 \pi^2 n} \right)^{\frac{1}{3}}, 
\label{eq:Fermi_momentum}
\end{equation}
  where
  \begin{eqnarray}
    x_{\sigma} \equiv (1+ \sigma \zeta)^{\frac{1}{3}},
    \label{spin_scale}
    \end{eqnarray}
and $\sigma = +1,-1$ for spin-up and spin-down respectively.

In order to calculate the contributions to the
ground-state correlation energy we use the following general expression for
the ground-state expectation value of the bare Coulomb interaction term, i.e., 
$\langle \Psi_0 |\hat{V}|\Psi_0 \rangle $, where $|\Psi_0\rangle$ is the
interacting ground state, for the
electron gas of spin-polarization $\zeta$:
\begin{equation}
  \langle \Psi_0| \hat{V} | \Psi_0 \rangle = -\frac{i\hbar V}{2} \int_0^1\frac{d\lambda}{\lambda} \int \frac{d^4k}{(2\pi)^4}e^{ik^0 \eta} \Tr\left[\Sigma^{*\lambda} G^{\lambda} \right],
\label{eq:Cluster_Expansion}
\end{equation}
where $\Sigma^{*\lambda}=\Sigma^{*\lambda}_{\alpha \beta}(\Vec{k},k^0)$ and $G^{\lambda}=G^{\lambda}_{\alpha \beta}(\Vec{k},k^0)$ are the proper self-energy and the fully-interacting fermion propagator tensors, respectively, under the rescaling of the coupling of the interaction: $e^2 \rightarrow \lambda e^2$, where $\lambda$ guarantees the interaction to be turned on very slowly in the system. The $\alpha$ and $\beta$ are spin indices (with values of 1 or -1) and $\eta$ is an infinitesimal factor with time dimension) that preserves the correct time-order of the field operators when evaluating expectation values. Both tensors are diagonal in this system since the electron's spin is conserved at the vertex level in the Coulomb interaction.

\subsection{Issues and history of the perturbative expansion}
 
     \begin{figure}[htp]
       \begin{center}
         \subfigure[]{
            \includegraphics[scale=0.2]{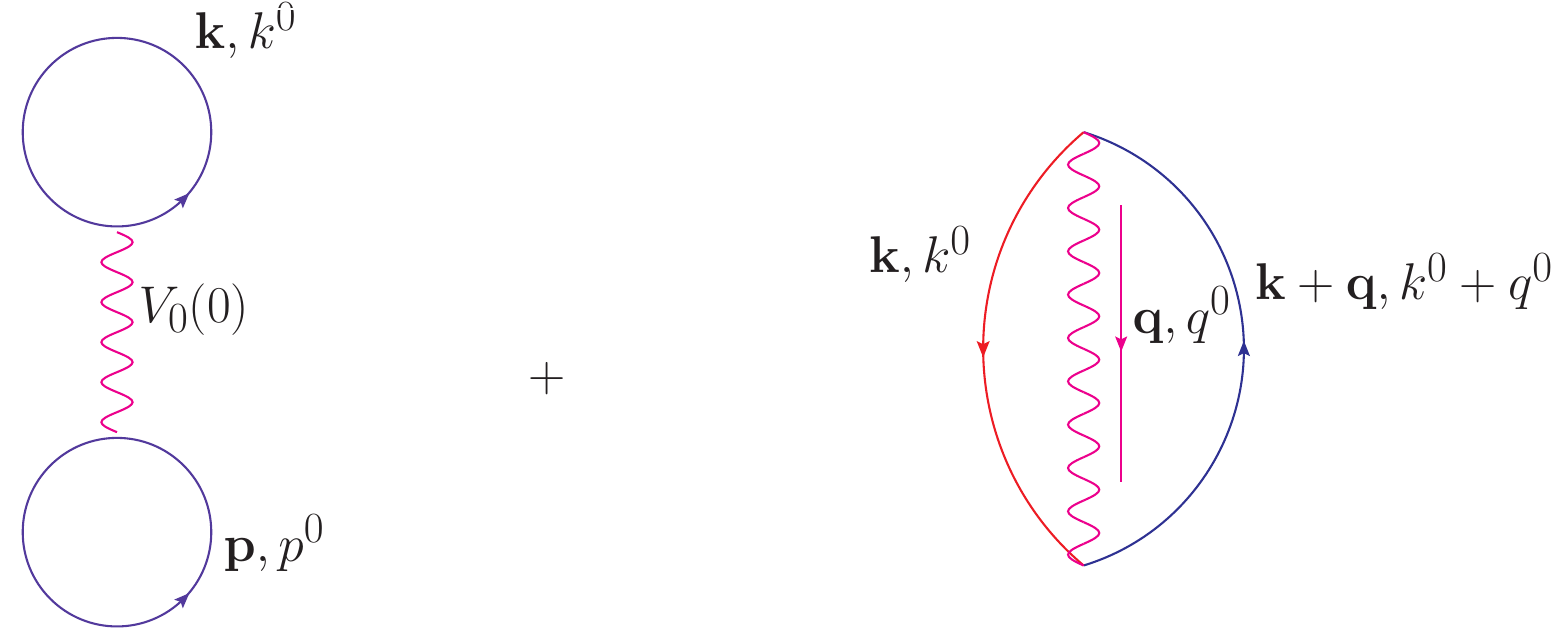} 
   \label{first-order}
         }
         \subfigure[]{
               \includegraphics[scale=0.2]{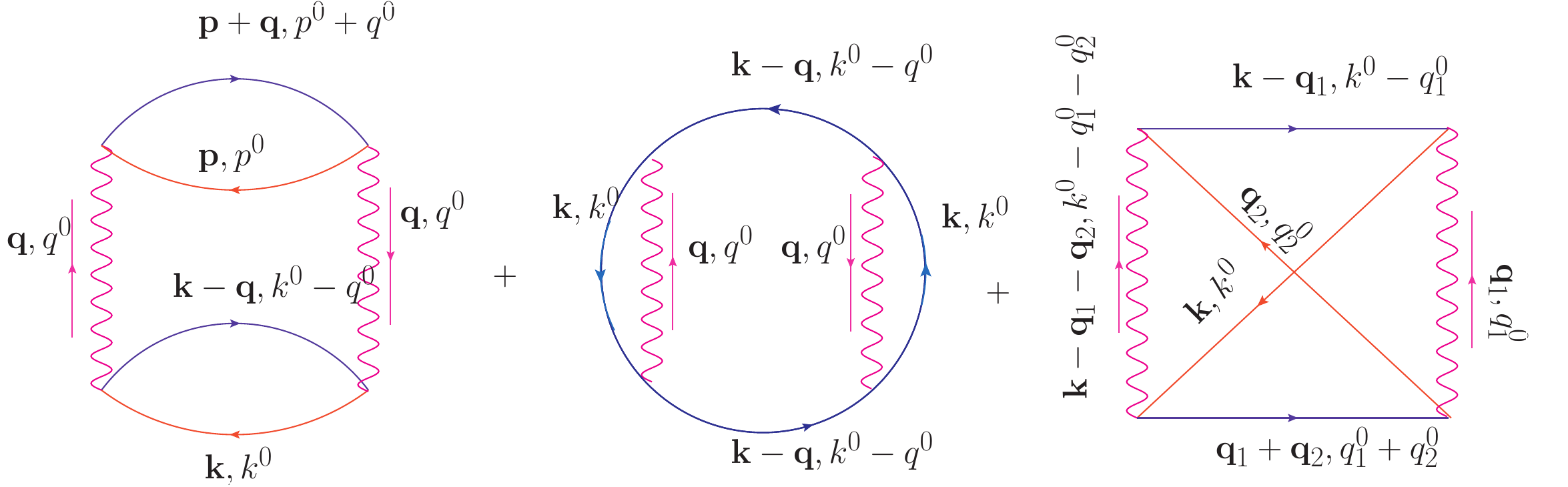} 
   \label{second-order}
         }
   \end{center}
  \caption{(a) First order in the bare Coulomb interaction.
    (b) Second order, i.e., the correlation energy correct up to leading order in bare Coulomb interaction.}
  \label{ecor-bare}
     \end{figure}
     Fig.~\ref{ecor-bare} gives the ground-state energy correct up to leading order in the bare-Coulomb interaction.
Fig.~\ref{first-order} is the Hartree-Fock approximation
where the first term is the Hartree term, which, in the case of
our Jellium model, is canceled by the uniform background of positive charge.
Fig.~\ref{second-order} gives the next-order correction, which is the
correlation energy in leading order in the bare Coulomb interaction.
The first diagram of Fig.~\ref{second-order} alone is infinite at
any value of $r_s$ because of
the singular nature of the Coulomb interaction in the infrared.
The second diagram of Fig.~\ref{second-order} is zero and the third has been first estimated by
Gellmann and Brueckner\cite{PhysRev.106.364} by means of Monte Carlo integration
and was later calculated exactly by Onsager\cite{Onsager1966IntegralsIT}.
In the rest of this paper, we will refer to this latter diagram as the ``kite''
diagram.

In order to ``bypass'' the problem of the infinite character of
the first diagram of Fig.~\ref{second-order},  what has been done was to replace
one of the interacting lines by the
renormalized interaction-line given by the RPA sum of ring-diagrams\cite{PhysRev.92.626,PhysRev.92.609,PhysRev.106.364} 
shown in Fig.~\ref{rpaline} and this makes this renormalized diagram
  finite at any finite value
of $r_s$ and it diverges as $~\ln(r_s)$ in the  $r_s \to 0$ limit.

The above series of all the correlation
diagrams is the RPA, which leads to dressing one of the interaction lines as shown in Fig.~\ref{rpaline}. This effective
interaction-line introduces an emerging regulator  which endows the new diagram with a finite value at finite $r_s$.

    Hence, 
    a perturbation expansion in the bare Coulomb-interaction
    leads to infinities and it is necessary to renormalize the Coulomb
    interaction. Therefore, the entire perturbation expansion needs to be
    systematically reorganized in terms of such a renormalized interaction.
    
\subsection{Summation of the ring-diagrams}

    \begin{figure}[htp]
      \vskip 0.3 in
      \begin{center}
            \includegraphics[scale=0.25]{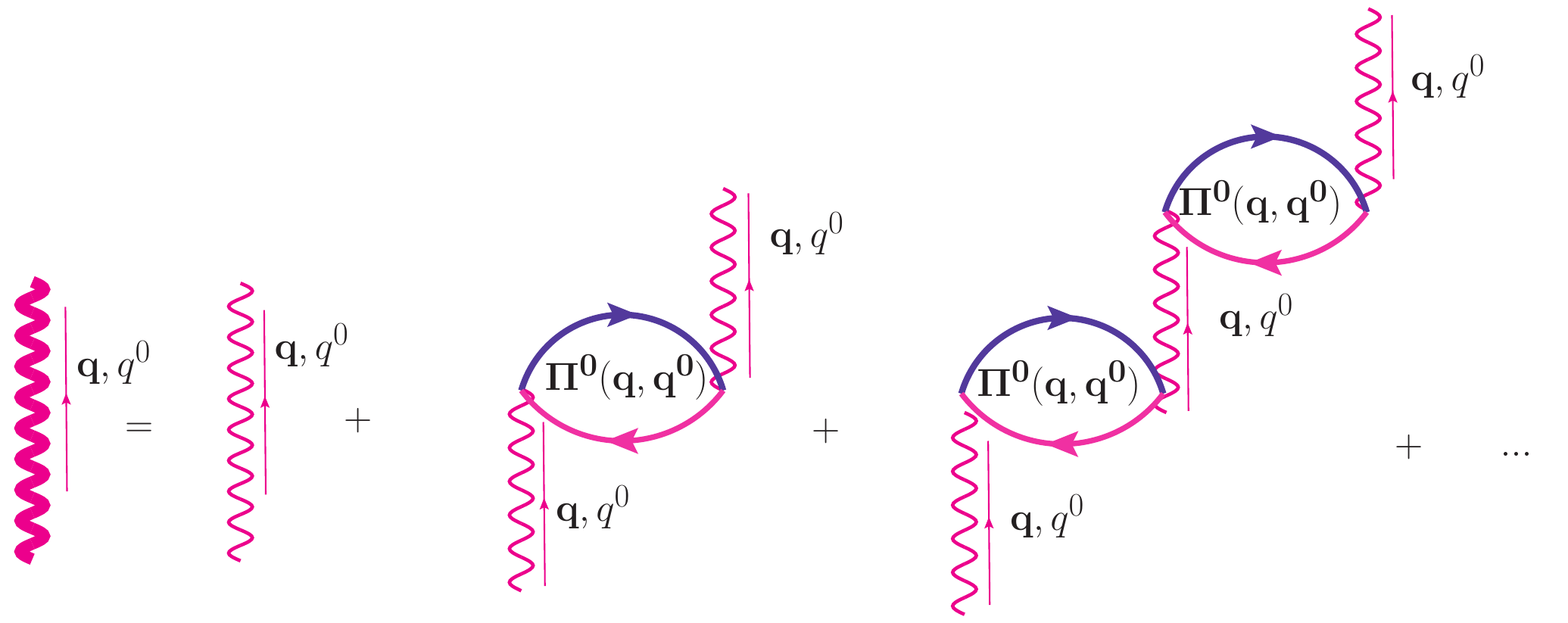}
    \end{center}
    \caption{The bold interaction line represents
      the sum of the terms included in the RPA and each loop ($\Pi^0(q,q^0)$) represents the spin-trace of the polarization tensor. For more details see text.}
            \label{rpaline}
\end{figure}

    First, we summarize the calculation for the well-known RPA
    approach\cite{PhysRev.92.609,PhysRev.92.626,PhysRev.106.364}.
The fundamental aspect of this theory is that it renormalizes the
bare Coulomb interaction by adding together the selected terms of
the perturbation expansion illustrated in Fig.~\ref{rpaline} which defines
an effective interaction with a Fourier transform given as follows:
    \begin{eqnarray}
      \tilde {V}_{eff}(q, q^0) \equiv {{4\pi e^2} \over {q^2 \epsilon(q,q^0)}},
      \label{effective}
      \end{eqnarray}
where the dielectric function $\epsilon(q,q^0)$ is found by calculating the spin-trace of the polarization $\Pi^0_{\sigma \sigma'}(q,q^0)$
   (bubble or particle-hole diagram):    
\begin{equation}
  \epsilon(q,q^0) = 1-\frac{4 \pi e^2}{q^2}\sum_{\{ \sigma \}=\pm} \Pi^0_{\sigma}(q,q^0),
\label{eq:dielectric}
\end{equation}
where one spin index $\sigma$ is dropped in the polarization tensor since it is diagonal, and $\Pi^0_{\sigma}(q,q^0)$ can be calculated by a 4-dimensional integral of a product of two Green's functions, which the result yields an even function in frequency $q^0$. The integral expression of $\Pi^0_{\sigma}(q,q^0)$ is given by:
\begin{equation}
    \Pi^0_{\sigma}(q,q^0) = \frac{-i}{\hbar}\int \frac{d^4 k}{(2 \pi)^4} G^0_{\sigma}(k^{\mu}+q^{\mu}) G^0_{\sigma}(k^\mu),
\label{polarization_function}    
\end{equation}
where the $\mu$ superscript is used only as a shorthand notation for $q^{\mu}=(\vec{q},q^0)$ (not to be confused as a covariant quantity) and the non-interacting Green's function has the usual integral expression that takes into consideration the energy dispersion $\epsilon_{\vec{k},\sigma}$ of the electron measured relative to the chemical potential with spin $\sigma$. At zero temperature we have:
\begin{equation}
G^{0}_{\sigma}(k^{\mu}) = \frac{\Theta(k-k_{F \sigma})}{k^0-\frac{\epsilon_{\vec{k},\sigma}}{\hbar}+i\eta} + \frac{\Theta(k_{F \sigma}-k)}{k^0-\frac{\epsilon_{\vec{k}, \sigma}}{\hbar}-i\eta}.
\label{free_Green_function}
\end{equation}

The calculation of the polarization tensor is similar to the spin-unpolarized case, where the latter is solved in Ref.~\onlinecite{Fetter,Mahan,Gorkov}. The difference in the calculation is at the level of rescaling the wavevector $\vec{q}$ and frequency $q^0$ into unitless variables, but instead of using $k_F$ for the rescaling, we use $k_{F \sigma}$. The result for $\Pi^0_{\sigma}(q,q^0)$ depends on the Lindhard's function where its real and imaginary parts can be found in Ref.~\onlinecite{Fetter,Mahan}. In the following discussion, we will involve the expression of its analytic continuation in imaginary frequency (see our Eq.~\ref{g_function}). 

\begin{figure}
    \vskip 0.3 in \begin{center}
            \includegraphics[scale=0.3]{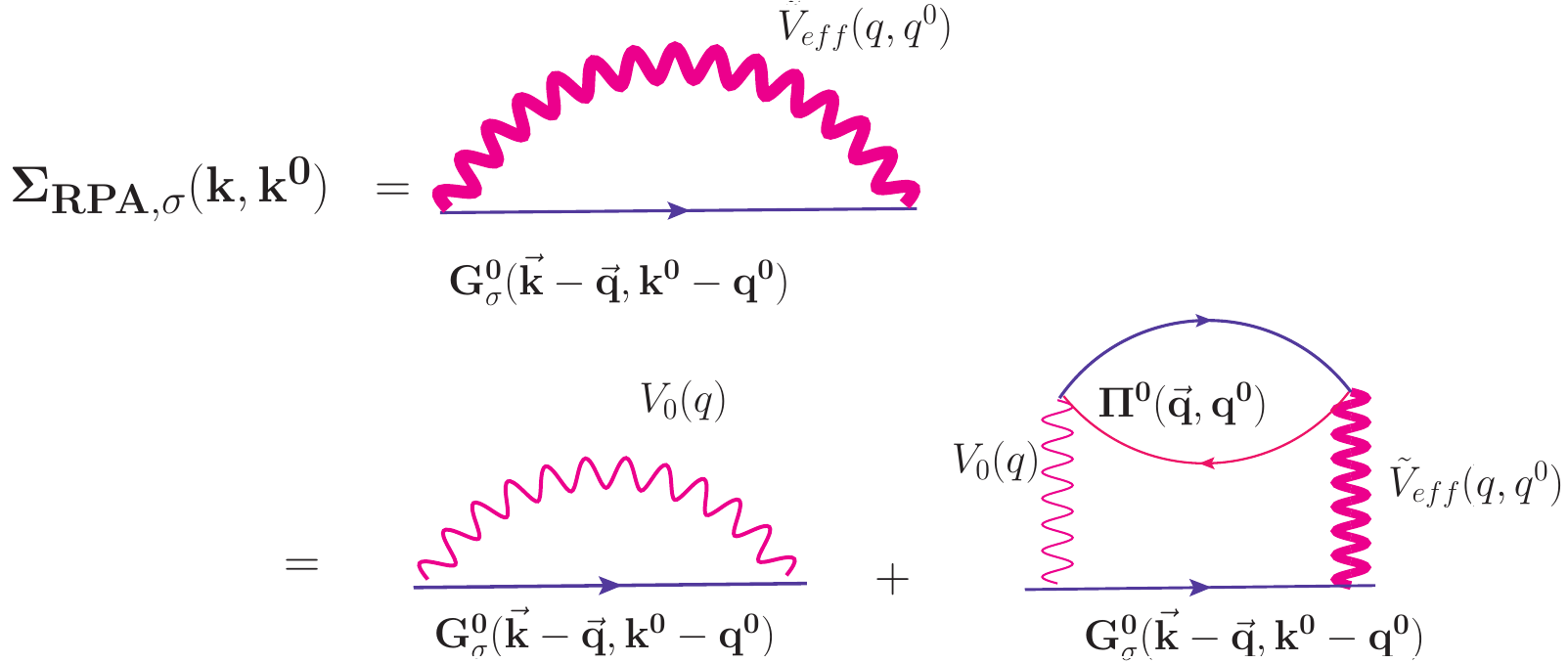}
    \end{center}
    \caption{The self-energy within the RPA. We have not included the
      Hartree term as it is canceled by the contribution of the
      uniform positive background in the Jellium model.}
            \label{self} 
      \vskip 0.2 in
\end{figure}

The self-energy $\Sigma_{RPA,\sigma}(k,k^0)$
within the RPA is illustrated in Fig.~\ref{self} and
when multiplied by the non-interacting Green's function, i.e., $G^0_{\sigma}(k,k^0)$,
as in expression given by Eq.~\ref{eq:Cluster_Expansion}, and integrated over the external leg variables frequency ($k^0$), and
wavevector (${\bf k}$) it gives the ground-state expectation-value of the
interaction energy.
The first term is the exchange (Fock) diagram and only the second term contributes
to the correlation energy.
    \begin{figure}
    \vskip 0.3 in \begin{center}
            \includegraphics[scale=0.21]{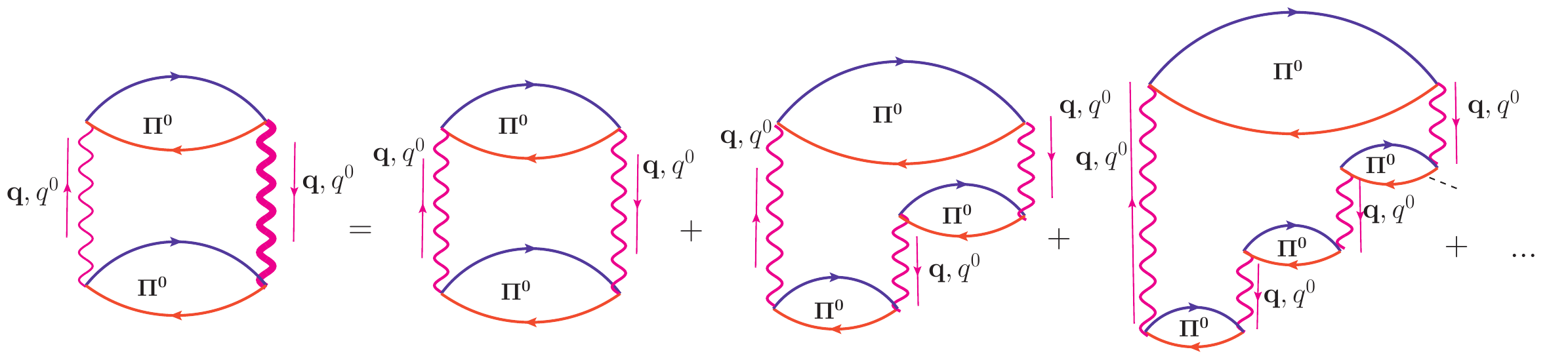}
    \end{center}
    \caption{The sum of the series of diagrams depicted  in the right-hand-side of the equation in the figure is the correlation energy within RPA and it corresponds to
      replacing one of the interaction lines in the first diagram of Fig.~\ref{ecor-bare} 
      with a bold interaction line which is symbolically  depicted by the diagram in the
      left-hand-side of the equation. In this figure we labeled $\Pi^0$ as the spin-trace of the polarization tensor.}
            \label{rpacor} 
      \vskip 0.2 in
\end{figure}
Once this step is done, this yields the sum of the ring diagrams with partial spin-polarization ($\zeta$), which consists of summing all of the first-order Goldstone diagrams within the RPA, as listed in Fig.~\ref{rpacor}. The calculation of these ring diagrams consists of carrying out a calculation of a 5-dimensional integral, one integration over the $\lambda$ parameter, and the other 4-dimensional integrals along the frequency variable $q^0$ and the wavevector $\vec{q}$. After performing a simple integral along the $\lambda$ parameter on the integral expression of the ring diagram series, and by rescaling the wavevector and frequency variables by the mapping $\vec{q}\rightarrow k_f \vec{\kappa}$, $q^0 \rightarrow \hbar k_F^2 \nu/m$, one obtains the general expression for the contribution to the ring diagram series $E_r(r_s,\zeta)$, as follows:
\begin{equation}
E_r(r_s,\zeta) = \frac{N e^2}{2 a_0}\epsilon_r(r_s,\zeta),
\label{ring}    
\end{equation}
where the contribution to the ground-state energy per particle in Ry, i.e., $\epsilon_r(r_s,\zeta)$, is given  which is given by the following 4-dimensional integral:
\begin{equation}
\epsilon_r(r_s,\zeta) = \frac{-3 i}{16 \alpha^2 \pi^2 r_s^2} \int d^3\kappa \int^{\infty}_{-\infty} d\nu \left[ \ln(\epsilon_d) + 1 - \epsilon_d \right],
\label{ring2}    
\end{equation}
    \begin{figure}
    \vskip 0.3 in \begin{center}
            \includegraphics[scale=0.4]{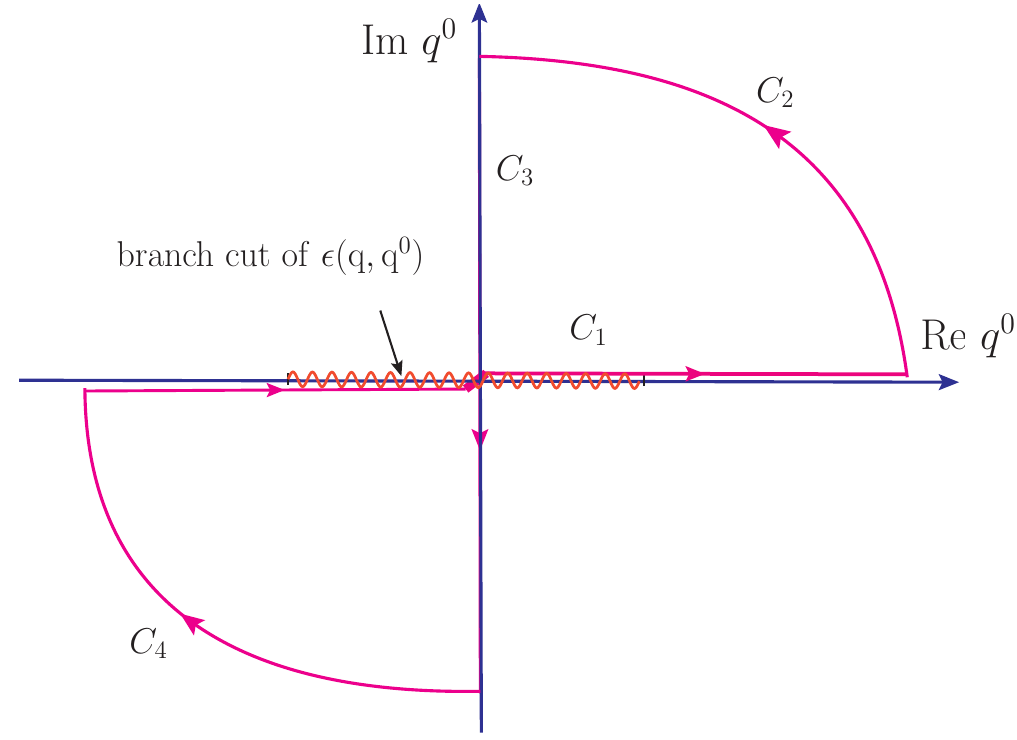}
    \end{center}
    \caption{The path on the complex frequency plane chosen for
      the calculation of the contribution to the ring diagrams series
      (see text).}
            \label{path-ring} 
      \vskip 0.2 in
    \end{figure}
    where $\alpha = (4/(9\pi))^{1/3}$ and
$\epsilon_d =\epsilon\left(k_F \kappa,  \frac{\hbar k_F^2 \nu}{m}\right)$.
We can proceed to choose a particular complex contour, to map the rescaled frequency integral along the real line into a pure imaginary frequency integration $i \nu$ in a way that doesn't enclose the branch cuts that arise from the logarithmic terms that come from the dielectric function, as seen in  Fig.~\ref{path-ring}. In this case, we use the analytic continuation of the Lindhard function where the expression is more simplified since this yields a real function. After some simplification of the expression of the integrand for the ring diagram series per particle by using $x = \nu/\kappa $, we are left with an imaginary frequency integration given by:
\begin{eqnarray}
  \epsilon_r(r_s,\zeta) &=& \gamma \int^{\infty}_0 d\kappa \kappa^3 \int^{\infty}_0  dx [\ln(1+{\tilde \Pi}) -{\tilde \Pi})],\label{ring4}\\
          {\tilde \Pi} &\equiv& \frac{\alpha r_s}{\pi \kappa^2}\sum_{\sigma}x_{\sigma}(\zeta) g\left({\kappa \over {x_{\sigma}(\zeta)}}, i{{\kappa x} \over {x^2_{\sigma}(\zeta)}}\right),\label{Pi_tilde}
\end{eqnarray}
where $\gamma = \frac{3}{2 \pi \alpha^2 r_s^2}$, and the analytic continuation of the Lindhard function is given by the following expression:
\begin{eqnarray}
  g(q_1,i\nu) &=& 1+ \frac{\nu^2+\kappa_+ \kappa_-}{2q_1^3} \ln\left( \frac{\kappa_+^2+\nu^2}{\kappa_-^2 +\nu^2} \right)\nonumber\\
  &-&\frac{\nu}{q_1}\left(\tan^{-1}(\frac{\kappa_{+}}{\nu} )+ \tan^{-1} (\frac{\kappa_{-}}{\nu}) \right),
\label{g_function}   
\end{eqnarray}
where, $\kappa_{\pm} = q_1 \pm q_1^2/2$
We used the integral expression given by Eq.~\ref{ring4} to calculate numerically the ring diagram series per particle $\epsilon_r(r_s,\zeta)$ by using integration by quadrature. As in the ring diagram series of the unpolarized case, $\epsilon_r(r_s,\zeta)$ also has a $\ln(r_s)$ divergent part which we were able to calculate analytically by keeping track of the integral on regions of integration of very small $q$. Such calculation gives rise to the generalized coefficient of this divergent part, given by:
\begin{equation}
\epsilon_{r}(r_s,\zeta) = c_L(\zeta) \ln(r_s) + c_0(\zeta),
\label{ring_small-rs}
\end{equation}
where $c_{L}(\zeta)$ is the coefficient of the $\ln(r_s)$ term given by:
\begin{eqnarray}
  c_{L}(\zeta)&=&\frac{1}{\pi^2} \Bigl [(1-\ln 2) + {{x_+x_-}\over 2}
    \chi - \ln(\chi)\nonumber \\
    &+& {1 \over{2}} \sum_{\sigma} x_{\sigma}^{{3}} \ln (x_{\sigma}) \Bigr ],
\label{c_log}    
\end{eqnarray}
where 
\begin{eqnarray}
  \chi(\zeta) = \sum_{\sigma} x_{\sigma}(\zeta), \label{chi}
  \end{eqnarray}
and $x_{\sigma}(\zeta)$ is given by Eq.~\ref{spin_scale}.
We will use this expression in our functional.

\subsection{RPA-renormalized perturbation expansion}
We wish to carry out a perturbation expansion
   in the RPA-renormalized
  Coulomb interaction as in Hedin\cite{PhysRev.139.A796}.   We are
  not regarding this expansion as an $r_s$-expansion. We regard it
  as an expansion in powers of the renormalized interaction.

\begin{figure}[htp]
           \subfigure[]{
            \includegraphics[scale=0.2]{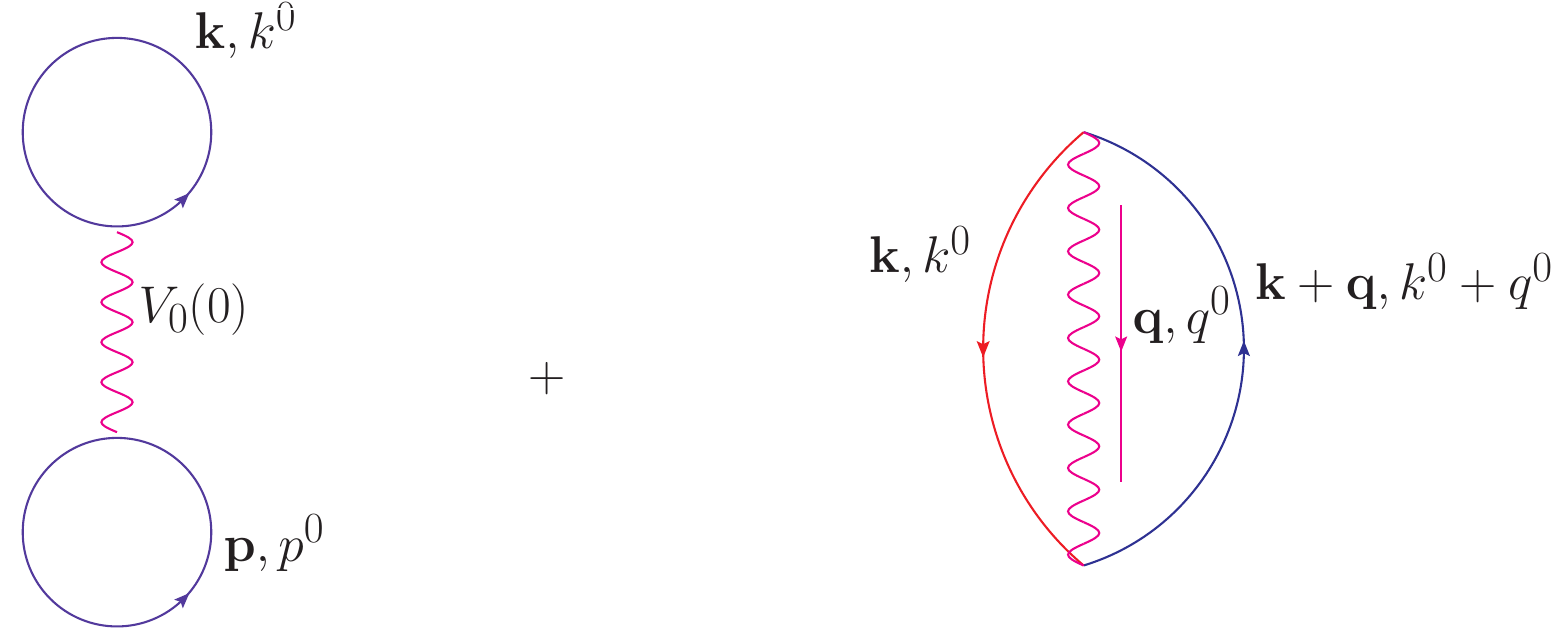} 
   \label{zeroth-order-R}
         }
           \subfigure[]{
            \includegraphics[scale=0.2]{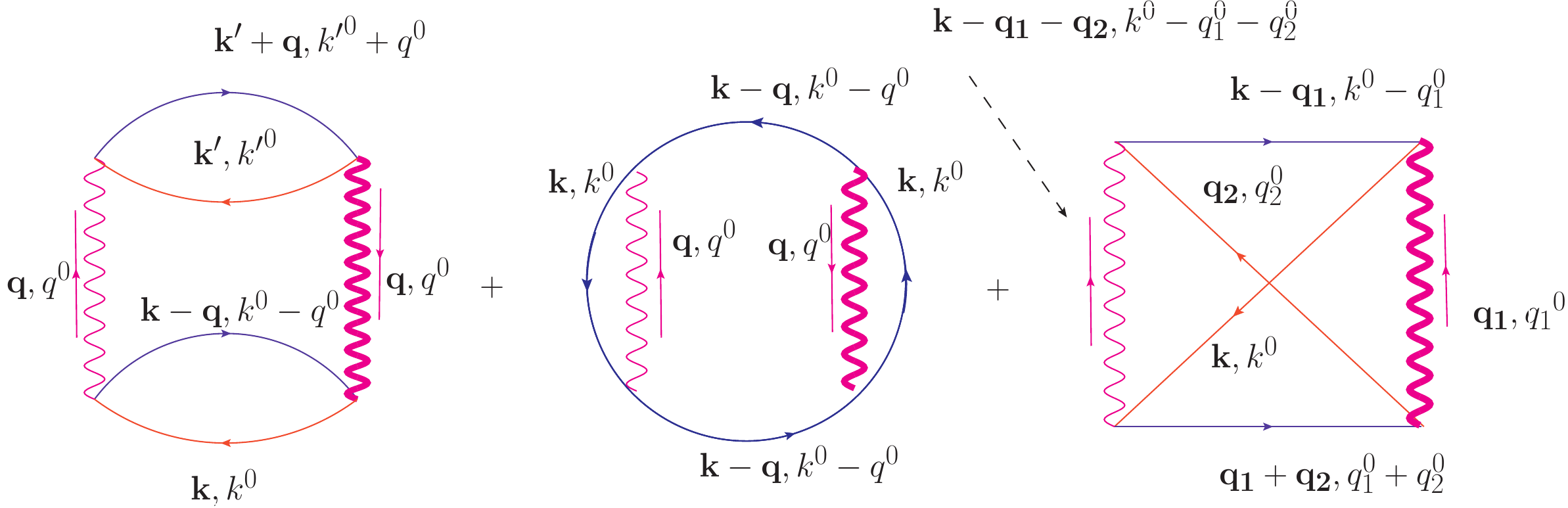}
   \label{First-order-R}
   }     
           \caption{(a) The zeroth order diagrams to the total ground-state energy in the RPA-renormalized interaction.
             (b) The set of all the diagrams contributing to the first-order
            in the RPA-renormalized interaction.}
   \label{leading-order-correlation-energy}
\end{figure}

  As stated earlier, our goal is to compute by a perturbative
  expansion the total ground-state expectation-value of the
  interaction energy, i.e., the expectation value of the bare
  Coulomb-interaction (as it appears in the original many-body Hamiltonian)
  in the interacting ground-state wavefunction, i.e.,
  \begin{eqnarray}
    \langle\Psi_0|\hat  V |\Psi_0 \rangle = \langle \Psi_0| \sum_{i<j} {{e^2} \over {r_{ij}}} | \Psi_0 \rangle,
  \end{eqnarray}
  where $| \Psi_0 \rangle$ is the interacting ground-state. This was our starting computational goal (as set originally by Eq.~\ref{eq:Cluster_Expansion}).
  This series 
  up to the first-order in the RPA-renormalized interaction-line
  are shown in Fig.~\ref{leading-order-correlation-energy}.
  Fig.~\ref{leading-order-correlation-energy}(a) lists all the zeroth
  order terms contributing to the ground-state interaction-energy.
  The presence of the bare-Coulomb interaction-line should not be confused
  with the first order, it is the operator of which we compute the expectation value
  of, it is not coming from the expansion of the interacting ground-state.
    The diagrams contributing to the ground-state interaction-energy (and correlation energy) to leading order in the RPA-renormalized interaction are shown in Fig.~\ref{leading-order-correlation-energy}(b). Notice that one bare Coulomb interaction line is unavoidable
and cannot be dressed by the RPA effective-interaction. The reason is
that the interaction energy is
calculated as the  expectation value in the interacting ground-state of the bare
Coulomb-interaction. Namely, this bare interaction-line is the observable
to which the ground-state expectation-value is computed within perturbation
theory.
The other lines can be renormalized because they come from the Goldstone
expansion of the ground-state.
Now, we note that the second diagram in Fig.~\ref{leading-order-correlation-energy}(b) is identically equal to zero. The reason is because there is the
product of two bare fermion propagators, a particle of
four-momentum $({\bf k},k^0)$ and a hole propagator with the same
four-momentum.

The two zeroth order diagrams are the familiar Hartree (which cancels the
interaction of the electron gas with the uniform positive background) and the bare exchange
terms contributing to the total density-functional. The exchange term
is a well-known contribution to the interaction energy per particle, which in Ry is given by
\begin{eqnarray}
V_0(r_s,\zeta) = -3 {{\sum_{\sigma}x^4_{\sigma}} \over {4\pi\alpha r_s}},
\end{eqnarray}
with $x_{\sigma}$ defined by Eq.~\ref{spin_scale}.

The first diagram of Fig.~\ref{leading-order-correlation-energy}(b) is the sum of all the ring diagram series which was computed in
the the previous Section. The third diagram, which we call it ``the kite-diagram'' because of its shape, is calculated in the following Section.

\begin{figure}[htp]
   \begin{center}
            \includegraphics[scale=0.3]{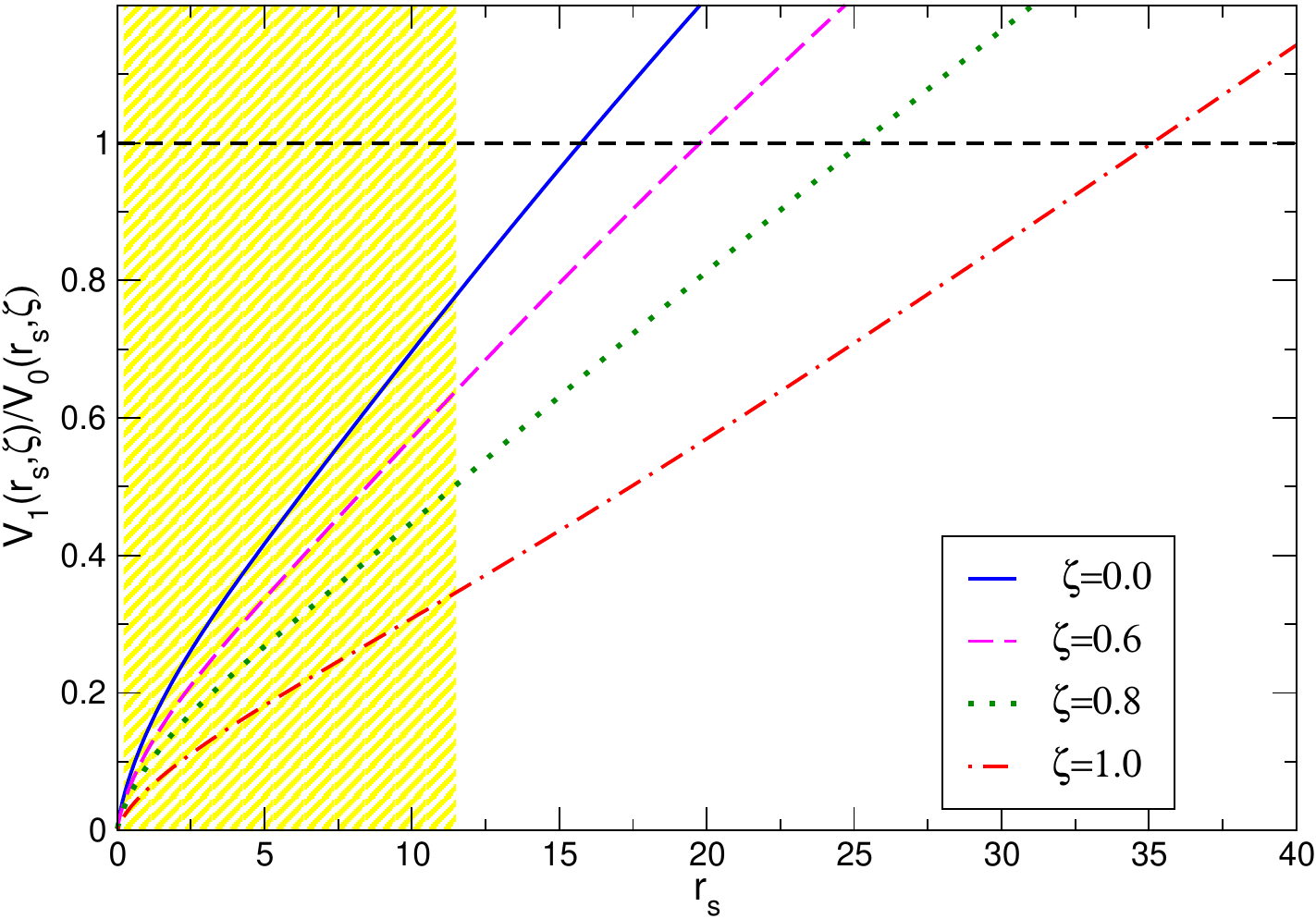} 
   \end{center}
   \caption{Comparison of the first-order contribution to the zeroth order
     contribution to the interaction energy in
   RPA-renormalized interaction-line.}
   \label{rsrange}
       \end{figure}

Our series expansion of the interaction energy, i.e., the quantity which we calculate can be written as
\begin{eqnarray}
  \langle \Psi_0 | \hat V | \Psi_o \rangle = V_0(r_s,\zeta) + V_1(r_s,\zeta) +
  V_2 (r_s,\zeta) + ..., \end{eqnarray}
where $V_n(r_s,\zeta)$ denotes the $n^{th}$ order in the RPA-renormalized interaction.
Namely, $V_n(r_s,\zeta)$ contains only $n$ of such lines.
Such a series will converge fast as long as $|V_{n+1}(r_s,\zeta)| << |V_n(r_s,\zeta)|$. Using the final results of our calculation we can provide
a justification of the validity of this expansion using the known results
for $n=0$ and our results for $n=1$. Fig.~\ref{rsrange} provides
the ratio $V_{1}(r_s,\zeta)/V_0(r_s,\zeta)$ for various values of the spin-polarization parameter $\zeta$. Given that this ratio
is significantly smaller than unity in the region of $r_s$ realized
in most real materials, we should expect to have a converging series of
our expansion in such region and our functional to be a good approximation for direct application to real materials. We note that we checked in the 
charge density output file of the
QE calculation and we did not find any value of density listed in the file which corresponds to a value of $r_s$ which lies
outside the yellow-shaded region of this figure.

Therefore, we conclude that using the
screened (and, therefore weaker) RPA-renormalized interaction vertex should lead to
a converging perturbative expansion in the
region of $r_s$ and $\zeta$ realized in most real materials.

\subsection{Summation of the kite-diagram series}
This section will briefly explain how to obtain an integral expression for the kite-like-diagram series that contributes to the ground-state (GS) energy of the homogeneous spin-polarized electron gas, where more details concerning the calculation are given in the Appendix~\ref{Calculation-of-kite}. This term
is a first-order in terms of the cluster expansion, where the integral expression of such a term is found by calculating the expectation value of the bare Coulomb potential operator $\langle \Psi_0| \hat{V}| \Psi_0 \rangle$ (our calculation begins from Eq.~\ref{eq:Cluster_Expansion}).
Therefore,  the bare interaction-line represents the operator and
the other line is due to the expansion. Our expansion re-sums
a selected series of diagrams illustrated in Fig.~\ref{rpaline}
resulting in the
renormalization of the second interaction-line to become an RPA-dressed
interaction-line.

\begin{figure}[htp]
  \includegraphics[scale=0.2]{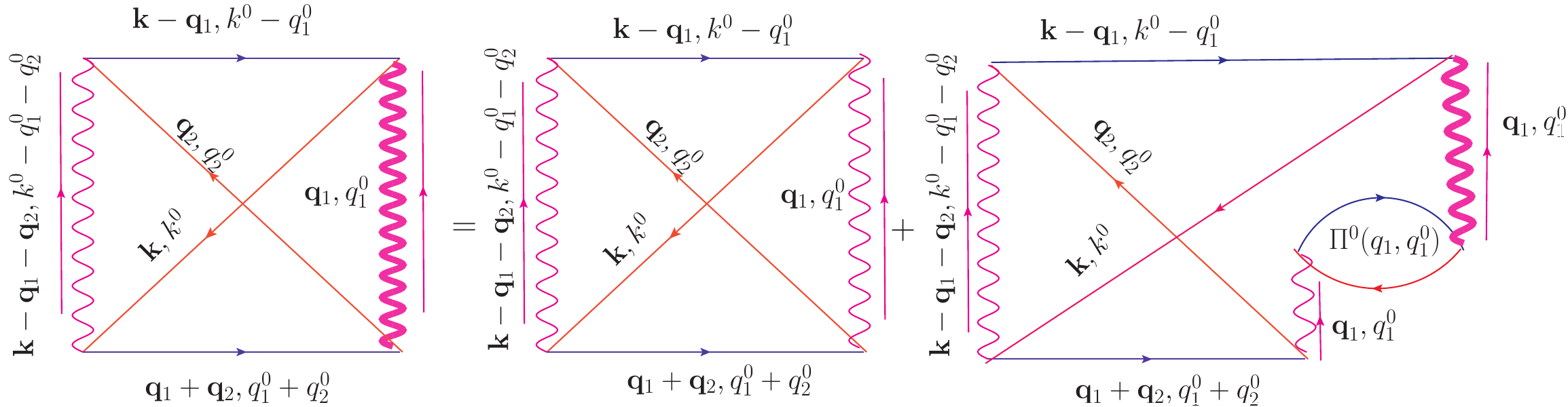}
  \caption{The kite-diagram series can be broken into two parts. The first
    consists of using the bare Coulomb interaction. This diagram was first
    estimated
    by Gellmann-Brueckner by means of Monte-Carlo integration
    and it was later calculated
    exactly by Onsager\cite{Onsager1966IntegralsIT}. The second, which is $r_s$ and $\zeta$ dependent,
    is calculated in this paper.}
   \label{kite}
\end{figure}

    \begin{figure*}
   \begin{center}
            \includegraphics[scale=0.45]{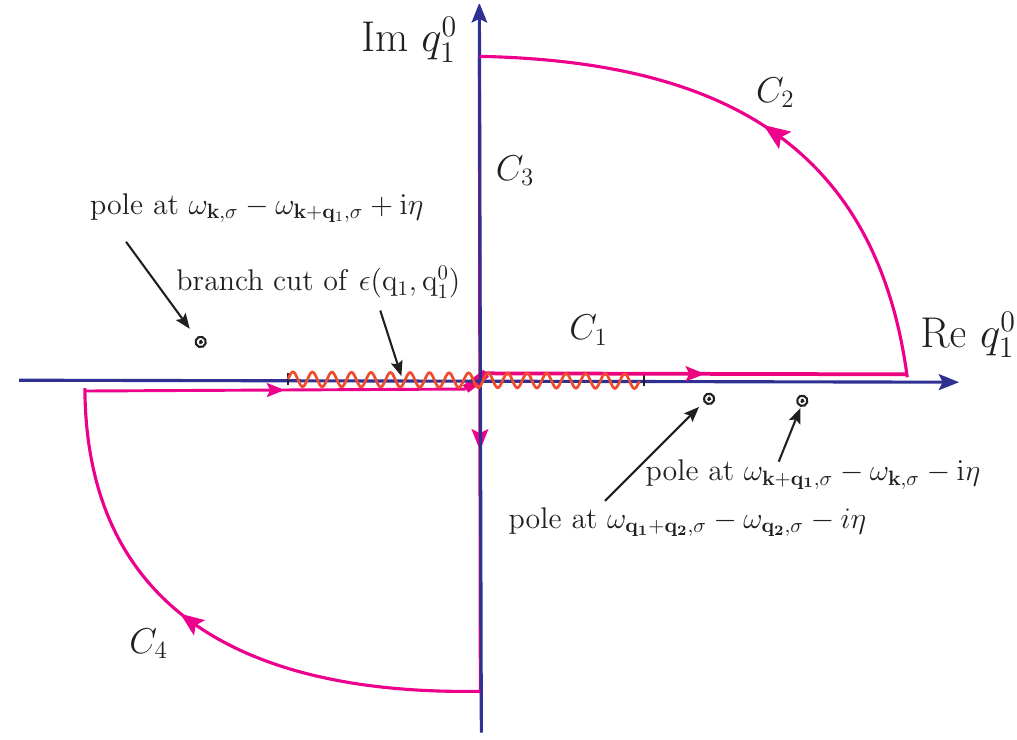} \hskip 0.2 in
            \includegraphics[scale=0.45]{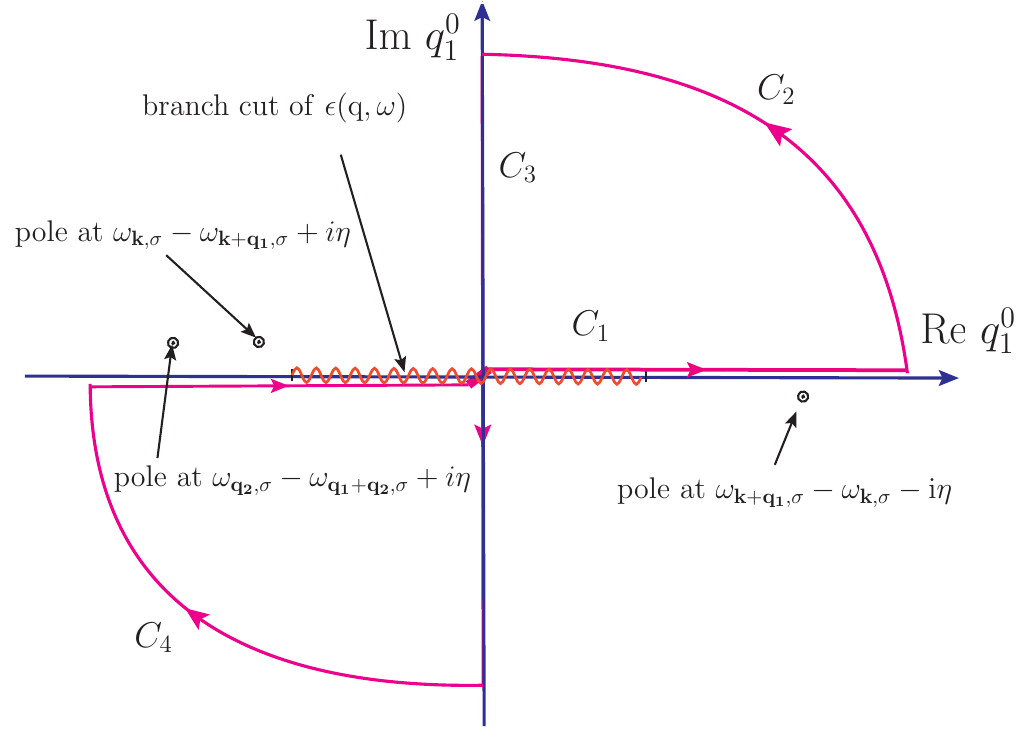}
            \caption{The paths we chose to calculate the two contributions
              $\Delta_1$ and $\Delta_2$ to the kite-diagram series. Both
              avoid enclosing the branch-cut of the dielectric function
              and the single-poles of the non-interacting Green's functions.}
      \label{paths}
    \end{center}
    \end{figure*}

The calculation of the kite diagram involves computing the diagram-series illustrated in Fig.~\ref{kite}. As seen in this figure, we separate the zeroth order of the expansion of the renormalized interaction line (first diagram), from the correction to the kite diagram which carries a $r_s$ dependence given by the dielectric function included in the Goldstone diagram (second diagram).  We used Eq.~\ref{eq:Cluster_Expansion} to calculate both contributions to the Goldstone diagram within renormalization, as explained in Appendix~\ref{Calculation-of-kite}. The expression for the complete kite diagram within renormalization is then written as:
\begin{equation}
    E_{2b}[n_{\uparrow},n_{\downarrow}] = E^0_{2b}[n_{\uparrow},n_{\downarrow}] + \Delta E_{2b}[n_{\uparrow},n_{\downarrow}],
\label{E2b_rpa}
\end{equation}
where the first term corresponds to the kite-diagram having two bare interaction lines which comes from the leading term of the effective Coulomb interaction that comes from the renormalization in RPA, while the second term corresponds to the kite diagram correction given by the single renormalized interaction line. Their corresponding expressions are given below:
\begin{widetext}
\begin{eqnarray}
E^0_{2b}\left[n_{\uparrow},n_{\downarrow}\right]& = & \frac{iV}{2\hbar} \int_0^1 d\lambda \lambda \int \frac{d^4k}{(2 \pi)^4}\int \frac{d^4q_1}{(2 \pi)^4}\int \frac{d^4q_2}{(2 \pi)^4} e^{i k^0 \eta} V_0(\vec{k}-\vec{q_1}-\vec{q_2})V_0(\vec{q_1})
\tilde D, \label{E2b_Onsager}  \\  
\Delta E_{2b}\left[n_{\uparrow},n_{\downarrow}\right] &=& \frac{iV}{2\hbar} \int_0^1 d\lambda \lambda^2 \int \frac{d^4k}{(2 \pi)^4}\int \frac{d^4q_1}{(2 \pi)^4}\int \frac{d^4q_2}{(2 \pi)^4} \frac{e^{i k^0 \eta} V_0(\vec{k}-\vec{q_1}-\vec{q_2}){V^2_0(\vec{q_1}) \Pi^0(q_1,q^0_1)} \tilde D}{{\epsilon_{\lambda}(q_1,q^0_1)}},
\label{E2b_correction}    
\end{eqnarray}
\end{widetext}
where  
\begin{widetext}
\begin{eqnarray}
 \tilde D \equiv  \sum_{\sigma =\pm} G^0_{\sigma}(\vec{q_1}+\vec{q_2},q_1^0+q_2^0) G^0_{\sigma}(\vec{q_2},q_2^0)G^0_{\sigma}(\vec{k}-\vec{q_1},k^0-q_1^0)G^0_{\sigma}(\vec{k},k^0),
\end{eqnarray}
\end{widetext}
and $\epsilon_{\lambda}(q_1,q^0_1)$ is the dielectric function given by Eq.~\ref{eq:dielectric} with the interaction coupling rescaled by $e^2\rightarrow \lambda e^2$. 

By means of a dimensional analysis of the contribution to the ground-state energy given by Eq.~\ref{E2b_Onsager}, it can be shown that it is independent of $r_s$, which implies that it is independent of the relative spin polarization $\zeta$. Its contribution is just a constant independent of $r_s$ and it has the same value as in the unpolarized case, which Onsager previously obtained and it is
given  by\cite{Onsager1966IntegralsIT}:
\begin{equation}
    E^0_{2b}[n_{\uparrow},n_{\downarrow}] = \frac{N e^2}{2 a_0}\epsilon^0_{2b},
\label{E2b_factor}    
\end{equation}
where the energy per particle is given in Ry as:
\begin{equation}
 \epsilon^0_{2b}=\frac{\ln(2)}{3}-\frac{3}{2 \pi^2} \zeta(3)\approx 0.04836.
\label{Onsager_value} 
\end{equation}

The correction to the kite diagram is the piece that contains the $\zeta$ and $r_s$ dependence and this Goldstone diagram cannot be computed accurately
by using the Monte-Carlo integration-technique directly by using the expression given by Eq.~\ref{E2b_correction} due to the so-called ``sign'' problem that comes at the level of the frequency integrals of the product of the Green's function. It is possible to get around this problem by isolating the momentum and frequency associated with the renormalized interaction line and doing the integrals of the remainder frequency variables, which the dielectric function is independent of, by using Cauchy's theorem. Also, as explained in Appendix~\ref{Calculation-of-kite}, to make the integrand more compact, we had to rely on a sequence of
transformations of the dummy wavevector integration variables which
helped us reduce the number of integrals to calculate numerically using the Monte Carlo integration technique. The integral also becomes easier to handle when
applying similar transformations that we did in the case of the ring diagram series (check appendix~\ref{Calculation-of-kite} for more details). We chose the same complex contour path on the complex plane (See Fig.~\ref{paths}) which avoids
enclosing the branch cuts of the dielectric function\cite{PhysRev.112.812} and the poles that come from the denominator part of the non-interacting Green's function, we can map the frequency integral contained in the argument of the dielectric function along the real line into an integral along the imaginary line. The last step is to do a dimensional analysis, as done for the ring-diagram series which allows us to factor out the total number of electrons in the correction of the kite-diagram series, which allows us to write this ground-state energy contribution in terms of an energy-per-particle factor:
\begin{equation}
  \Delta E_{2b}[n_{\uparrow},n_{\downarrow}] = \frac{N e^2}{2 a_0} \Delta \epsilon_{2b}[n_{\uparrow},n_{\downarrow}],
\label{kite_factor_2b_correction}    
\end{equation}
where the energy per particle of the correction of the kite diagram within renormalization is given by the sum of two major contributions which are given as:
\begin{widetext}
\begin{eqnarray}
\Delta \epsilon_{2b,1} =  \int_0^1 d\lambda \int d^3q_1 \int_{k \leq 1} d^3k \int_{q_2 \leq 1} d^3q_2\sum_{\sigma} \int_0^{{{\pi}\over 2}} du {\cal L}\frac{f^{\sigma}_{\lambda}\left(\vec{q_1},a(\vec{q_1},\vec{q_2})\tan(u),\zeta\right)-f^{\sigma}_{\lambda}\left(\vec{q_1},a(\vec{q_1},\vec{k})\tan(u),\zeta\right)}{|\vec{k}-\vec{q_2}|^2\left(a(\vec{q_1},\vec{q_2})-a(\vec{q_1},\vec{k})\right)}, 
\label{e2b_1} \\    
  \Delta \epsilon_{2b,2}= -\int_0^1 d\lambda \int d^3 q_1 \int_{k \leq 1} d^3{k} \int_{q_2 \leq 1} d^3{q_2}
  \sum_{\sigma} \int_0^{{{\pi}\over 2}} du {\cal L}\frac{f^{\sigma}_{\lambda}\left(\vec{q_1},a(\vec{q_1},\vec{q_2})\tan(u),\zeta \right)+f^{\sigma}_{\lambda}\left(\vec{q_1},a(\vec{q_1},\vec{k})\tan(u),\zeta \right)}{|\vec{k}+\vec{q_1}+\vec{q_2}|^2\left(a(\vec{q_1},\vec{q_2})+a(\vec{q_1},\vec{k})\right)},
\label{e2b_2}    
\end{eqnarray}
\end{widetext}
where
\begin{eqnarray}
 {\cal L} \equiv \frac{3 \alpha r_s}{16 \pi^7} \frac{\lambda^2}{q_1^4} \Theta\left(|\vec{q_1}+\vec{q_2}|-1\right) \Theta\left(|\vec{k}+\vec{q_1}|-1\right),
\label{L_mathcal} 
\end{eqnarray}    
and the term $a(\vec{q_1},\vec{q_2})$ is related with the relative differences between the energy dispersion of the electrons:
\begin{eqnarray}
  a(\vec{q_1},\vec{q_2}) = \frac{|\vec{q_1}+\vec{q_2}|^2-q_2^2}{2}.
\label{a_b}    
\end{eqnarray}
 
A last integration mapping had to be done to smooth out the frequency integrand to get more precise numerical results for the correction to the kite diagram $\Delta \epsilon_{2b}\left[n_{\uparrow},n_{\downarrow} \right]$. Such mapping is what yields an expression for the integrands in terms of the function $f_{\lambda,\sigma}\left(\vec{q_1},\nu,\zeta \right)$, which is defined as:
\begin{equation}
  f^{\sigma}_{\lambda}\left(\vec{q_1},\nu,\zeta \right) \equiv \sum_{\sigma'} \frac{x_{\sigma}^3 x_{\sigma '} g(\theta q_1, \theta^2 z)}{x_{\sigma}^2+\frac{\lambda \alpha r_s}{\pi q_1^2} \sum_{\sigma''} x_{\sigma''} g(\theta_{{2}} q_1, \theta_{{2}}^2 z) },
\label{f}    
\end{equation}
where $z=i \nu$, $\theta\equiv x_{\sigma}/x_{\sigma '}$ and $\theta_2 \equiv x_{\sigma}/x_{\sigma''}$.
The expressions given by Eq.~\ref{e2b_1} and Eq.~\ref{e2b_2} are the definite expressions that we used to do the Monte Carlo integration. The number of sample points used was $N_b=10^9$ to reduce the statistical uncertainty given by the stochastic method used. The data that was obtained was later combined along with the numerical data obtained for the ring diagram series and Onsager's value $\epsilon^0_{2b}$ to obtain the numerical data for the correlation energy $\epsilon_c(r_s,\zeta)$ for several values of $\zeta$ and $r_s$. This was important to obtain an interpolation function for the coefficients of the expansion in $r_s$ of the correlation energy.

\subsection{Checking our Calculation}
\label{check}

Our calculations of the ring-diagram series agree well with Hedin's reported
results within his large error bars. However,  our results are of much higher precision as  we have more powerful
computational resources  compared to  those available at the time of
Hedin's paper\cite{PhysRev.139.A796}.
 
We have verified the results of the full-kite diagram series using various numerical checks discussed in this subsection. In addition, we used analytical results obtained 
at certain extreme limits to check that the results of our
Monte Carlo calculations
are correct.

  
First, to verify our numerical results from our Monte Carlo code, we asked another member of our research group (who joined our research group briefly in the summer of 2024,
named in the acknowledgments) to independently
calculate the full-kite diagram by using Eqs.~\ref{e2b_1} and~\ref{e2b_2}
and by writing a different Monte Carlo code. The results agree with ours within the Monte Carlo error. 

We have also derived a different expression for the kite-diagram-series
of $\epsilon_{2b}(r_s,\zeta)$ starting from the full expression without separating  the
contribution  of the Onsager-kite (i.e., the $r_s \rightarrow 0$ limit) and 
  carrying out the frequency integral by avoiding the branch cuts from the dielectric function to obtain the different expression in the same way as for the correction to the full-kite diagram series explained in Appendix~\ref{Calculation-of-kite}. The new expression of the full-kite diagram series that we found is given by:
\begin{equation}
  \epsilon_{2b}(rs,\zeta) = \epsilon^1_{2b}(rs,\zeta) + \epsilon^2_{2b}(r_s,\zeta),
\label{e2b_another}    
\end{equation}
where the expressions for $\epsilon^1_{2b}(rs,\zeta)$ and $\epsilon^2_{2b}(rs,\zeta)$ are given by:
\begin{widetext}
\begin{eqnarray}
\epsilon_{2b,1} = -\int_0^1 d\lambda \int d^3q_1 \int_{k \leq 1} d^3k \int_{q_2 \leq 1} d^3q_2\sum_{\sigma} \int_0^{{{\pi}\over 2}} du {\cal A}\frac{M^{\sigma}_{\lambda}\left(\vec{q_1},a(\vec{q_1},\vec{q_2})\tan(u),\zeta\right)-M^{\sigma}_{\lambda}\left(\vec{q_1},a(\vec{q_1},\vec{k})\tan(u),\zeta\right)}{|\vec{k}-\vec{q_2}|^2\left(a(\vec{q_1},\vec{q_2})-a(\vec{q_1},\vec{k})\right)}, 
\label{e2b_1_full} \\    
\epsilon_{2b,2}= \int_0^1 d\lambda \int d^3 q_1 \int_{k \leq 1} d^3{k} \int_{q_2 \leq 1} d^3{q_2}
  \sum_{\sigma} \int_0^{{{\pi}\over 2}} du {\cal A}\frac{M^{\sigma}_{\lambda}\left(\vec{q_1},a(\vec{q_1},\vec{q_2})\tan(u),\zeta \right)+M^{\sigma}_{\lambda}\left(\vec{q_1},a(\vec{q_1},\vec{k})\tan(u),\zeta \right)}{|\vec{k}+\vec{q_1}+\vec{q_2}|^2\left(a(\vec{q_1},\vec{q_2})+a(\vec{q_1},\vec{k})\right)},
\label{e2b_2_full}    
\end{eqnarray}
\end{widetext}
where ${\cal A}=({\cal L}\pi q_1^2)/( \lambda \alpha r_s)$, ${\cal L}$ was given in Eq.~\ref{L_mathcal}, while the ${\cal M}^{\sigma}_{\lambda}(\vec{q_1},\nu,\zeta)$ is expressed as:
\begin{equation}
  M^{\sigma}_{\lambda}\left(\vec{q_1},\nu,\zeta \right) \equiv \frac{x_{\sigma}^5}{x_{\sigma}^2+\frac{\lambda \alpha r_s}{\pi q_1^2} \sum_{\sigma'} x_{\sigma''} g(\theta_{{2}} q_1, \theta_{{2}}^2 z) },
\label{M_for_e2b_full}    
\end{equation}
where $z=i \nu$ and $\theta_2 \equiv x_{\sigma}/x_{\sigma''}$. The numerical results that we have obtained by using Eqs.~\ref{e2b_1_full} and~\ref{e2b_2_full} agree within the Monte Carlo error for various values of $\zeta$ and $r_s$ with the results obtained as explained in the previous section. We have also used
these expressions to calculate the kite-diagram series at $r_s=0$ and
they yield Onsager's value within the $1\%$ error. In addition, we have worked on a code that uses the importance sampling algorithm with the goal of reducing our current numerical error by using the Monte Carlo algorithm, and it yields similar values for $\epsilon_{2b}(r_s,\zeta)$ reported in Table~\ref{table_kite_bare} within the error bars.

A consistency check of our analytical expressions for $\epsilon_{2b}(r_s,\zeta)$  is obtained by using Onsager's value $\epsilon^0_{2b}$ and both expressions from Eqs. \ref{e2b_1}, \ref{e2b_2}, and calculating the limits of $r_s \rightarrow 0$ and $r_s \rightarrow \infty$. We have also done an internal
consistency-check using a different expression for the full-kite diagram series given by Eqs.~\ref{e2b_another},~\ref{e2b_1_full} and~\ref{e2b_2_full} for these two limits. 

Our analytical result for the correction to the full-kite diagram for every value of partial spin-polarization $\zeta$, obtained from the sum of the two terms from Eq.~\ref{e2b_1} and Eq.~\ref{e2b_2}, also has the correct asymptote at $r_s \rightarrow \infty$. In this limit, the function $f_{\lambda}^{\sigma}(\vec{q},\nu,\zeta)$ from Eq.~\ref{f} yields the following expression:
\begin{equation}
f_{\lambda}^{\sigma}(\vec{q_1},\nu,\zeta) \overset{r_s \rightarrow \infty}{\approx} \frac{\pi q_1^2 (1+\sigma \zeta)}{\lambda \alpha r_s},
\label{f_large_rs}    
\end{equation}
where the sum over the spin variable $\sigma$ on this function is independent of $\zeta$.
By inserting Eq.~\ref{f_large_rs} in Eqs.~\ref{e2b_1} and~\ref{e2b_2}, we find that $\Delta \epsilon_{2b,1}$ yields zero, while $\Delta \epsilon_{2b,2}$ is the surviving term that yields the same expression as Onsager's integral expression of $\epsilon^0_{2b}$, but with the opposite sign. In the limit $r_s \rightarrow \infty$, we have:
\begin{equation}
    \Delta \epsilon_{2b,2} \overset{r_s \rightarrow \infty}{\approx} -\epsilon^0_{2b},
\end{equation}
where this limit implies that the full-kite diagram yields zero in the limit $r_s \rightarrow \infty$ when Onsager's result is added.
  
It is trivial to check that both corrections to the full-kite diagram series, at $r_s=0$, are also zero since both terms from Eqs.~\ref{e2b_1} and~\ref{e2b_2} have a $r_s$ global factor. The expression from Eq.~\ref{M_for_e2b_full} is equal to $x_{\sigma}^3$ at $r_s=0$, which is substituted in both expressions Eqs.~\ref{e2b_1_full} and~\ref{e2b_2_full}. After doing the substitution, only the term $\epsilon_{2b,1}$ is equal to zero, while $\epsilon_{2b,2}$ yields the Onsager's integral expression\cite{Onsager1966IntegralsIT} after doing the integral over the variable $u$ and the sum over the spin $\sigma$.


We also extracted the coefficient of $r_s \ln{r_s}$ in the small $r_s$ limit
to the value reported analytically\cite{loos2011correlation} (a value of 0.01304 Ry) as follows. We fit the full-kite diagram Monte Carlo data from Table \ref{table_kite_bare} to the following small $r_{s}$ expansion
\begin{equation}
\epsilon_{2b}(r_s) = \epsilon^0_{2b} + C_1 r_s \ln{r_s} + C_2 r_s + C_3 r_s^2 \ln(r_s).
\label{fit_kite_sosex}    
\end{equation}
The results of the fit are given in Table~\ref{sosex_table}. Notice that
the coefficient $C_1$ is in reasonable agreement with the exact
value\cite{loos2011correlation}. As we will discuss in the following subsection
and in Sec.~\ref{comparison-sosex} that SOSEX and PW give a much different (by about a
factor of 2) value for this coefficient. We feel that this is a strong
test that our calculation of the kite-diagram series was done
correctly.


\begin{table}
  \begin{center}
    \vskip 0.2 in
\begin{tabular}{|c|c|c|c|} 
\hline
& $C_{1}$ & $C_{2}$ & $C_{3}$\\
\hline
$\epsilon_{2b}$ & 0.01212 & -0.02083 & -0.00562 \\
\hline
\end{tabular}
\caption{Coefficients of $\epsilon_{2b}$ fitted to Eq.~\ref{fit_kite_sosex} in Ry. The coefficient $C_{2}$ for $\epsilon_{2b}$ is close to the coefficient reported in Ref.~\onlinecite{loos2011correlation}. The fit was done for values of $r_{s}$ from 0 to 1.} 
\label{sosex_table}
\end{center}
\end{table}

\subsection{Comparison of the contributions}

\begin{figure}[htp]
   \begin{center}
     \includegraphics[scale=0.35]{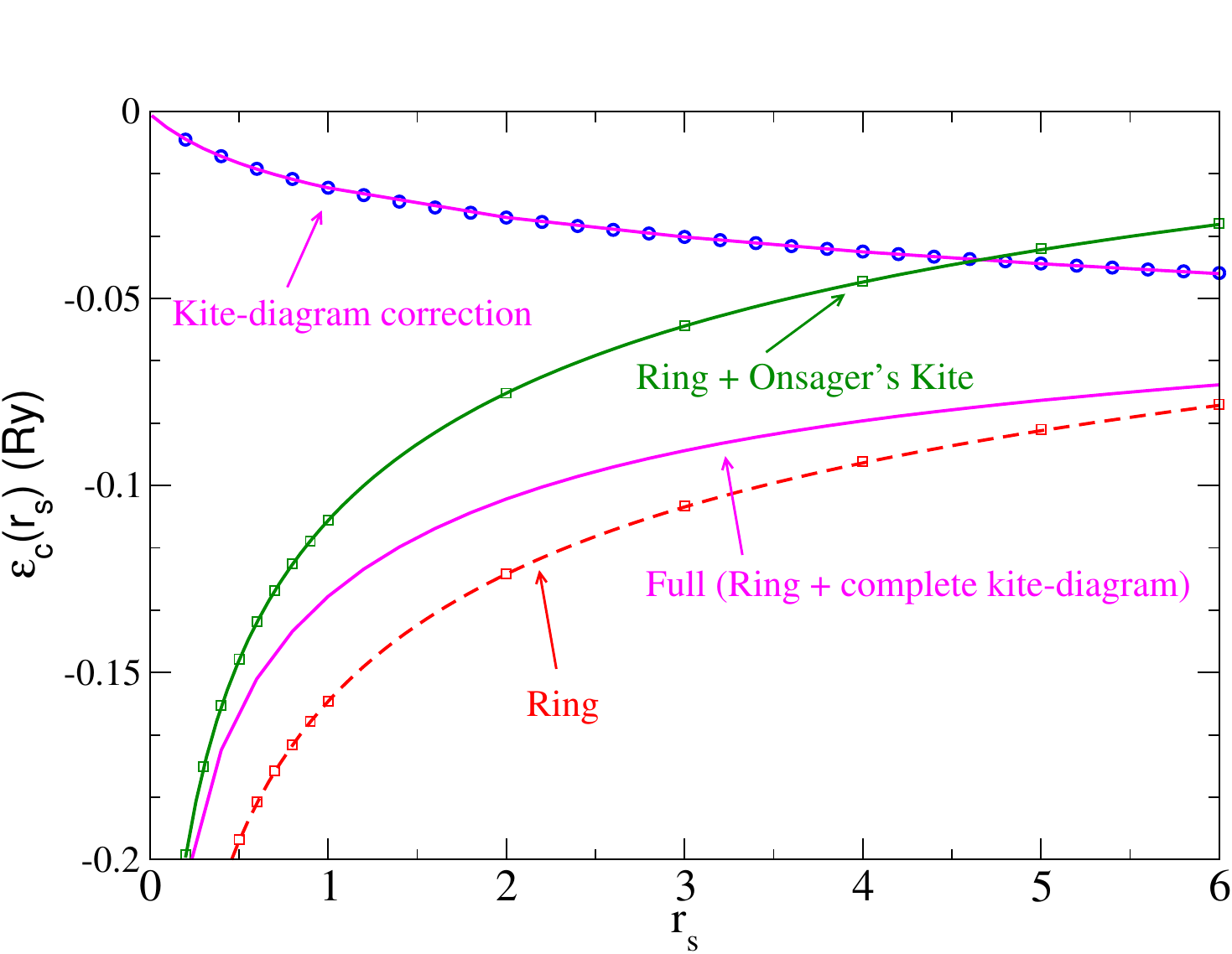}
     \caption{The contribution of the various terms to the
       correlation energy as a function of $r_s$.}
     \label{comparison-of-terms}
    \end{center}
    \end{figure}

Here we wish to compare the various contributions to the correlation energy
at the leading order of the RPA renormalized interaction.
Namely, the sum of the ring diagrams, the simple contribution of
the bare-kite diagram calculated by Onsager and the full
kite diagram as calculated here numerically.

The various levels of approximation and the relative significance of
 each term are demonstrated in Fig.~\ref{comparison-of-terms} as a function of $r_s$ for $r_s<6$.
 The correlation energy obtained
 by including only the ring-diagrams  is shown by the red-dashed line.
 Adding
       the $r_s$-independent contribution of the kite-diagram,
       as calculated by Onsager exactly, to the sum of the ring diagrams
       gives rise to the solid green solid-line. The full calculation
       is shown by the purple solid-line where we have added the $r_s$ dependent correction to the kite (open-circles).

\section{Our functional}
\label{our-functional}
In this section we present our functional form obtained
by fitting the results discussed in the previous Sections
to analytic forms for a straightforward inclusion in the
present implementations of the DFT. Because our results
are based on a order by order expansion in the number of RPA-renormalized
interaction lines, we name this functional and will refer to it
in the future as RPA based Functional (RPAF).

The correlation energy per particle $\epsilon_c(r_s,\zeta)$ calculated within the RPA is given by the sum of the two Goldstone diagrams within the first order of the reorganized
perturbative expansion: $\epsilon_{2b}(r_s,\zeta)$ and $\epsilon_r(r_s,\zeta)$.

The data for the ring diagram series per particle $\epsilon_r(r_s,\zeta)$ have been obtained by integration by adaptive quadrature, which yields a more precise calculation in contrast to the numerical results of the integrals of the correction of the  kite-diagram series within the RPA, which was obtained with Monte Carlo integration technique and it is explained in Appendix~\ref{Calculation-of-kite}.

It is very important to have an analytic expression for the
coefficient of the $\ln(r_s)$ for any value of $\zeta$
because it diverges for $r_s \to 0$ and it continues to be a large part of
the functional for values of $r_s$ in the physically realizable region.
This is done in Appendix~\ref{logarithm} briefly as follows:
Each of these diagrams has one renormalized interaction line which sums up the
series of all the polarization bubbles and one bare interaction line.
Due to this, if we use the expression of the integrand for each diagram that is being integrated along the imaginary frequency $i \nu$, one can separate the regions of the integration as we did for the calculation of the $c_{L}(\zeta)$ in Appendix~\ref{logarithm}. By doing this, we can keep track of the coefficients of the expansion in $r_s$ of $\epsilon_r(r_s,\zeta)$. Since the diagram has one renormalized interaction line, this gives a dependence on $r_s$ in the integrand inside a logarithm of the dielectric function.

\vskip 0.8 in

\subsection{Fit Equations for the Ring Diagram}

\subsubsection{Small r$_s$ limit}

 \begin{table}[htp]
  \vskip 0.2 in
\begin{center}
\begin{tabular}{|c|c|c|c|c|} 
\hline
$\zeta$&$c_{0}$ (Ry)&$e_{1}$ (Ry)&$a_{2}$&$b_{2}$\\
\hline
0.00 & -0.1423 & 0.8822 &  90.76 &  54.55 \\
\hline
0.02 & -0.1408 & 0.8893 &  91.67 &  54.64 \\
\hline
0.40 & -0.1365 & 0.9111 &  95.14 &  55.38 \\
\hline
0.60 & -0.1287 & 0.9488 & 104.51 &  58.17 \\
\hline
0.80 & -0.1165 & 1.0055 & 132.21 &  68.05 \\
\hline
0.90 & -0.1079 & 1.0431 & 168.80 &  82.18 \\
\hline
0.91 & -0.1069 & 1.0474 & 174.99 &  84.60 \\
\hline
0.92 & -0.1059 & 1.0517 & 182.03 &  87.39 \\
\hline
0.93 & -0.1049 & 1.0562 & 190.14 &  90.62 \\
\hline
0.94 & -0.1038 & 1.0608 & 199.66 &  94.45 \\
\hline
0.95 & -0.1027 & 1.0656 & 211.07 &  99.04 \\
\hline
0.96 & -0.1016 & 1.0705 & 225.43 & 104.97 \\
\hline
0.97 & -0.1004 & 1.0755 & 244.48 & 112.94 \\
\hline
0.98 & -0.0993 & 1.0808 & 272.55 & 124.94 \\
\hline
0.99 & -0.0983 & 1.0864 & 323.50 & 147.28 \\
\hline
1.00 & -0.0998 & 1.0925 & 593.52 & 283.05 \\
\hline
\end{tabular}
\caption{First (Second) column: The values of the coefficient $c_0$ ($e_1$) obtained by fitting the
  small (large) $r_s$ part of our data to the form Eq.~\ref{eq:small-rs-fit}
  (Eq.~\ref{eq:large-rs-fit})  for various values of $\zeta$. Third and fourth column gives
  the values of the coefficients $a_2$ and $b_2$ 
  entering the functional form for $\epsilon_r(r_s,\zeta)$ (Eq.~\ref{ring-fit-form}) obtained by fitting our data to this form under the constraints
  (Eqs.~\ref{eq:constr1},\ref{eq:constr2},\ref{eq:constr3},\ref{eq:constr4}). See text for details.}
	\label{table:5}
\end{center}
  \vskip 0.2 in
 \end{table}
 
For a given value of
$\zeta$, we fit the data as a function of $r_s$ in a small region of $r_s$ ($0<r_s<1$) to the function
\begin{eqnarray}
  \epsilon_r(r_s,\zeta)= c_0 + c_L \ln(r_s) + c_1 r_s + c_2 r_s \ln(r_s),
  \label{eq:small-rs-fit}
\end{eqnarray}
 using the exact coefficient
$c_L$ as a function of $\zeta$.
This fit yields the values of the coefficient $c_0(\zeta)$ given in the
first column of Table~\ref{table:5} for the values of $\zeta$ for which
we have calculated diagram. We are not going to use the values of the
other coefficients obtained this way in our functional; these other
coefficients were only necessary in order to extract the correct value of $c_0$.
As explained in the following, we are only going to force the contribution of the ring-series to our functional to have the correct  $c_0$ and $c_L$ coefficients in the $r_s \to 0$ limit.

\subsubsection{large r$_s$ limit}

  For the large $r_s$ limit ($100<r_s<1000000$), we have found that the numerical results
for $\epsilon_r(r_s,\zeta)$ can be very accurately fit to the form
\begin{eqnarray}
  \epsilon_r(r_s,\zeta) = {{e_0} \over {r_s^{\frac{3}{4}}}} +
          {{e_1} \over {r_s}}.\label{eq:large-rs-fit}
\end{eqnarray}
We know that at very large $r_s$ values, when the $1/r_s^{3/4}$ term
is the dominant term\cite{PhysRevB.44.13298}, the coefficient $e_0$ should be
$e_0=-0.803$ Ry as calculated
{\it exactly} in Appendix~\ref{large-rs}. It was also found to be $\zeta$
independent.  
We have  verified that,
 when we fit the results for $r_s > 1000$,  $e_0$ approaches the calculated
 value. Therefore,  we adopt this value of $e_0$ for our functional
 for all values of $\zeta$.
The values of $e_1$ found by fitting
this large $r_s$ ($r_s>100$) behavior is given
in the second column of Table~\ref{table:5} for
data corresponding to the calculated
values of $\zeta$.

  \subsubsection{Our functional for all values of r$_s$}

  We will need a compact functional form to describe our data for the
  series of the ring diagrams which satisfy the above discussed small
  $r_s$ and large $r_s$ behavior, and at the same time it describes
  accurately our numerical results in the entire region, especially the
  region of $r_s$ realized in the real materials.
  We found that the following form accomplishes these requirements.
\begin{widetext}
  \begin{eqnarray}
  \epsilon_r(r_s,\zeta) = (a_0+a_1 r_s) \ln\Bigl (1 + {{a_2} \over {r^2_s}}\Bigr )
  + (b_0+b_1 r_s) \ln\Bigl (1 + {{b_2} \over {r^{7/4}_s}}\Bigr ),
           \label{ring-fit-form}
  \end{eqnarray}
\end{widetext}
where the coefficients $a_n=a_n(\zeta),b_n=b_n(\zeta)$ ($n=1,2,3$) are
functions of $\zeta$.

  The small $r_s$ limit  given by Eq.~\ref{ring_small-rs} imposes the following constraint on the coefficients:
  \begin{eqnarray}
    a_0 \ln(a_2) + b_0 \ln(b_2) = c_0, \label{lowrs1}\\
    2a_0 + {\frac{7}{4}} b_0 = -c_L, \label{lowrs2}
  \end{eqnarray}
  where, for simplicity, we used the notation $c_0$ simply for $c_0(\zeta)$ and
  $c_L$ for $c_L(\zeta)$.
  There are no spurious terms, such as ${\sim} \sqrt{r_s}$, which exist in the
  PBE functional in the small $r_s$ limit. The $\zeta$ dependence of $c_L$ is analytically known (Eq.~\ref{c_log}).
   \begin{table}
\begin{tabular}{|c|c|c|c|} 
    \hline
    $n$ & 0 & 1 & 2\\ [0.5ex] 
    \hline
$c_{0n}$ & -0.1423 & 0.0036 & 0 \\
\hline
${\bar c}_{0n}$ & 0.1971 & -0.0326 & -0.0177\\
    \hline
\end{tabular}
\caption{Coefficients for $c_{0}$ entering in Eq.~\ref{c0zeta2} (in Ry).}
\label{tablec0}
   \end{table}
\begin{table}
\begin{tabular}{|c|c| c | c|} 
    \hline
    $n$ & 1 & 2 \\ [0.5ex] 
    \hline
$e_{1n}$  & 0.1648 & 0.0432 \\
\hline
$a_{2n}$  & 192.62 & -3956.38 \\
\hline
$b_{2n}$  & 149.46 & -2070.06 \\
    \hline
\end{tabular}
\caption{Coefficients for $e_{1}$ (in Ry), $a_{2}$, and $b_{2}$ entering in Eq.~\ref{a2b2zeta}.}
\label{tablee1a2b2}
\end{table}

     \begin{figure*}[htp]
       \begin{center}
         \subfigure[]{
            \includegraphics[scale=0.3]{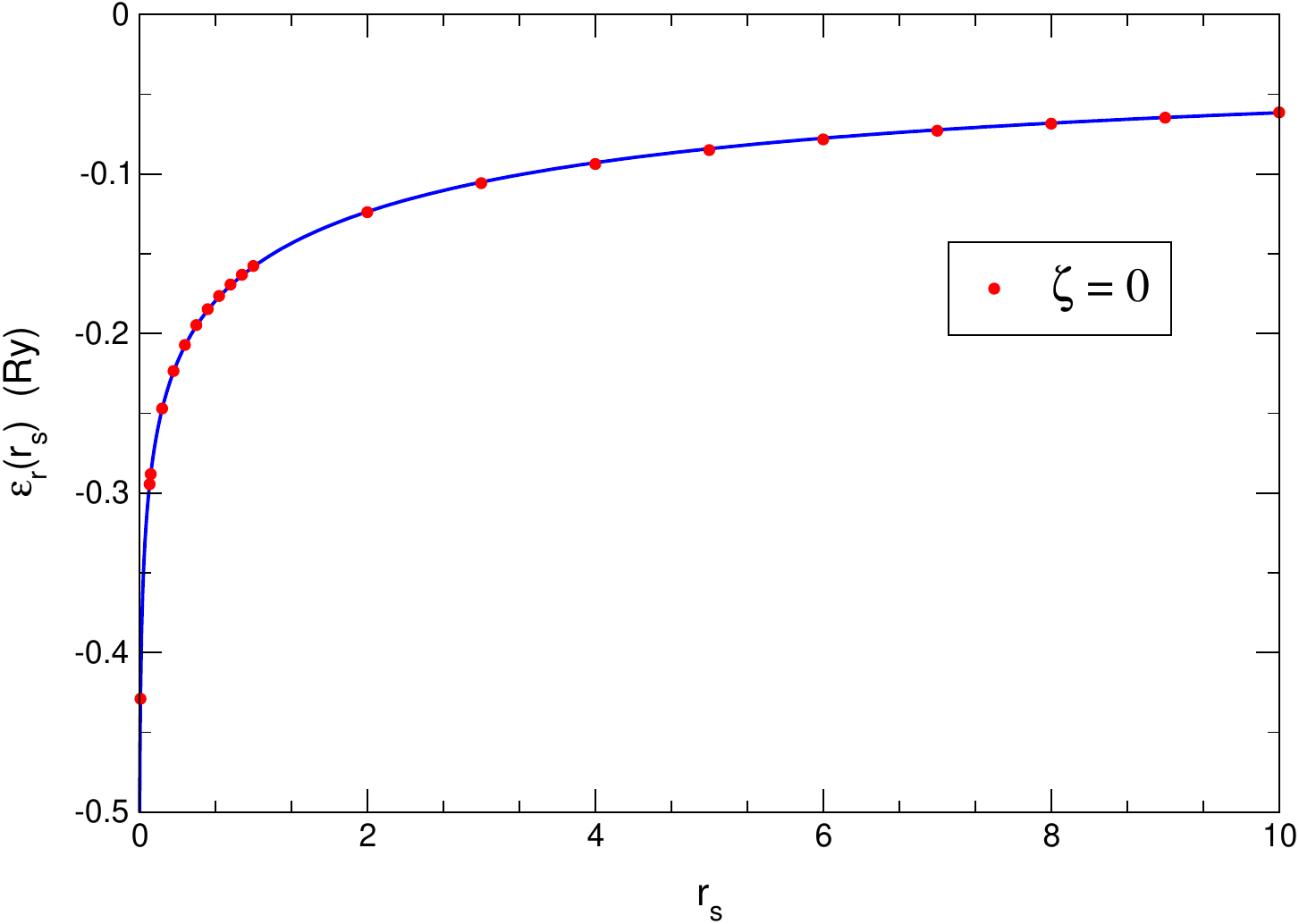} 
   \label{fit-ring-small}
         }
         \subfigure[]{
               \includegraphics[scale=0.3]{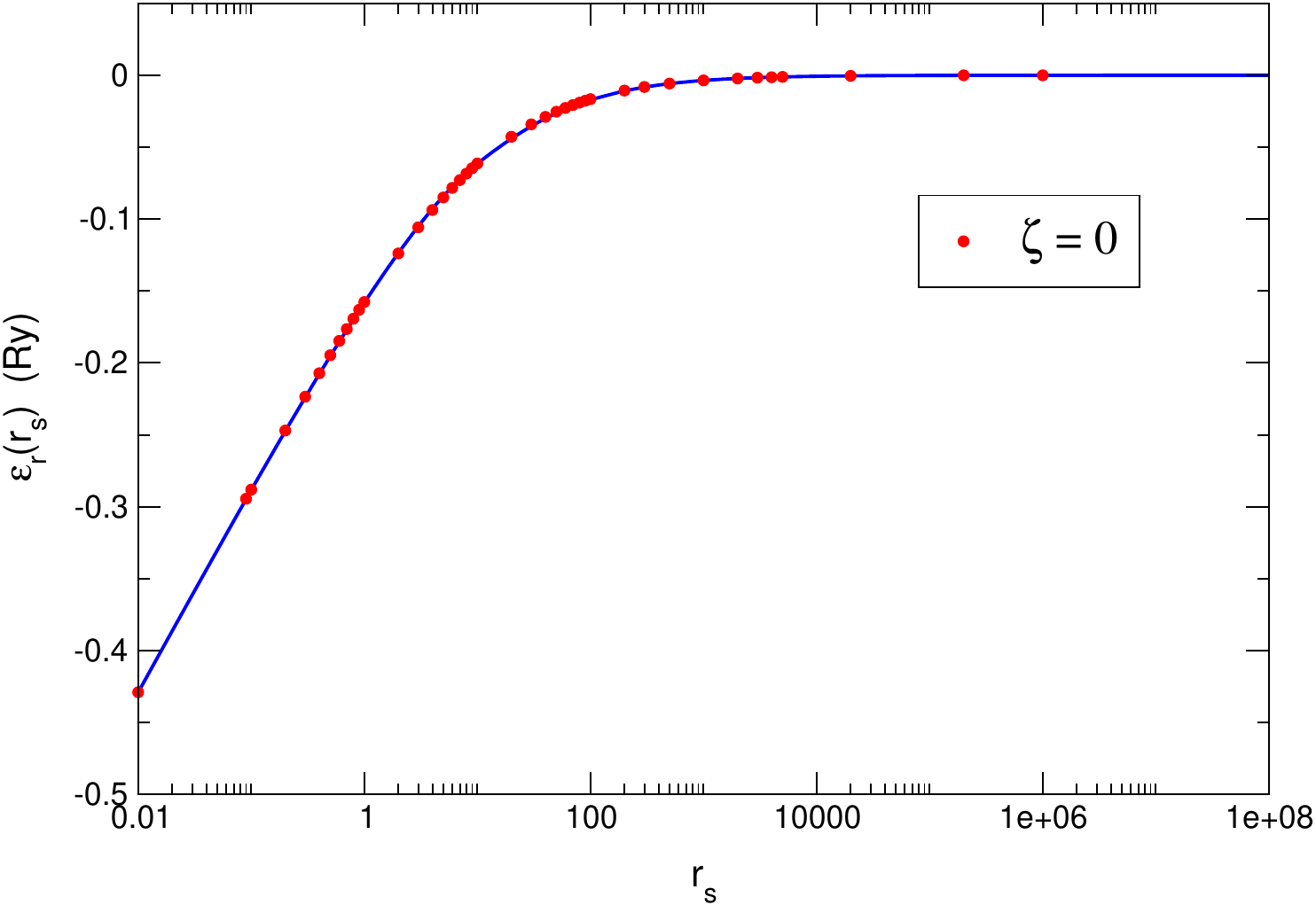} 
   \label{fit-ring-all}
         }
   \end{center}
   \caption{Fit of the form given by Eq.~\ref{ring-fit-form} of our numerical results by numerical integration of the expression given by
         Eq.~\ref{ring4} for the sum of the
         ring-diagrams. (a) Small $r_s$ region. (b) the entire region.}
       \end{figure*}

The next step of our fitting procedure,  we fit our results
to the functional form given by Eq.~\ref{ring-fit-form} in the range $0<r_s<100$ with the constraints given by Eqs.~\ref{lowrs1},\ref{lowrs2} and the following
additional constraints:
\begin{eqnarray}
  a_1 a_2 = e_1,\label{rargers1}\\
  b_1 b_2 = e_0,\label{largers2}
\end{eqnarray}
where $e_1$ and $e_0$ are the known coefficients already discussed and
$e_0=-0.803$ and $e_1$ is 
tabulated in the second column of Table~\ref{table:5}.

For any given value of $\zeta$ we fit the data to the form
given by Eq.~\ref{ring-fit-form}. There are 6 unknowns and the 4  equations.
The other parameters can be expressed in terms of
$a_2$ and $b_2$ by solving the above 4 equations. We find
\begin{eqnarray}
  a_0 &=& - {1 \over 2} \Bigl [ c_L + {7 \over 4} {{2 c_0 + c_L \ln(a_2)}
      \over {2 \ln(b_2) - {7 \over 4} \ln(a_2)}} \Bigr ],\label{eq:constr1}\\
  b_0 &=&   {{2 c_0 + c_L \ln(a_2)}
    \over {2 \ln(b_2) - {7 \over 4} \ln(a_2)}},\label{eq:constr2}\\
  a_1 &=& {{e_1} \over {a_2}},\label{eq:constr3}\\
  b_1 &=& {{e_0} \over {b_2}}.\label{eq:constr4}
\end{eqnarray}
Using these expressions for the coefficients $a_0,a_1,b_0,b_1$ in terms
of $a_2$ and $b_2$, we can do a two-parameter fit for each calculated value
of $\zeta$.
Namely, the same procedure is repeated for any given value of $\zeta$ and the
the results of the fits are given in Table~\ref{table:5}.

\begin{figure}
  \vskip 0.2 in
   \begin{center}
\includegraphics[scale=0.3]{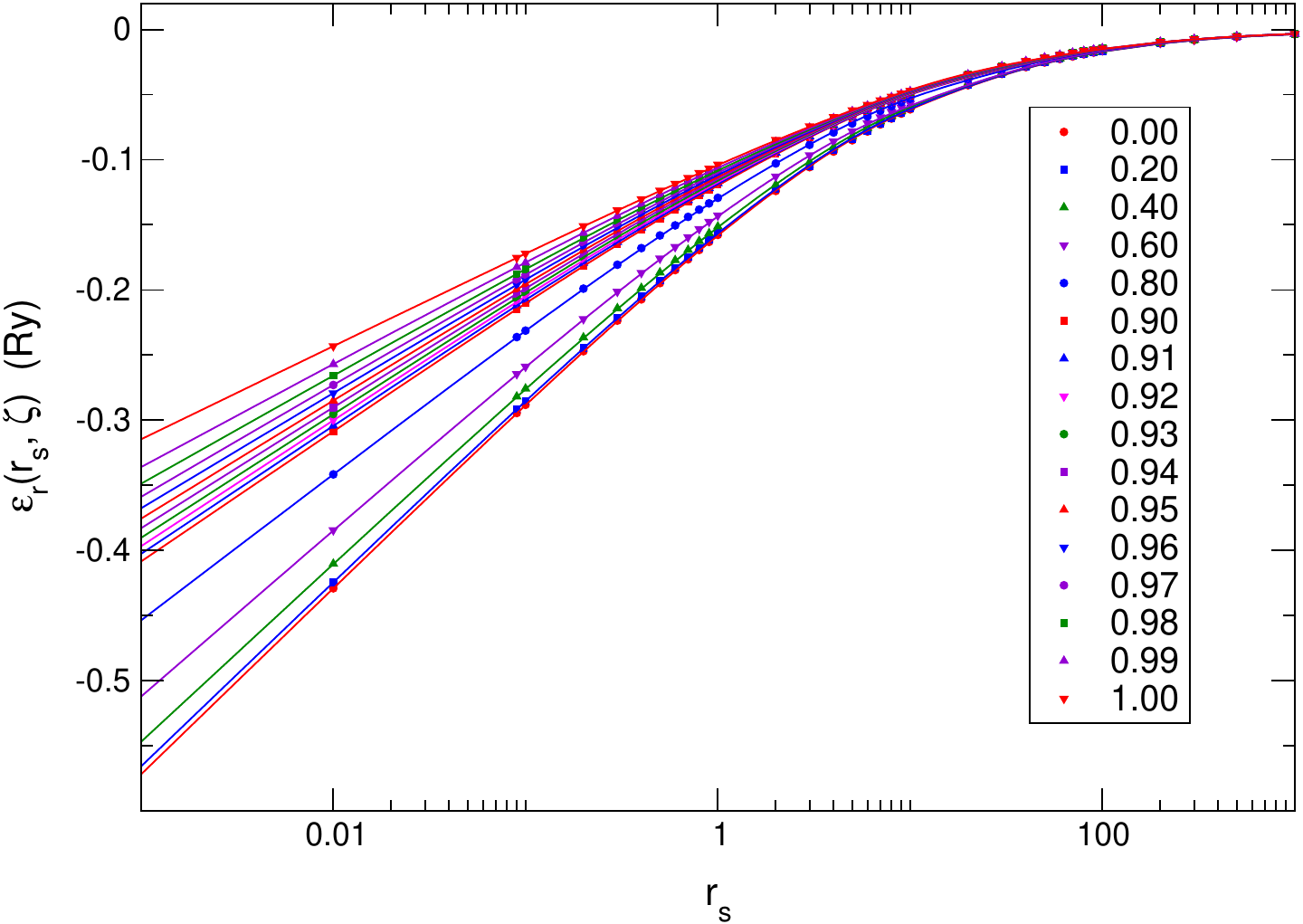}
    \end{center}
   \caption{The data points are the ring diagram data from Table \ref{table_ring_bare} for different values of $\zeta$ and the solid lines are the fit of the ring diagram
     using the functional given by Eq.~\ref{ring-fit-form}.}
   \label{ring-all-zeta}
  \vskip 0.2 in
    \end{figure}

The quality of the
fit to the formula given by Eq.~\ref{ring-fit-form} for $\zeta=0$ is
illustrated in Fig.~\ref{fit-ring-small} for small values of $r_s$
and in Fig.~\ref{fit-ring-all} for the whole range of $r_s$.
For all other values of $\zeta$ the fits are shown in Fig.~\ref{ring-all-zeta}.

  \subsubsection{Functional dependence on spin-polarization $\zeta$}

\begin{figure}[htp]
   \begin{center}
     \includegraphics[scale=0.35]{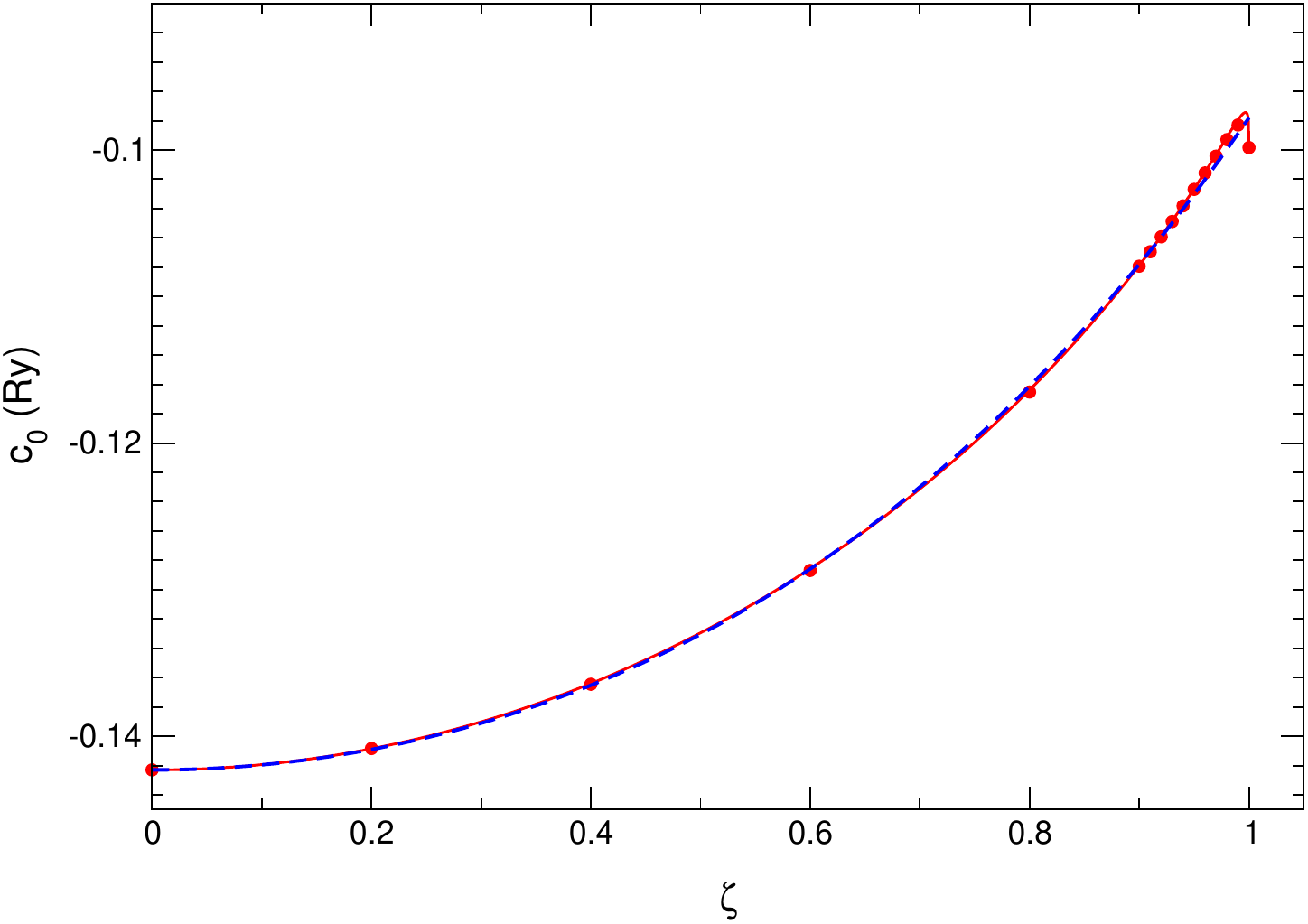}
     \caption{$\zeta$ dependence of the constant $c_0$ of
       the functional.}
     \label{anomaly}
    \end{center}
    \end{figure}
Now, we wish to find interpolation formulas to describe the
    $\zeta$ dependence of the coefficients in Table~\ref{table:5}.
     First, the values of the constant $c_0(\zeta)$ is plot
     in Fig.~\ref{anomaly}. Notice that the form
  \begin{eqnarray}
    c_0(\zeta) = \sum_{n=0}^2 c_{0n} \zeta^{2n}, \label{c0zeta}
  \end{eqnarray}
  fits the data very well with exception of the region near $\zeta=1$
  as illustrated in Fig.~\ref{anomaly} by the blue dashed-line.
  This behavior near $\zeta=1$ was first discussed in
  Ref.~\onlinecite{PhysRevB.45.8730}.
  We were able to obtain a reasonable
  fit in the whole region of $\zeta$ using the following interpolation
  formula:
  \begin{eqnarray}
    c_0(\zeta) = \sum_{n=0}^1 c_{0n} \zeta^{2n} + \sum_{n=1}^3
    {\bar c}_{0n} (\chi^{n}-2^n), \label{c0zeta2}
  \end{eqnarray}
  where $\chi(\zeta)$ is given by Eq.~\ref{chi}. The reason for the $2^n$ is because $\chi(0)=2$, i.e., 
  we forced the fit to the
  constraint that the value of $c_0(\zeta=0)$ should be that of Table~\ref{table:5}, namely, $c_{00}=c_0(0)$.
  This form captures the singular behavior
  near at $\zeta=1$ by the fact that the variable $\chi$ has
  a singular dependence on $\zeta$ near $\zeta=1$.
  The fit is very good and it is given by
  the red solid-line in Fig.~\ref{anomaly}. 
  The coefficients $c_{0n}$ ($n=0,1$) and ${\bar c}_{0n}$ ($n=1,2,3$) are given in  Table~\ref{tablec0}.

  It is well-known\cite{PhysRev.140.A1645} that there is an additional constraint linking 
  $\epsilon_r(r_s,\zeta)$ for $\zeta=0$ and $\zeta=1$:
  \begin{eqnarray}
    \epsilon_r(r_s,1) = {1 \over 2} \epsilon_r(r_s',0),
    \label{misawa1}
  \end{eqnarray}
  where $r_s'$ is the rescaled Wigner-Seitz radius given by $r_s'=r_s/2^{4/3}$. We did not impose this constraint, because we found that our results,
  given in the Appendix~\ref{tables} for the ring-series, satisfy this
  relationship for all values of $r_s$ within error bars.
  The above expression, which leads to the following relationship
  between $c_0(0)$ and $c_0(1)$:
  \begin{eqnarray}
    c_0(1)  = {1 \over 2} c_0(0) - {4 \over {3 \pi^2}} \ln(2) (1-\ln(2)),
    \label{misawa2}
  \end{eqnarray}
  is found to be satisfied within the errors of the
  fitting procedures (a)  in Table~\ref{table:5}, where the values
  listed were found by fitting to the small $r_s$ formula given by Eq.~\ref{eq:small-rs-fit} and (b) in the results of the fitting procedure
  to the expression given by Eq.~\ref{c0zeta2}, where Eq.~\ref{misawa2}
  was not imposed.

\begin{figure}
   \begin{center}
     \includegraphics[scale=0.3]{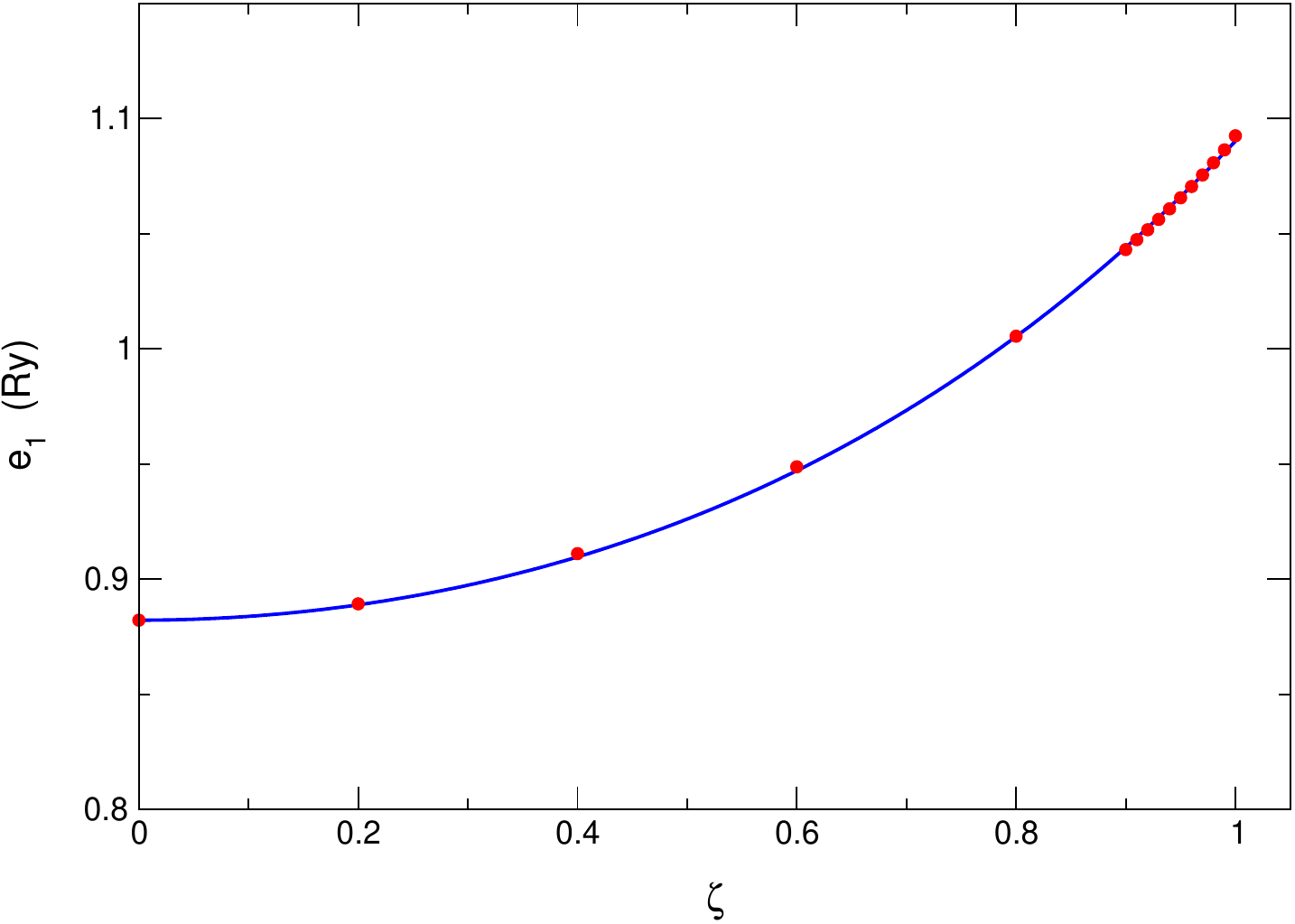} \\
     \vskip 0.2 in
\includegraphics[scale=0.3]{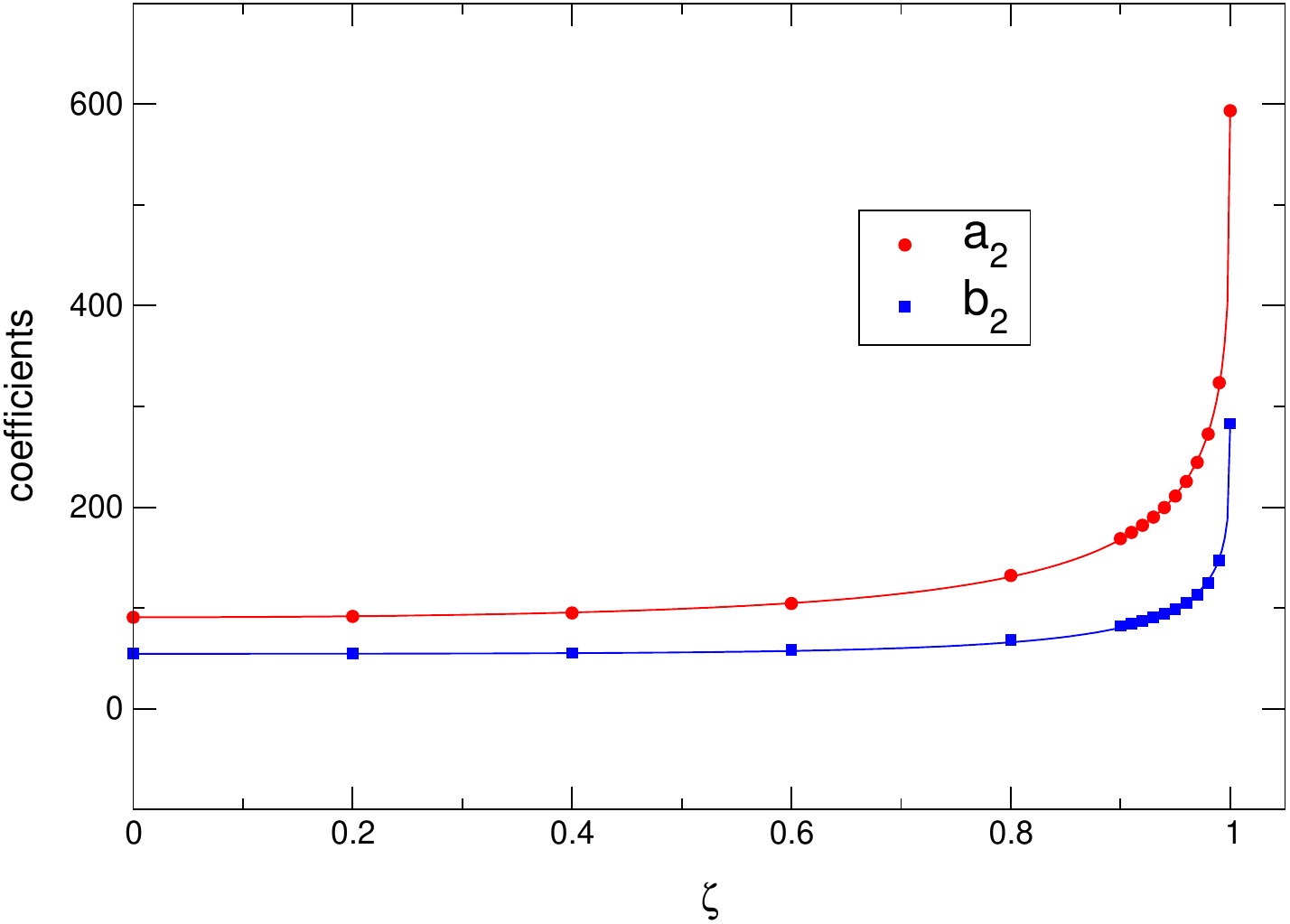}
    \end{center}
   \caption{The data points are the values of Table \ref{table:5} and the solid lines are the coefficients of 
     Table~\ref{tablee1a2b2}  fitted to the form Eq.~\ref{e1-fit}
     for $e_{1}$ (Top) and Eq.~\ref{a2b2zeta} for $a_{2}$ and $b_{2}$ (bottom).}
   \label{fit-e1-a2-b2}
\end{figure}

We used the form, 
  \begin{eqnarray}
    e_1(\zeta) = \sum_{n=0}^2 e_{1n} \zeta^{2n},\label{e1-fit}
  \end{eqnarray}
to parametrize the $\zeta$ dependence of $e_1$.
We forced the fit to the
constraint that the value of $e_1(\zeta=0)$ is that of Table~\ref{table:5},
i.e., $e_{10}=e_1(0)$.
The other coefficients $e_{1n}$  are given in the second row of
Table~\ref{tablee1a2b2} and a graph illustrating the quality of the
fit is given in Fig.~\ref{fit-e1-a2-b2}.

The values of the coefficients $a_2$ and $b_2$ as a function of $\zeta$ are
fit to the form 
\begin{eqnarray}
  a_2 &=&a_{20} + a_{21}(\chi-2) + a_{22}\Bigl ({{\ln(\chi)} \over {\chi}}-{{\ln(2)} \over 2}\Bigr ), \\
  b_2 &=&b_{20} + b_{21}(\chi-2) + b_{22}\Bigl ({{\ln(\chi)} \over {\chi}}-{{\ln(2)} \over 2}\Bigr ).
  \label{a2b2zeta}
\end{eqnarray}
Again we forced the fit to the
constraint that  $a_{20}=a_2(\zeta=0)$ and $b_{20}=b_2(\zeta=0)$ given in
Table~\ref{table:5} determined earlier.
The other coefficients are  given in
the last two rows of Table~\ref{tablee1a2b2} and a graph illustrating the
quality of the fit is given in Fig.~\ref{fit-e1-a2-b2}.

     \begin{figure*}[htp]
       \begin{center}
         \subfigure[]{
            \includegraphics[scale=0.3]{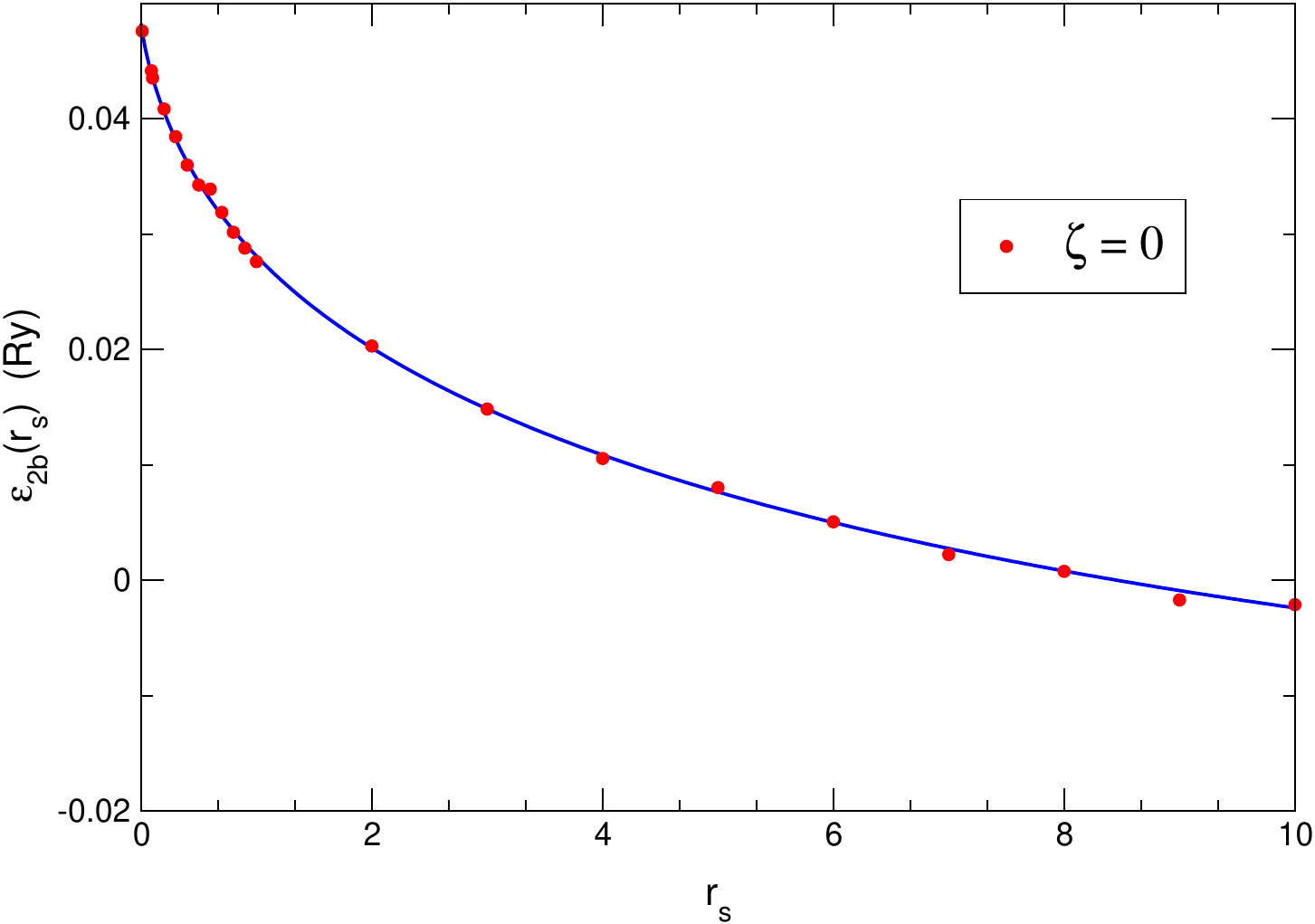}
   \label{fit-kite-small}
         }
         \subfigure[]{
               \includegraphics[scale=0.3]{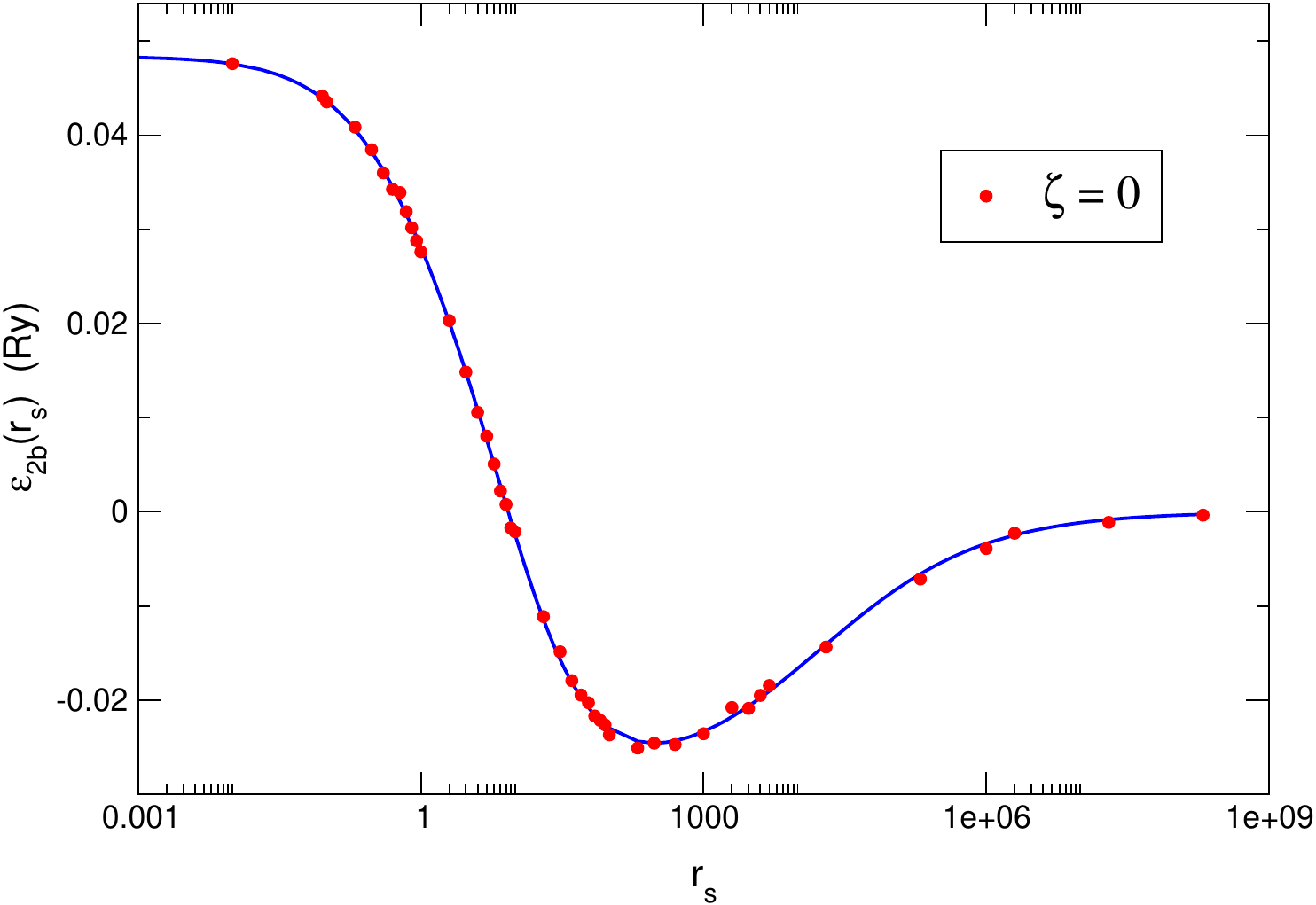}
   \label{fit-kite-all}
         }
   \end{center}
       \caption{Fit of the form given by Eq.~\ref{kite-fit-form} to our numerical results obtained by Monte Carlo integration of the expressions given by
         Eq.~\ref{e2b_1} and Eq.~\ref{e2b_2} for the
         kite-diagram series. (a) Small $r_s$ region. (b) the entire region.}
       \end{figure*}

\subsection{Fit Equations for the Kite Diagram}

\begin{table}
\begin{center}
\begin{tabular}{|c|c| c |c | c|} 
\hline
$\zeta$&$a_{1}$&$a_{2}$ (Ry)&$a_{3}$&$a_{4}$\\
\hline
0.0 & 0.10215 & -0.01382 & 0.46529 & 0.00364 \\
\hline
0.2 & 0.10876 & -0.01317 & 0.45202 & 0.00329 \\
\hline
0.4 & 0.10456 & -0.01305 & 0.46978 & 0.00291 \\
\hline
0.6 & 0.07801 & -0.01265 & 0.4441 & 0.00277 \\
\hline
0.8 & 0.06266 & -0.01023 & 0.36888 & 0.00179 \\
\hline
0.9 & 0.05106 & -0.00835 & 0.29562 & 0.00148 \\
\hline
1.0 & 0.04143 & -0.00571 & 0.19503 & 0.00088 \\
\hline
\end{tabular}
\caption{Coefficients of the fit of the kite diagram for different values of $\zeta$ obtained by fitting the data in Table~\ref{table_kite_bare}.}
\label{table_kite_zeta_fits}
\end{center}
\end{table}

  \begin{figure}
   \begin{center}
\includegraphics[scale=0.3]{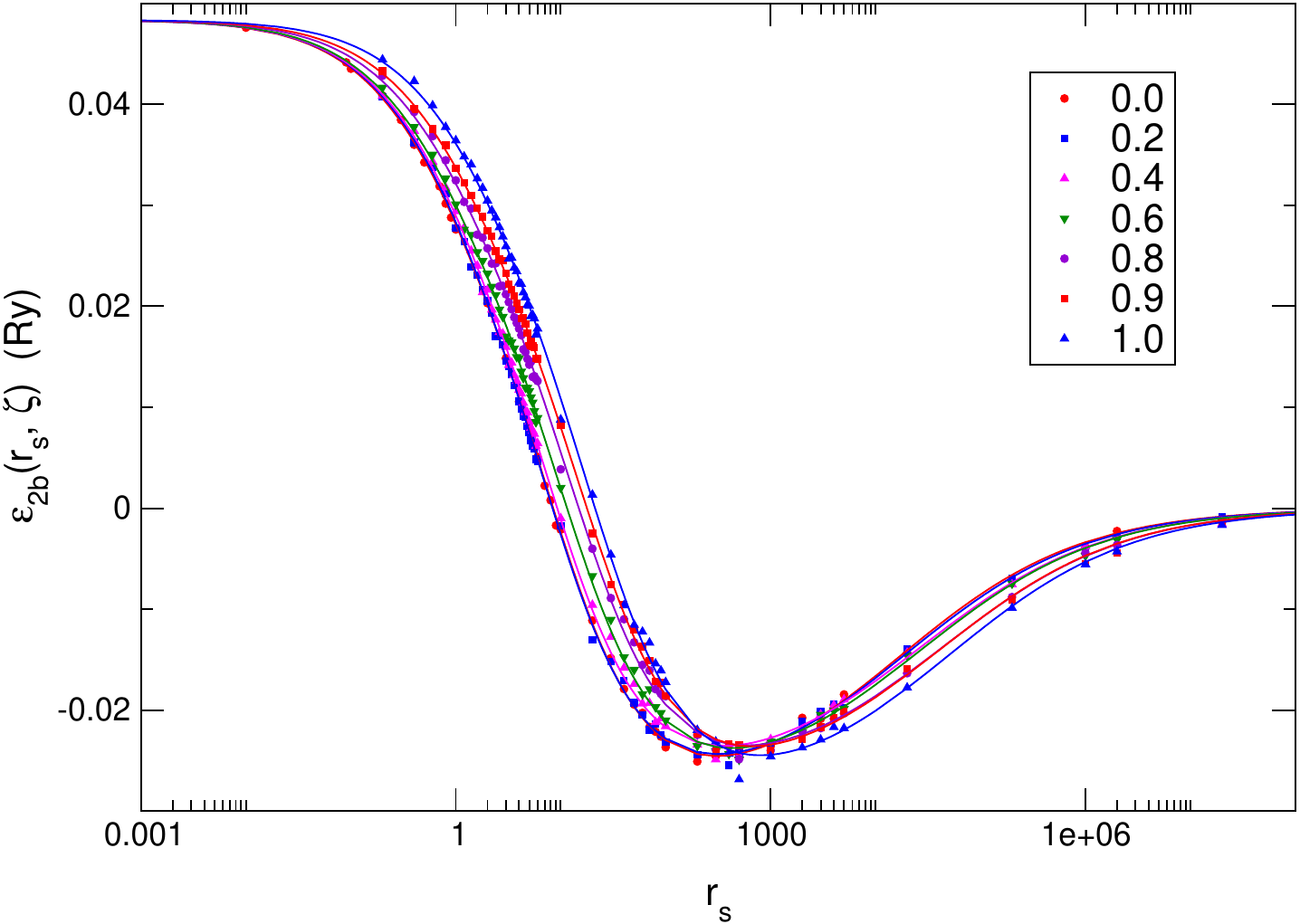}
\caption{The data points are the kite diagram data from Table \ref{table_kite_bare} and the solid lines the fit of the kite diagram presented in Eq.~\ref{kite-fit-form}.}
\label{kite-all-zeta}
    \end{center}
    \end{figure}

     We also need to represent the calculated data for the kite-diagram series
     as a function of $r_s$ and polarization $\zeta$ with a functional form
     which accurately reproduces the Onsager result at the static and $r_s \to 0$ limit and fit
     accurately the data for all the calculated values of $r_s$.
     It would have been nice to have other constraints obtained analytically to impose on the functional form. However, handling analytically the 11-dimensional integral
     in any limit has turned out to be difficult. Nevertheless, we have found numerically that the kite-diagram series decay as $1/\sqrt{r_s}$ at large values of $r_s$ and we chose a functional form that has this asymptotic behavior.
     The following form can fit accurately the results
     for the kite-diagram series:
\begin{eqnarray}
   \epsilon_{2b}(r_s) =
           {{a_0} \over {1 + a_1r_s}} +  a_2 r_s \ln \Bigl (1 +  { 1 \over {a_3r_s+a_4r_s^{3/2}}} \Bigr ),
          \label{kite-fit-form} 
  \end{eqnarray}
where to obtain the $r_s \to 0$ limit found by Onsager, we enforce $a_0=0.04836$ Ry,
so this is only a four-parameter fit.
Repeating the fit to our results for all the calculated values
of $\zeta$, we obtain the values listed in Table~\ref{table_kite_zeta_fits}
for the values of these four parameters
for the calculated values of $\zeta$ and the quality of the fit is
illustrated in Fig.~\ref{kite-all-zeta}.

We use the following interpolation functional for the $\zeta$-dependence of these four coefficients:
\begin{equation}
\begin{aligned}
    a_{n} &= \sum_{m=0}^{2}a_{nm}\zeta^{2m}, \quad {\mathrm{for}} \quad n=1,2,3,4,
\end{aligned}
\label{a1zeta}
\end{equation}
where we forced the fit to go through the value of each coefficient
obtained for $\zeta=0$ as in the case of the ring diagram series, i.e.,
$a_{n0}=a_n(0)$.
The other two coefficients $a_{n1}$ and $a_{n2}$ for each value of
values of $n$ obtained by fitting the
results listed in Table~\ref{table_kite_zeta_fits} are given
in Table~\ref{tablekite}.

\begin{table}
\begin{center}
    \begin{tabular}{|c|c|c|}
    \hline
    $n\backslash m$ & 1 & 2 \\ [0.5ex] 
    \hline
1 & -0.05028 & -0.01283 \\
    \hline
2 & 0.00016 & 0.00808 \\
    \hline
3 & 0.05868 & -0.32923 \\
    \hline
4 & -0.00259 & -0.00021 \\
    \hline
\end{tabular}
\end{center}
\caption{{Coefficients for $a_{1}$, $a_{2}$, $a_{3}$, and $a_{4}$ (in Ry) as function of $\zeta$ obtained by fitting the kite diagram data.}}
\label{tablekite}
\end{table}

    \begin{figure}
   \begin{center}
\includegraphics[scale=0.3]{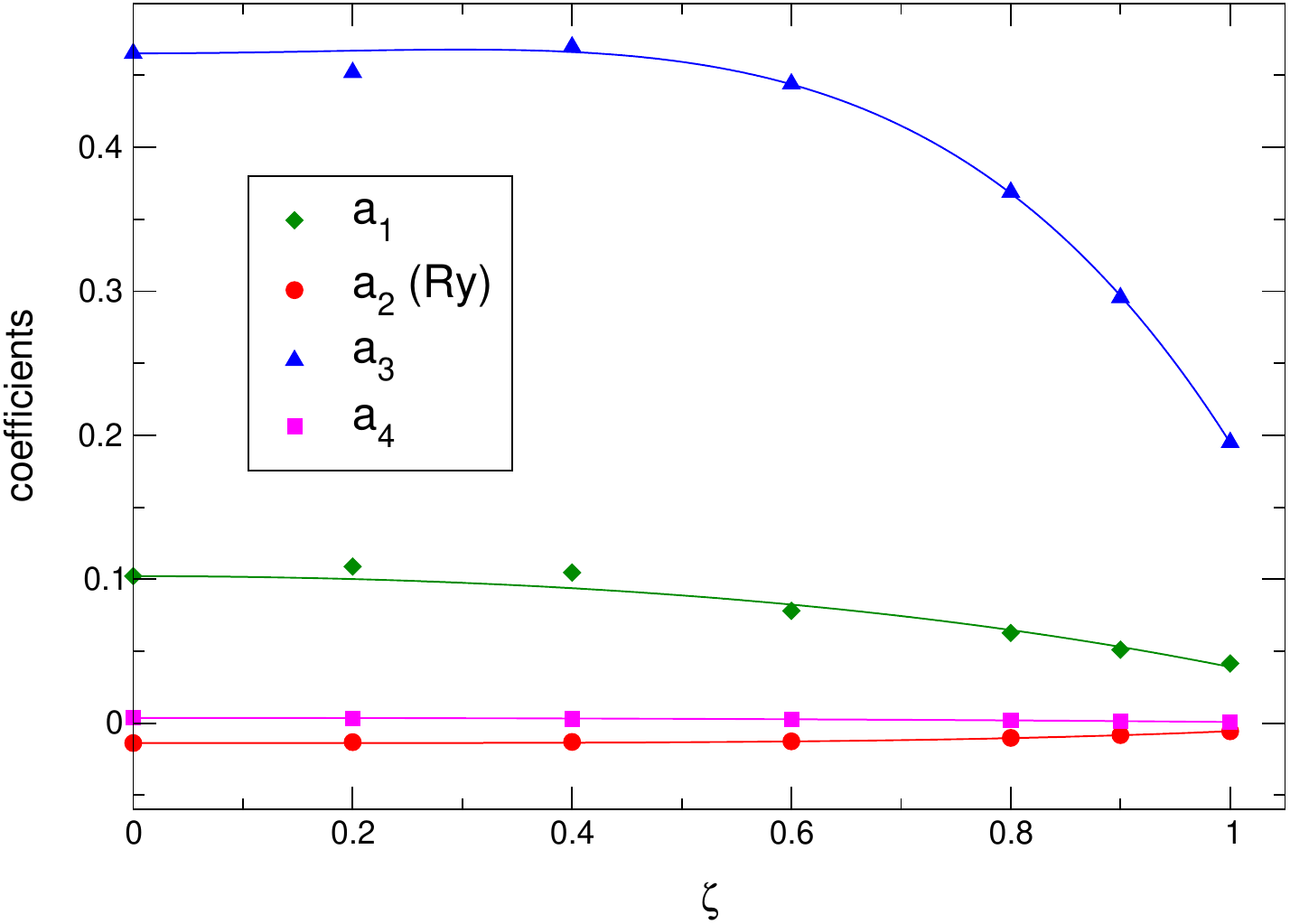}
   \end{center}
\caption{The data points are the values of Table~\ref{table_kite_zeta_fits} and the solid lines are the result of the fit to the form in Eq.~\ref{a1zeta} for the the coefficients of Table~\ref{tablekite}.}
 \label{kite-coeff-fig}
    \end{figure}

The quality of the fit for $\zeta=0$ is as shown in Fig.~\ref{fit-kite-small} for small $r_s$ values and in Fig.~\ref{fit-kite-all} for a range of $r_s$ which spans several
orders of magnitude. The results for other values of $\zeta$ are
illustrated in the Fig.~\ref{kite-coeff-fig}.

\section{Comparison of the RPAF functional with other functionals}
\label{comparison-with-other}

\subsection{Comparison with the PW functional}
\label{comparison-with-PW}
\begin{figure}[htp]
   \begin{center}
     \includegraphics[scale=0.35]{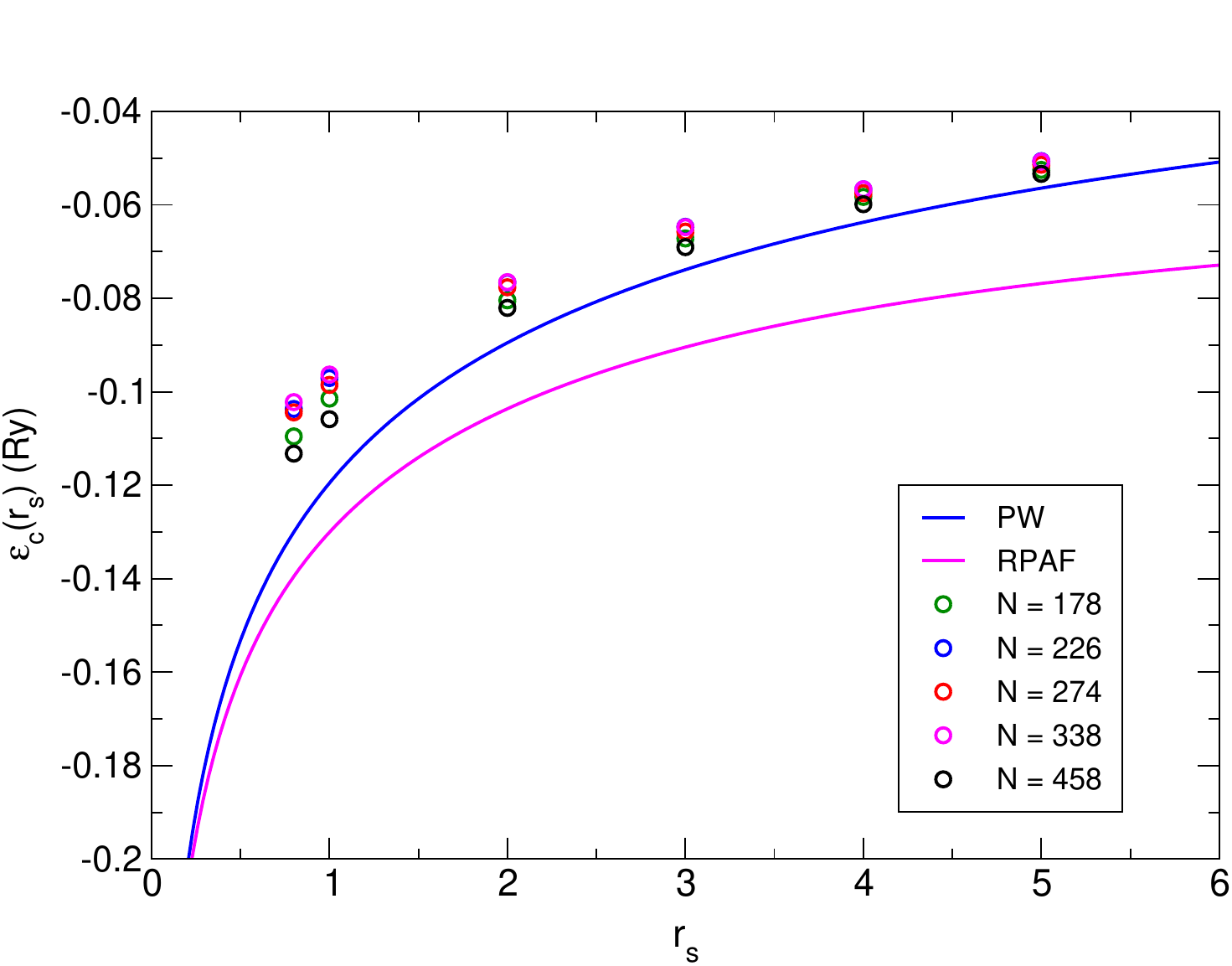}
     \caption{Comparison of the RPAF functional correlation energy (magenta line) as a function of $r_s$ with the PBE functional (blue line) and the MC results on finite-size lattices.}
     \label{comparison-with-PW}
    \end{center}
    \end{figure}

\begin{figure}[htp]
   \begin{center}
     \includegraphics[scale=0.4]{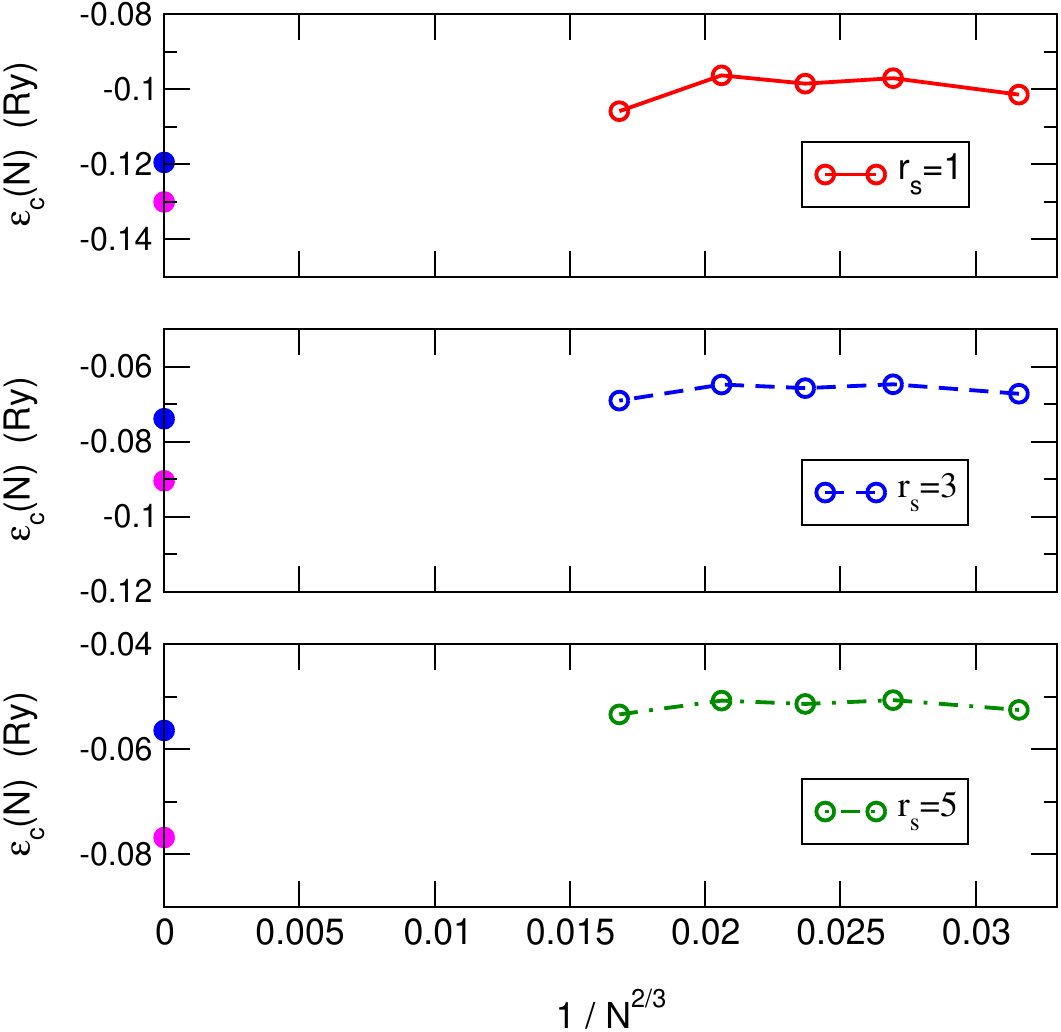}
     \caption{We plot the correlation energy as calculated by
       Monte Carlo in Ref.~\onlinecite{PhysRevB.50.1391} (open circles) for
       different number of electrons $N$ as
       a function of $1/N^{2\over 3}$  for various values of $r_s=1,3,5$.
       The blue solid circle at the origin is
       the value for the PW functional, while the magenta is the corresponding
       value of the RPAF functional.}
     \label{comparison-with-MC}
    \end{center}
    \end{figure}In Fig.~\ref{comparison-with-PW} the RPAF functional is compared with
the PW functional as a function of $r_s$ for the spin-unpolarized case.
We also plot the Monte Carlo results from Ref.~\onlinecite{PhysRevB.50.1391}
for various values of $N$. Note that the raw-data obtained in
Ref.~\onlinecite{PhysRevLett.45.566}, i.e., for any given value of $N$ as
not available. Only the $N \to \infty$ ``extrapolated'' results are provided.
For the purpose of the following discussion the $N$-dependence of the
results is needed. Nevertheless, the results of
Ref.~\onlinecite{PhysRevB.50.1391}
are even more valuable, as they are for larger values of $N$.

In Fig.~\ref{comparison-with-MC} we plot the  Monte Carlo results from
Ref.~\onlinecite{PhysRevB.50.1391} 
for various values of $N$ vs $1/N^{2 \over 3}$ as this is the reading term in $N\to \infty$
       limit in the forms used in Refs.~\onlinecite{PhysRevB.50.1391,
       PhysRevLett.45.566,PhysRevB.18.3126}.
From this Figure, it becomes evident that,
unless the asymptotic form of the correlation energy per
site $\epsilon_c(N)$ for very large $N$ is known,
the extrapolation process cannot distinguish between
the PW values (blue solid-circle at the origin) and the RPAF values (magenta solid-circle at the origin). They both seem to be within the  error
of the extrapolation. Note that the PW functional was obtained by fitting its form to the
extrapolation results of Ref.~\onlinecite{PhysRevLett.45.566}, which are for values of $N$ even smaller than the ones
used in our Fig.~\ref{comparison-with-MC}.
Notice, that the $N$-dependence is neither smooth  nor monotonic.
A non-monotonic behavior is a clear sign that the results of the
finite-$N$ Monte Carlo calculations have not
reached the $N \to \infty$ asymptote for a clear and unambiguous $N \to \infty$
extrapolation.
Moreover, the formula for extrapolation used
in these Monte Carlo studies was not based on any detailed knowledge of
the actual form, which is based on the structure of the interaction.

In the low $r_s$ region there is the $\ln(r_s)$ singularity which is obtained
within RPA due to the emergence of the infrared cutoff 
$q_{IR}=\sqrt{2 \alpha r_s/\pi}$ (which the proportional to the
Thomas-Fermi screening length
in units of the Fermi wavevector). However, when $N$ is not infinite
there is a second competing emergent cutoff given by
\begin{eqnarray}
  q_{N} ={{2\pi\sqrt{3}} \over {(3\pi^2 N)^{1 \over 3}}},
\end{eqnarray}
which interferes with the extraction of the well-known $c_{L} \ln(r_s)$ term
given by Eq.~\ref{ring_small-rs}.
Therefore, one has to be in the region where $q_N << q_{IR}$ to obtain the
asymptotic ($N\to \infty$) value given by Eq.~\ref{c_log}.
However, for a value of $N\sim 246$ (which is the
maximum value calculated in Ref.~\onlinecite{PhysRevLett.45.566})
the equation $q_{IR}=q_N$ (where one begins to feel the existence of $q_N$) leads to a value of $r_s$ of the order of unity.
Moreover, for larger $r_s$ this modification  of
the dependence of $\epsilon_c(N)$ on $N$ should cross-over to a different
one {\it smoothly} as a function or $r_s$; therefore, the extrapolation forms
discussed in Refs.~\onlinecite{PhysRevB.50.1391,PhysRevLett.45.566,Mihm2021,PhysRevB.18.3126} are not valid
for even larger values of $r_s$.
Therefore, one deals with relative large finite-size
effects especially for small values of $r_s$. 

In conclusion, the belief that the PW functional is based on results of
accurately extrapolated Monte Carlo results is not quite valid, at least
for values
of $r_s \simeq 2$ which were used to fit the PW functional.
\vskip 0.5 in

\subsection{Comparison with SOSEX}
\label{comparison-sosex}
In this subsection we compare the results of our method
  to SOSEX~\cite{freeman1977coupled}, which stands for second
  order screened exchange, a computational method used to calculate the correlation energy for the homogeneous electron gas based on the coupled cluster expansion.

The coupled cluster expansion method\cite{coester1960short,kummel1971theory,kummel1972equations}  specifically solves the time-independent Schr\"odinger's equation by writing the ground state of the many-electron system
as an operator of the form $\hat R \equiv e^{-\hat S}$ acting on a non-interacting
ground-state. The ansatz involves the operator $\hat R$, which is not unitary, typically applied to the ground-state solution from the Hartree-Fock problem. The exponential contains the so-called ``cluster'' operator, where the order of the cluster expansion depends on up to what ``$n$-th'' body operator has been used. The order of the approximation that has been used for the homogeneous electron gas by SOSEX is for $n=2$, where the cluster operator depends on the combination of two-body and a single-particle operators\cite{freeman1977coupled}. The focus of this method is to use the ansatz to find the expectation value of the bare-Coulomb potential operator $\hat{V}$. By rearranging the terms that reproduce the correlation energy, conveniently, one can isolate a specific normal ordering between the annihilation and creation operators that reproduce the RPA results within a good degree of accuracy. By subtracting this term to the whole expression that contributes to the correlation energy by the RPA, the leftover is what is commonly known as SOSEX.

The SOSEX contribution, has a specific normal ordering of the annihilation and creation operators that contributes to the correlation energy. Such rearrangement of operators can be represented as a diagram that has the same topology as the full-kite diagram (from Fig.~\ref{kite}), which we calculated in this work. The difference in the approaches for calculating the correlation energy is that the coupled cluster starts from a time-independent perturbation
expansion, while in our case, we include time-dependent effects.
Namely, when calculating the full-kite diagram series, we take into consideration the collective behavior of the screening interaction by performing the frequency integrals. Based on such differences in the approaches, we suspect that the results of the correlation energy between these two approaches should be different. 

More specifically, the expression of the full-kite diagram has a total of four integrals that were calculated by using Onsager's constant $\epsilon^0_{2b}$ and the correction to the kite diagram, given by the sum of Eq.~\ref{delta1,3} and Eq.~\ref{delta2,3}. When the branch cuts of the dielectric function are ignored, two out of the four integrals remain for the correction to the kite diagram, since in the integral of the first term in the bracket of Eq.~\ref{delta1,3} all the complex poles are below the real line, while the integral from the second term of the bracket of Eq.~\ref{delta2,3}, all the complex poles are above the real line. Hence, we think the major disagreement comes from the extra contribution that arises by avoiding the branch cuts from the dielectric function when doing the contour complex integration in the frequency variable, which is ignored in the time-independent approach (SOSEX). To test this explanation for the disagreement, we carry out an approximation to the frequency variable of the dielectric function by setting it to zero, and compare the results with SOSEX.

If we set $\omega=0$ in the expression for the kite-diagram series
and carry out all frequency integrals (of the free-interacting Green's functions) we obtain
the following expression:
\begin{widetext}
  \begin{eqnarray} 
    \epsilon_{2b}^{qs}(r_s,0)=\frac{3}{8 \pi^5} \int_0^1 d\lambda \int_{k \leq 1} d^3k \int_{q_2 \leq 1} d^3q_2 \int d^3q_1 \frac{ \lambda \Theta(|\vec{q_1} + \vec{q_2}|-1) \Theta(|\vec{k}+\vec{q_1}|-1)}{\epsilon_{\lambda}(q_1,0) q_1^2 [q_1^2 + (\vec{q_2}+\vec{k})\cdot \vec{q_1}]|\vec{k}+\vec{q_2}+\vec{q_1}|^2 },
  \label{e2b_qs}
  \end{eqnarray}     
\end{widetext}
where the superscript ``qs'' stands for quasi-static.

\begin{figure*}[htp]
   \begin{center}
            \includegraphics[scale=0.3]{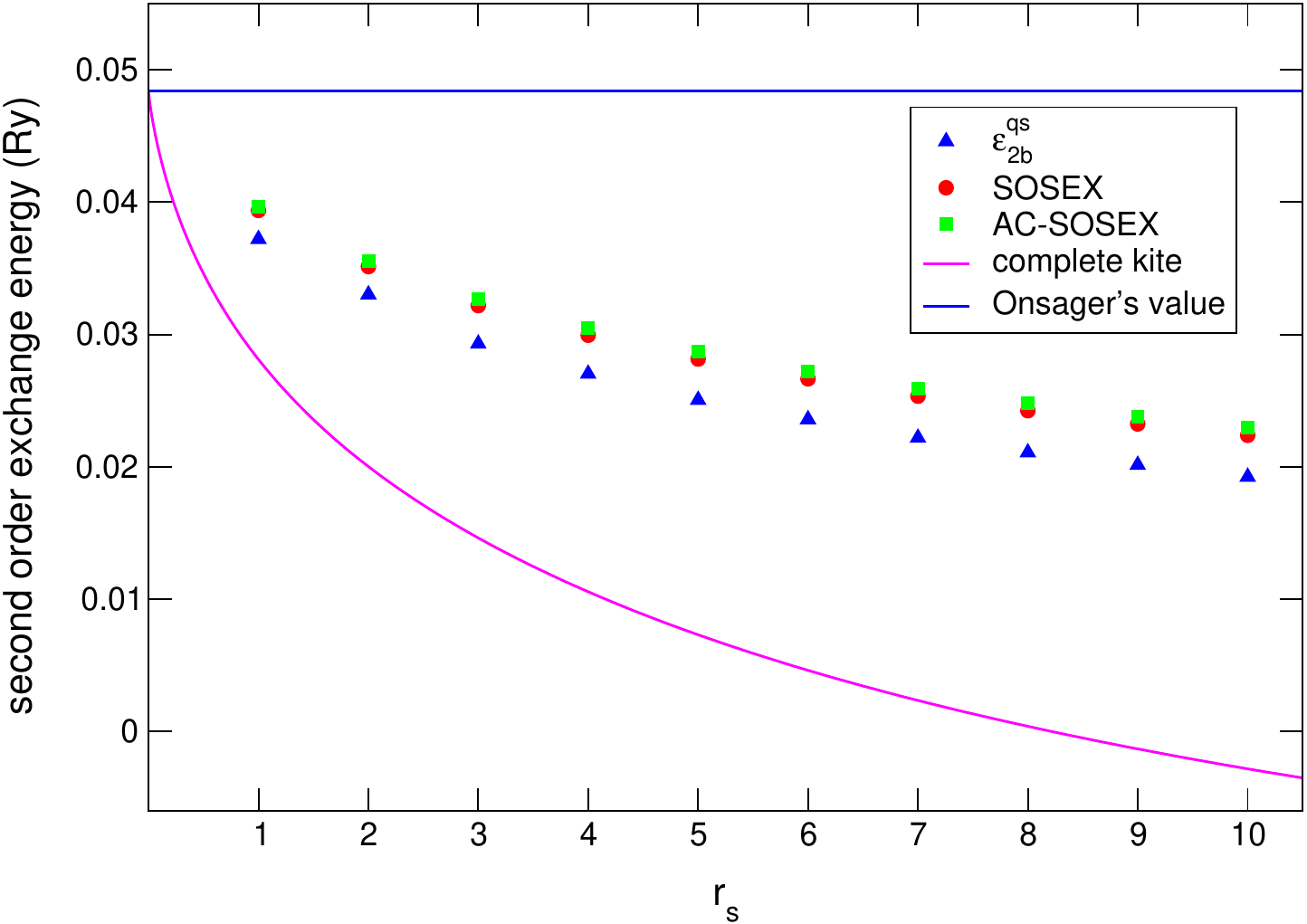} 
   \end{center}
   \caption{Comparison between the results of the full-kite diagram series from Table~\ref{table_kite_bare} at $\zeta=0$, the values obtained in the quasi-static limit for the full-kite diagram series $\epsilon^{qs}_{2b}(r_s,0)$, Onsager's value,  the SOSEX\cite{freeman1977coupled} and the AC-SOSEX values.}
   \label{sosex}
\end{figure*}

In Fig.~\ref{sosex}, we have made a comparison between the results of the full-kite diagram at $\zeta=0$ by using the data reported in Table \ref{table_kite_bare} and the SOSEX data given in Ref.~\onlinecite{freeman1977coupled}. There is a large numerical discrepancy between these two results, as expected since the two approaches  differ in a very important way as discussed earlier.
The results obtained in the quasi-static limit
(Eq.~\ref{e2b_qs}) and Onsager's value are
shown to compare the SOSEX results obtained by Freeman\cite{freeman1977coupled}. Notice that the results obtained by SOSEX are close to the results
obtained using the additional (and unnecessary) approximation given by
Eq.~\ref{e2b_qs} to the full kite-diagram series. 

Additionally, the data obtained from Eq.~\ref{e2b_qs} by Monte Carlo integration was fitted to the equation given by Eq.~\ref{fit_kite_sosex}. The value of the coefficient found for the $r_s \ln{r_s}$ term in the long wavelength limit is given by $0.00572$ Ry, which almost agrees with the $C_1$ value of $0.00641$ Ry, found for PW contribution (i.e., the difference
of PW from the sum of the ring-diagram series) by the same fitting
procedure reported in Sec.~\ref{check}. The numerical results of the correlation energy obtained by SOSEX, almost agree with the correlation energy from the PW functional. This means that the hypothetical coefficient of the $r_s \ln{r_s}$ obtained from the SOSEX data should almost agree with the $C1$ obtained through the PW functional reported in Table~\ref{sosex_table}.
Therefore, while a subset of the terms that occur in the coupled cluster
expansion agrees well with the RPA results obtained by Hedin\cite{PhysRev.139.A796}, there is a major disagreement with the $C_1$ value reported by Ref.~\onlinecite{loos2011correlation}.

Therefore, the SOSEX contribution to the correlation energy
corresponds to approximating the dielectric function with
its frequency independent limit.
However, this is a  poor approximation to this important
response function because it
 ignores the retardation effects\cite{PhysRev.112.812}
which characterize the response of the electronic system to
not-only external time-dependent electric fields, but more
importantly to the effective interaction between electrons inside
the Jellium system. When one ignores such effects, this interaction
gives the Thomas-Fermi result, which fails to produce the
 Friedel oscillations.

 AC-SOSEX\cite{jansen2010equivalence,ren2013renormalized,hummel2019screened} is a method which attempts  to include the frequency dependence in the second order exchange diagram contribution to the correlation energy using the adiabatic connection-dissipation formalism (ACFD). AC-SOSEX has been used on several materials\cite{ren2013renormalized} and the homogeneous electron gas\cite{hummel2019screened}.

 Within the ACFD formalism, the RPA part of the correlation energy,  $E^{RPA}_c$, has been calculated and can be written as:
\begin{eqnarray}
E^{RPA}_c = -\frac{1}{2} \int_{-\infty}^{\infty} \frac{d \nu}{2 \pi} \sum_{ijab} W^{ab}_{ij}(i \nu) F^a_i(i \nu) F^b_j(i \nu) V^{ij}_{ab}, 
\label{RPA_AC_SOSEX}    
\end{eqnarray}
where $i,j$ ($a,b$) correspond to the quantum numbers of the occupied (unoccupied) states. $V^{ij}_{ab}$ ($W^{ab}_{ij}(i \nu)$) are the matrix element of the bare Coulomb (coupling-strength averaged RPA-renormalized)
interaction written in the particle/hole state basis
and  $F^a_i(i \nu)$ is the particle/hole propagator as defined in
Ref.~\onlinecite{ren2013renormalized} (See also Ref.~\onlinecite{hummel2019screened}).

This expression yields the correct expression for the ring-diagram series in the homogeneous electron gas problem. Within the AC-SOSEX method, the second-order exchange diagram is obtained from Eq.~\ref{RPA_AC_SOSEX} by doing an interchange between the occupied state indices in the $V^{ij}_{ab}$ term and by picking up an extra minus sign as a global factor, as explained in Refs.~\onlinecite{ren2013renormalized},~\onlinecite{hummel2019screened}.

\begin{figure*}[htp]
   \begin{center}
            \includegraphics[scale=0.3]{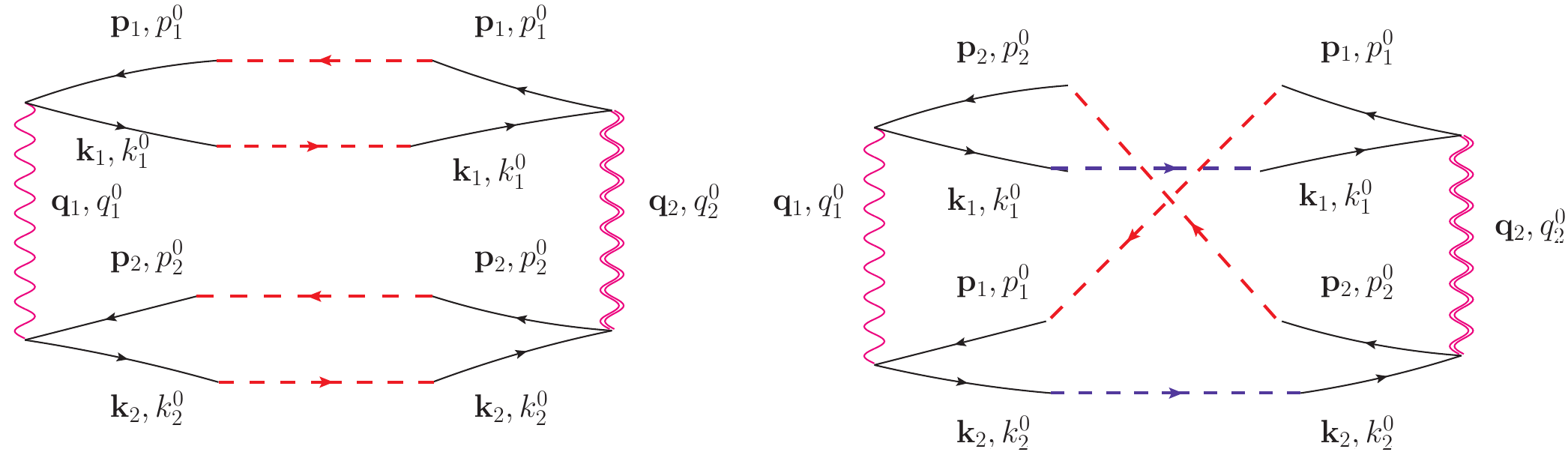}
   \end{center}
   \caption{Diagrammatic illustration that the kite-diagram series (right) can
     be obtained from the ring diagram series (left) by exchanging the frequency-momentum indices of two of the Fermionic propagators.}
  \label{diagram}
\end{figure*}

This is motivated
by the fact that the ring diagram series and the kite diagram series
are related by a similar exchange when one writes down their expressions using the Feynman rules of the
         many-body time-dependent perturbation approach. This is demonstrated in Fig.~\ref{diagram}. Namely, the diagram on the right (kite-series) can be obtained
         from the diagram on the left (ring-series) by simply exchanging indices of
         the two fermionic
         lines, which in this case represent two four-momenta (momentum and frequency) labels of the bare interaction term.
         Notice, however, that these are 4-momenta indices, which includes the frequency indices. However, in order to obtain the RPA result given by Eq.~\ref{RPA_AC_SOSEX}, two of these frequency integrations
         have been carried out. As a result, starting from this reduced
         equation the other two frequency indices are gone and, thus, there is no way to carry out the required exchange of 
         the full 4-vector indices in order to obtain the kite diagrams starting from the RPA.

The fact that  AC-SOSEX and the corresponding diagrams obtained by many-body perturbation are different was also argued in Ref.~\onlinecite{hummel2019screened}.

As illustrated in Fig.~\ref{sosex} the results of the AC-SOSEX are very close
to the SOSEX case and substantially different from our kite-diagram
contribution.

    \begin{figure*}[htp]
   \begin{center}
            \includegraphics[scale=0.3]{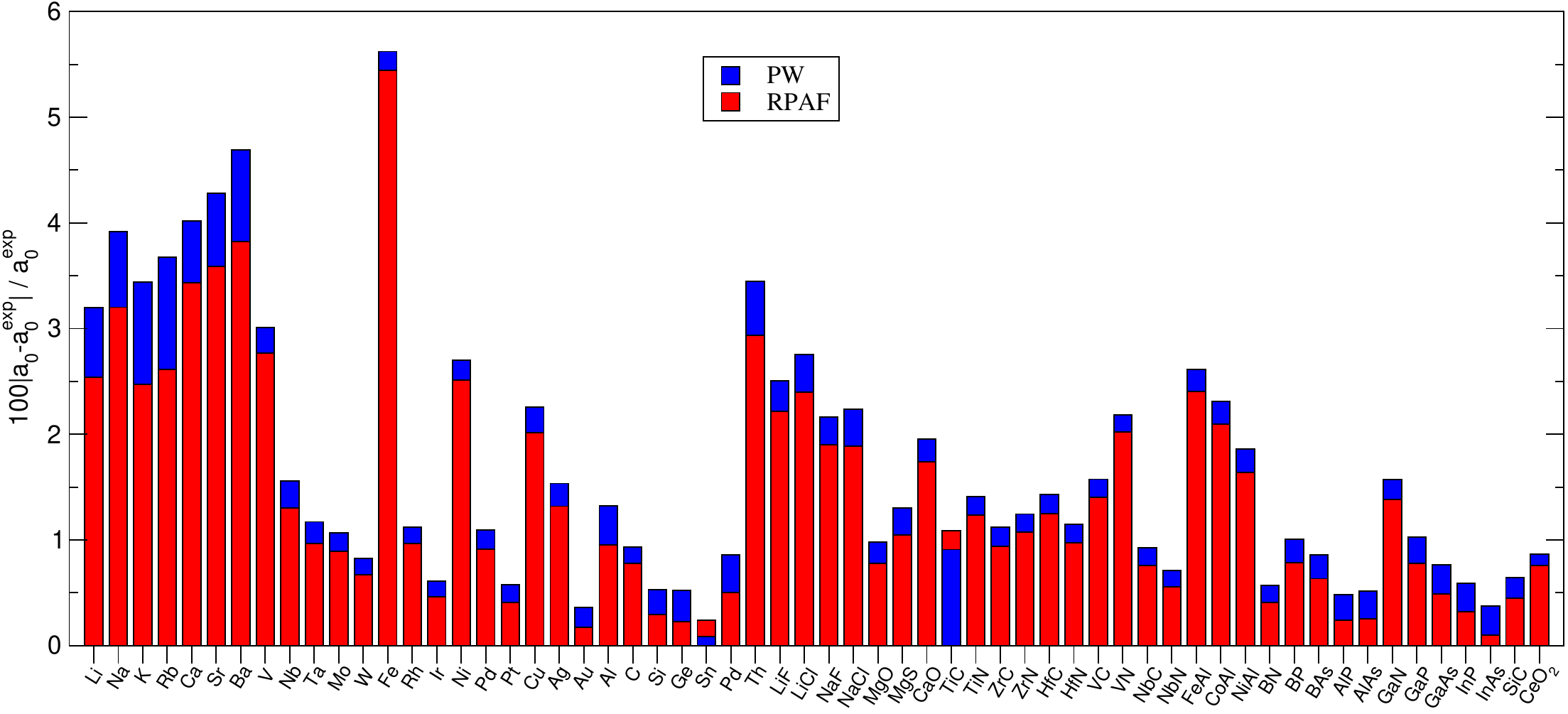}
   \end{center}
   \caption{Relative errors of equilibrium lattice constants ($a_{0}$) of
     the 60 crystals listed in Ref.~\onlinecite{PhysRevB.79.085104} using the PW functional (blue) and the RPAF functional presented in this work (red). The calculations were done in Quantum ESPRESSO. The values are listed in the Appendix~\ref{tables}.}
   \label{lattice-constant}
    \end{figure*}
    \begin{figure*}[htp]
   \begin{center}
            \includegraphics[scale=0.3]{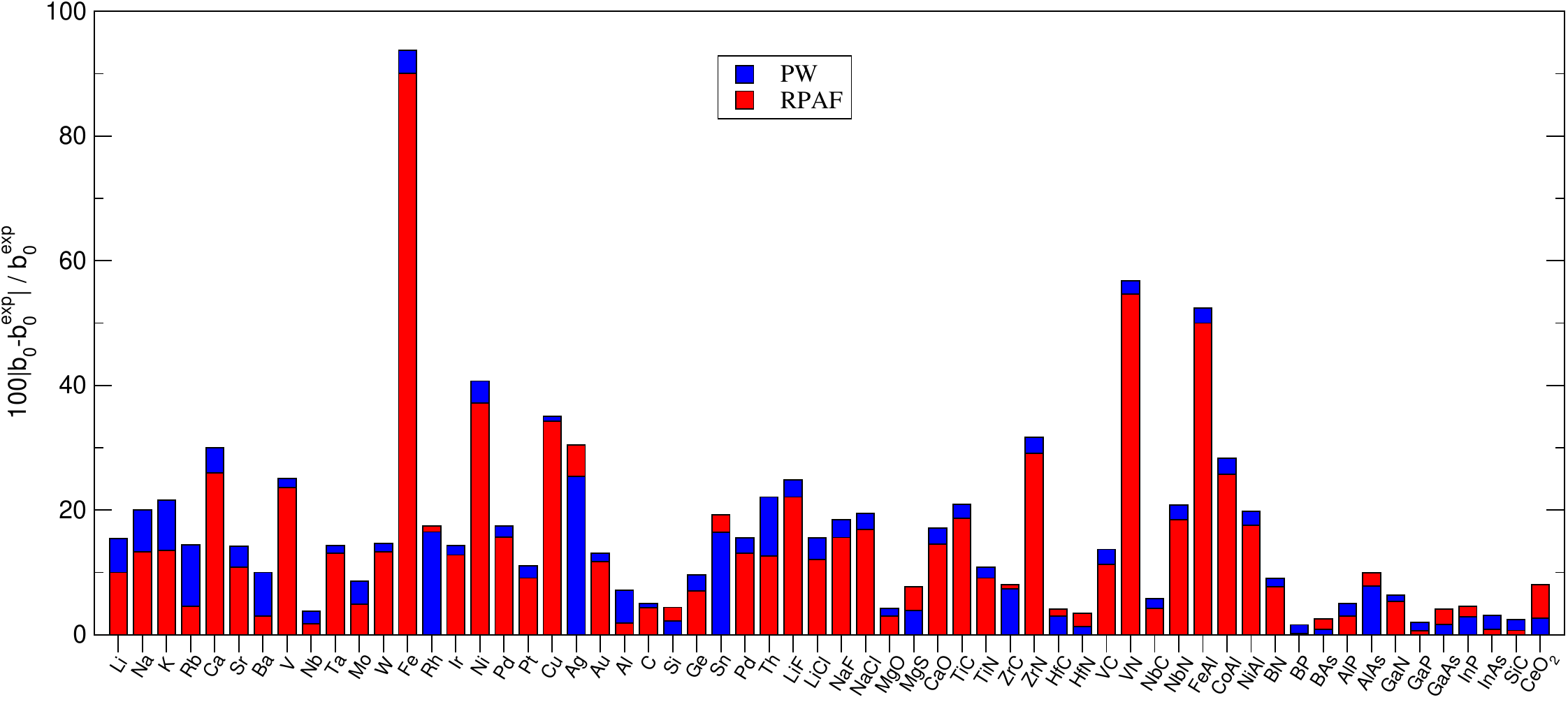} 
   \end{center}
   \caption{Relative errors of equilibrium bulk moduli ($B_{0}$) of 60 crystals listed in Ref.~\onlinecite{PhysRevB.79.085104} using the PW functional (blue) and the RPAF functional presented in this work (red). The values are listed in Appendix~\ref{tables}.}
   \label{bulk-modulus}
    \end{figure*}

      \section{Benchmarking our functional}
      \label{performance}
In order to assess the accuracy of our functional, we modified
the Quantum ESPRESSO v7.2 package\cite{QE-2017} to include our functional as an option
and we carried out DFT calculations within the local density approximation (LDA)
for the same entire list of crystalline
materials given in Refs.~\onlinecite{PhysRevB.79.085104,PhysRevB.75.115131,10.1063/1.5116025} where the performance of various other functionals
was evaluated.

The set of solids used in Ref.~\onlinecite{PhysRevB.79.085104} containing 60 crystals was used to compare the predicted lattice constant and bulk modulus of the RPAF functional with the popular Perdew-Wang (PW) LDA functional\cite{PhysRevB.45.13244}. The experimental lattice constants were obtained from Ref.~\onlinecite{PhysRevB.79.085104} and the bulk moduli from Ref.~\onlinecite{PhysRevB.75.115131}. 
Pseudopotentials were generated for the PW and RPAF functional using the ld1.x in Quantum ESPRESSO with the pslibrary1.0.0 library of inputs \cite{dal2014pseudopotentials}.
  A $10\times 10 \times 10$ Monkhorst-Pack mesh was used for the reciprocal lattice space for all the crystals. For the energy cutoff required for convergence, we found that, depending
  on the material, a range from 40 Ry up to 140 Ry was good enough.
  Self-consistent calculations were done for nine different lattice constants, four above and four below the equilibrium lattice constant, with a step of 0.01 Bohr radius. Afterward, the nine points were fitted using the third-order Birch-Murnaghan equation of state using the ev.x tool in Quantum ESPRESSO to obtain the equilibrium lattice constant and bulk modulus.

The Perdew, Burke and Ernzerhof (PBE) functional\cite{PhysRevLett.77.3865} shares
the same local part with the PW\cite{PhysRevB.45.13244} functional, i.e.,
PBE is PW plus
the correction based on the generalized gradient approximation (GGA).
The GGA is an additional feature included in the PBE functional, which can
be adopted or not, namely it has a different and independent
goal from the question of the local part of the functional. We
are mindful that the
focus of the present paper is precisely the local part of the functional.
Therefore, when we try to assess the accuracy of our functional
we compare it only to the PW functional, since the PBE and PW share the
same local part.

The results are compared with the PW functional 
in the bar-graph of Fig.~\ref{lattice-constant} for the lattice constants
and in Fig.~\ref{bulk-modulus} for the bulk moduli and the precise numerical values are given
in the tables of Appendix~\ref{tables}.
Our functional, which is represented by the red-color bars in both figures, are systematically better than
the results obtained with the PW functional.

We did not extend these LDA calculations to include
  GGA corrections to compare with the PBE functional because such an extension,
  albeit useful, will not provide a direct benchmarking of the
  novel functional proposed here. Namely, the present proposal only affects the
  local part of the density-functional. How to go beyond LDA is a useful
  future direction of our work.
\vskip 0.5 in
\section{Conclusions}
\label{conclusions}

The local part of most available functionals is fit to
    QMC calculations which are inaccurate and feel strong finite-size effects, especially when the electron gas is spin-polarized.

   We have argued that the perturbation expansion in an RPA-renormalized interaction-line
  not only removes the infinities introduced by the bare-Coulomb interaction
  but it also allows a fast convergence with respect to the number of RPA-renormalized interaction-lines.
  Therefore, we should expect such an expansion 
  to approach the correct functional  for a broad-range of value of $r_s$
  due to the screening effects without having to go to a high order in this expansion. 

  We computed the $r_s$ and spin-polarization dependence of the correlation
  energy to the leading order in the RPA-renormalized interaction, including the
  frequency dependence of the dielectric function.  This leads to a functional of the density and of spin-polarization expressed in an analytic form. 
  We modified the quantum ESPRESSO implementation of DFT to include our functional as one option and
  we benchmarked our functional on a list of materials selected and used in Refs.~\onlinecite{PhysRevB.79.085104,PhysRevB.75.115131,10.1063/1.5116025}.
  We demonstrated that the functional proposed here yields in general better relaxed lattice constants and bulk moduli (Figs.~\ref{lattice-constant} and \ref{bulk-modulus}) than
  the local part of the popular functionals PW\cite{PhysRevB.45.13244} (and, therefore, PBE\cite{PhysRevLett.77.3865} because they share the same local functional).

  For completeness, in Appendix~\ref{ferromagnetism} we studied the
  fate of ferromagnetism of the uniform electron fluid as implied
  by our functional. Our findings are 
  in qualitative agreement with other functionals
derived from QMC\cite{PhysRevLett.45.566} that the fully polarized electron gas becomes
favorable for values of $r_s$ well outside the region accessible by
crystalline solids and outside the region of validity of our expansion.

  This work introduces a systematic order-by-order approach
 in the number of RPA-renormalized interaction-lines, which can be improved by including the next order correction. This stands in contrast to the previously introduced functionals,
  which are based on ad hoc procedure and, thus,
  produce questionable results. The next order correction within this method, although cumbersome, is achievable and
  will be our future project.

\vskip 0.5 in
  \section{Acknowledgments}
  This work was supported by the U.S. National Science Foundation under Grant No. NSF-EPM-2110814.  We thank Marisa Langley who wrote a code to independently calculate the integrals given by Eq.~\ref{e2b_1} and Eq.~\ref{e2b_2} to verify the results of our calculation.

\appendix

\section{Small r$_s$ limit of the ring-diagram series}
\label{logarithm}
In this appendix section, we present the calculation of the coefficient of the $\ln(r_s)$ corresponding to the ring diagram series for the spin-polarized electron gas. The sum of all Goldstone diagrams illustrated in Fig.~\ref{rpacor} relies on the computation of the polarization tensor $\Pi^0_{\sigma}(q^{\mu})$, which is obtained from doing the integral given in Eq.~\ref{polarization_function}.

The calculation of the ring diagram series $E_r(r_s,\zeta)$ relies on finding an expression for the self-energy $\Sigma^{r}_{\sigma}(q,q^0)$:
\begin{eqnarray}
 && \Sigma^{r}_{\sigma}(q^{\mu},\lambda) =  \int \frac{d^4p }{(2 \pi)^4\hbar^2}\int \frac{d^4k }{(2 \pi)^4} \frac{\left( \lambda V_0(q)\right)^2}{\epsilon_{\lambda}(q,q_0)} \nonumber \\
   &\times& 
  G^0_{\sigma}(k^{\mu}-q^{\mu})   \sum_{\sigma '} G^0_{\sigma '}(p^{\mu}+q^{\mu}) G^0_{\sigma'}(p^{\mu}),
\label{self_energy_ring}    
\end{eqnarray}
which by using this expression in Eq.~\ref{eq:Cluster_Expansion} we obtain an expression for the contribution to the ground-state energy given by the ring diagram series. One can rewrite the expression of the integrated ring diagram series into a more compact form as a function of the polarization tensor:
\begin{equation}
  E_{r}(r_s,\zeta) = \frac{i\hbar V}{2} \int_0^1 d\lambda \int {{d^4q}
    \over {(2\pi)^4}}\frac{\lambda \left[ 1-\epsilon(q,q^0) \right]^2}{\epsilon_{\lambda}(q,q^0)},
\label{ring_compact}    
\end{equation}
where $\epsilon_{\lambda}(q,q^0)$ is the dielectric function given by Eq.~\ref{eq:dielectric} with the coupling constant rescaled $e^2 \rightarrow \lambda e^2$.
A useful change of variable can be performed at the integrand to convert the physical variables into dimensionless parameters as follows: $q \rightarrow k_F \kappa$, $q^0 \rightarrow \hbar k_F^2 \nu /m$. By doing this, the number of particles can be factored out and we can express the contribution to the ground-state energy in terms of the ring diagram series per particle $\epsilon_r(r_s,\zeta)$, as seen in Eq.~\ref{ring}. After performing the integral over the $\lambda$, we obtain the expression given in Eq.~\ref{ring2}.

Instead of integrating the frequency variable $\nu$ along the real line, we can exploit Cauchy's theorem. If we take the same path we used to calculate the $E_{2b}(r_s,\zeta)$ diagram, we can avoid the branch cuts from the logarithmic terms of the dielectric function which causes the integration over $\nu$ along the real line mapped into an integral along the imaginary axis of the complex plane. The parameterization of the new line integral is given by Eq.~\ref{ring4}, where we identify that the term in the argument of the logarithm is the dielectric function $\epsilon(k_F \kappa, i\hbar k_F^2 \nu/m)$:
\begin{equation}
\epsilon\left( k_F \kappa,i\frac{\hbar k^2_F \nu}{m}\right) = 1+\frac{\alpha r_s}{\pi \kappa^2}\sum_{\sigma=\pm} x_{\sigma} (\zeta) g\left(\frac{\kappa}{x_{\sigma}(\zeta)}, \frac{i \nu}{x^2_{\sigma}(\zeta)} \right),
\label{dielectric_appendix}
\end{equation}
where $g$ is the function of the Lindhard function given by Eq.~\ref{g_function}, which was derived by doing the high dimensional integral given in Eq.~\ref{polarization_function} at imaginary frequency.

By exploiting the fact that the integrand of $\epsilon_r(r_s,\zeta)$ is an even function in $\nu$, the expression picks up a factor of 2, and by performing the change of variable $x=\nu/q$ and integrating along the solid angle from the 3-dimensional integral, we obtain Eq.~\ref{ring4}.

This integral is not trivial to calculate analytically, but to make the integral more tractable when we separate the interval of integration into three different regions: $q \in [0,q_{IR}] \cup [q_{IR},q_c] \cup [q_c,\infty)$, where $q_c<<1$ and $q_{IR}$ is proportional to $r_s^{\frac{1}{2}}$, which gives the notion of an infrared cutoff for the screened interaction potential which arises from the dielectric function of the material in RPA. In the first region, the integration does not yield a $\ln(r_s)$ term, but rather contributions to the higher order terms in $r_s$, while the integral along the third region for $q$ yields a contribution to the constant term for the ring diagram series. We focus only on the second region of integration since it yields the coefficient of the $\ln(r_s)$ term, whose contribution to the ring diagram series is called  $\epsilon_{r,2}(r_s,\zeta)$. In  $q \in [q_{IR},q_c]$, we can do a Taylor expansion in terms of $(\alpha r_s)/(\pi q^2)$ at the level of the logarithm term of the integrand from Eq.~\ref{ring4} up to the second order, wherein such small $q$ region the Lindhard function can be approximated as:
\begin{eqnarray}
  g\left(\frac{q}{x_{\sigma}(\zeta)},i\frac{x q}{x^2_{\sigma}(\zeta)}\right) &=& 2 \left[1-\frac{x}{x_{\sigma}} tan^{-1}\left(\frac{x_{\sigma}}{x}\right) \right]\nonumber \\
  &-&\frac{q^2 x^2_{\sigma}}{6 (x^2+x^2_{\sigma})^2}.
\label{g_simplified}
\end{eqnarray}
The first term in parenthesis of Eq.~\ref{g_simplified} yields the $\ln(r_s)$ term when used in the second region of integration in the ring diagram per particle $\epsilon_{r,2}(r_s,\zeta)$. By ignoring the $q^2$ term in Eq.~\ref{g_simplified}, we obtain the following expression for $\epsilon_{r,2}(r_s,\zeta)$:
\begin{equation}
\epsilon_{r,2}(r_s,\zeta) = \frac{3}{\pi^3} \ln\left(\frac{q_{IR}}{q_c} \right)\sum_{i=1}^{3}I_i(\zeta),
\label{ring_simplified3}    
\end{equation}
with $q_{IR}=\gamma_2 r^{\frac{1}{2}}_s$, for some constant $\gamma_2$, and the integrals are along the ratio $x$ are $I_1(\zeta)$, $I_2(\zeta)$ and $I_3(\zeta)$, which are given by the following expressions:
\begin{eqnarray}
    I_1(\zeta) &=& \int_0^{\infty} dx h^2_{\uparrow}(x,\zeta) = \frac{\pi}{3}(\zeta_+)(1-\ln(2)),
\label{I1}\\
    I_2(\zeta) &=& \int_0^{\infty} dx h^2_{\downarrow}(x,\zeta) = \frac{\pi}{3}(\zeta_-)(1-\ln(2)),
\label{I2}\\
    I_3(\zeta) &=& 2\int_0^{\infty} dx h_{\uparrow}(x,\zeta) h_{\downarrow}(x,\zeta) \nonumber\\
     &=& \frac{\pi}{3}\left[\frac{{\zeta_+^{\frac{1}{3}} \zeta_-^{\frac{1}{3}} \chi}}{2}+\frac{1}{6}ln\left(\frac{\zeta^{\zeta_+}_+ \zeta^{\zeta_-}_-}{\chi^6}\right) \right],
\label{I3}
\end{eqnarray}
where $\zeta_{\pm} =1 \pm \zeta$, $\chi$ was defined in Eq.~\ref{chi} and the function $h_{\sigma}(x,\zeta)$ is defined as follows:
\begin{equation}
h_{\sigma}(x,\zeta) = x_{\sigma}(\zeta)-x \arctan\left(\frac{x_{\sigma}(\zeta)}{x}\right).
\label{h_sigma}
\end{equation}

 The three integrals have been calculated by using a simple change of variable $y=1/x$ and then integrating along the complex plane by avoiding the branch cuts that come from the logarithmic terms that come from integration by parts.

By using these results we obtain the $\ln(r_s)$ dependence of the ring diagram series per particle for small $r_s$:
\begin{equation}
\epsilon_{r}(r_s,\zeta) = c_{L}(\zeta) \ln(r_s) + c_0,
\label{ring_simplified4}
\end{equation}
where $c_{L}(\zeta)$ is the coefficient of the $\ln(r_s)$ term we previously discussed in Eq.~\ref{c_log}.

\vskip 1.0 in

\section{Large r$_s$ dependence of the sum of the Ring Diagrams}
\label{large-rs}

In this appendix section, we show how to obtain the asymptotic behavior of $\epsilon_r(r_s,\zeta)$ for large $r_s$. The change of variable $x = \nu/\kappa$ was done in Eq.~\ref{g_function} to study the large $r_s$ region. After this, it is convenient to do the following change of variables: $\kappa = 2 Q$ and after this mathematical step, we did the transformation $x = yQ$.

According to the behavior of the Lindhard function at imaginary frequency, as $Q$ and $y$ variables increase, this function decreases to values less than 1. For the purpose of making the integral expression of $\epsilon_r(r_s,\zeta)$ tractable given in Eq.~\ref{ring4}, we will find the region such as the condition for $\tilde{\Pi}(2 Q,Qy,\zeta,r_s)<1$ is satisfied. By considering, that this Lindhard function yields its maximum value when both variables $Q$ and $y$ are zero ($g(0,0)=2$), and the fact that $\tilde{\Pi}(2Q,yQ,\zeta,r_s)$ is proportional to the Wigner-Seitz radius $r_s$ but inversely proportional to $Q^2$, we have that $\tilde{\Pi}(2Q,yQ,\zeta,r_s)$ is less than unity only at some relatively large value of $Q$. We can expect that such value of $Q$ that satisfies the condition for $\tilde{\Pi}(2Q,yQ,\zeta,r_s)$ will depend on the variable $y$ due to the behavior of the function to decrease in value as the $y$ variable increases.

At the large $Q$ limit, the logarithm and the arc tangent functions in the expression of the Lindhard function can easily be expanded in powers of $1/Q$. We obtain:
\begin{equation}
g\left(\frac{2Q}{x_{\sigma}(\zeta)},i \frac{2yQ^2}{x^2_{\sigma}(\zeta)} \right) \overset{k \rightarrow \infty}{\approx} \frac{2 x^2_{\sigma}(\zeta)}{3 Q^2 (1+y^2)}.
\label{Lindhard_approx}    
\end{equation}
By replacing this approximated expression of the Lindhard function in Eq.~\ref{Pi_tilde} we obtain the approximated expression of $\Pi(2Q,yQ,\zeta,r_s)$ to find the value of $Q$ that guarantees that this function is less than unity. The approximated expression is:
\begin{equation}
\tilde{\Pi}\left(2Q,yQ,\zeta,r_s\right) \overset{Q \rightarrow \infty}{\approx} \frac{\alpha r_s}{3 \pi Q^4(1+y^2)},
\label{beta_approx}    
\end{equation}
where the condition $\tilde{\Pi}(2Q,yQ,\zeta,r_s)<1$ is only achieved if the inequality $Q>\overstar{k}(r_s,y)$, where $\overstar{k}(r_s,y)$ is given by:
\begin{equation}
\overstar{k}(r_s,y) = \left(\frac{\alpha r_s}{3 \pi (1+y^2)}\right)^\frac{1}{4}.
\label{k_cutoff}    
\end{equation}

We can now start approximating the integral expression of $\epsilon_r(r_s,\zeta)$ by separating the region of integration into two parts on the $Q$ variable:
\begin{equation}
\epsilon_r(r_s,\zeta) = \epsilon_{r1}(r_s,\zeta) + \epsilon_{r2}(r_s,\zeta),
\label{ring_diagram3}  
\end{equation}
where the first term integrates over $y\in [0,\infty)$ and the region of integration $Q \in \left(0,\overstar{k}(r_s,y)\right]$, which we call region 1. The second term is the contribution to $\epsilon_r(r_s,\zeta)$ in the second region of integration, which integrates on $y$ half of the real line, and $Q$ is integrated into $Q\in [\overstar{k}(r_s,y),\infty)$.

The integral in the second region was calculated by Perdew, where it relies on using the simplified expression of $\tilde{\Pi}(2Q,yQ,\zeta,r_s)$ we found in Eq.~\ref{beta_approx} to calculate all of the required integrals.

In the second region of integration, the integral is simple to calculate due to the condition on the function $\tilde{\Pi}(2Q,Qy,\zeta,r_s)< 1$, which makes it possible to do a Taylor expansion on the logarithm term of the integrand in small values of $\tilde{\Pi}(2Q,Qy,\zeta,r_s)$. To obtain the accurate contribution to the term that captures the asymptotic region of $\epsilon_r(r_s,\zeta)$ in the low-density regime, we must take into consideration all orders of expansion of the logarithm term. From such expansion, we have so far:
\begin{equation}
    \epsilon_{r2}(r_s,\zeta) \approx \frac{-24}{\pi (\alpha r_s)^2} \int_0^{\infty} dy \int_{\overstar{k}(r_s,y)}^{\infty} dQ Q^4 \sum_{n=2}^{\infty}S_n, 
\label{ring_2_step2}    
\end{equation}
where the term $S_n$ in the series is given by:
\begin{equation}
S_n = \frac{(-1)^n[\tilde{\Pi}(2Q,yQ,\zeta,r_s)]^n}{n}.
\label{Sn}
\end{equation}

By using the expression given in Eq.~\ref{beta_approx} in the integral above, the integral over $Q$ is simple to do, so we are left with an integral over $y$:
\begin{equation}
\epsilon_{r2}(r_s,\zeta) = -\frac{24}{\delta} \sum_{n=2}^{\infty} \frac{(-1)^n}{n(4n-5)} \int_0^{\infty} dy \frac{1}{(1+y^2)^{\frac{5}{4}}},
\label{ring_2_step3}    
\end{equation}
where $\delta \equiv 3^{\frac{5}{4}} \pi^{\frac{9}{4}} (\alpha r_s)^{\frac{3}{4}}$
and the sum over $n$ and the integral over the $y$ variable has been calculated with Wolfram Mathematica, which yield:
\begin{eqnarray}
M_1 &=& \sum_{n=2}^{\infty} \frac{(-1)^n}{n(4n-5)} \approx 0.133628,
\label{sum1}\\    
M_2 &=& \int_{0}^{\infty} \frac{{dy}}{(1+y^2)^{\frac{5}{4}}} = \frac{2 \sqrt{\pi}\Gamma(\frac{3}{4})}{\Gamma(\frac{1}{4})}.
\label{Integral1}    
\end{eqnarray}
By using these two results we obtain the asymptotic behavior of the ring diagram, where region 2 only has a partial contribution to the coefficient of the term $r_s^{-3/4}$. We obtained:
\begin{equation}
\epsilon_{r2}(r_s,\zeta) \overset{r_s \rightarrow \infty}{\approx} -\frac{2\Xi M_1}{r^{\frac{3}{4}}_s} \approx -\frac{0.120772}{r_s^{\frac{3}{4}}},
\label{er2_final}
\end{equation}
where $\Xi$ is a numerical factor that we will express in most of our calculations in this subsection, which is given by the following expression:
\begin{equation}
\Xi \equiv \frac{24 \Gamma(\frac{3}{4})}{\sqrt{\pi}(3\pi)^{\frac{5}{4}}\Gamma(\frac{1}{4})\alpha^{\frac{3}{4}}}.
\label{Chi_number}    
\end{equation}

In the first region of integration, we exploit the fact that the following condition for the function $\tilde{\Pi}(2Q,yQ,\zeta,r_s)\geq 1$ is satisfied. In the first region of integration of $\epsilon_r(r_s,\zeta)$, we expand the logarithm terms in the integrand in powers of $1/(\tilde{\Pi}(2Q,Qy,\zeta,r_s))$. Three main integrals have to be calculated in the first region :
\begin{equation}
\epsilon_{r1}(r_s,\zeta) = \epsilon^{1}_{r1}(r_s,\zeta) + \epsilon^{2}_{r1}(r_s,\zeta)+\epsilon^{3}_{r1}(r_s,\zeta), 
\label{ring_1_step2}
\end{equation}
where each of these terms are given by the following integrals:
\begin{eqnarray}
\epsilon^{1}_{r1}(r_s,\zeta) &=& \frac{\gamma_3}{r_s^2} \int_0^{\infty} dy \int_0^{\overstar{k}} dQ Q^4 \ln\left(\tilde{\Pi}\right),
\label{ring_11}\\    
\epsilon^{2}_{r1}(r_s,\zeta) &=& \frac{\gamma_3}{r_s^2} \int_0^{\infty} dy \int_0^{\overstar{k}} dQ Q^4 \ln\left(1+ \frac{1}{\tilde{\Pi}}\right),
\label{ring_12}\\    
\epsilon^{3}_{r1}(r_s,\zeta) &=& -\frac{\gamma_3}{r_s^2} \int_0^{\infty} dy \int_0^{\overstar{k}} dQ Q^4 \tilde{\Pi},
\label{ring_13}    
\end{eqnarray}
where $\tilde{\Pi}=\tilde{\Pi}(2Q,Qy,\zeta,r_s)$ due to the change of variables done in previous steps $\gamma_3=24/(\pi \alpha^2)$.

The first integral can be expressed as: 
\begin{widetext}
\begin{equation}
\epsilon^1_{r1}(r_s,\zeta) = \frac{24}{\pi (\alpha r_s)^2} \int_0^{\infty} dy \int_0^{\overstar{k}(r_s,y)} dQ Q^4 \left[ \ln\left(\frac{\alpha r_s}{4 \pi}\right) -2\ln(Q)+\ln\left(\sum_{\sigma} x_{\sigma}(\zeta)g\left(\frac{2Q}{x_{\sigma}(\zeta)},\frac{2yQ^2}{x^2_{\sigma}(\zeta)}\right)\right)  \right],
\label{ring_11_subparts}
\end{equation}
\end{widetext}
which can be further broken into a sum of three terms $\epsilon^1_{r1\alpha}(r_s,\zeta)$ (where $\alpha=1,2,3$ corresponding to the
three terms in the bracket),
where the first  and second  term can be integrated easily, which yields:
\begin{equation}
\epsilon^1_{r11}(r_s,\zeta) = \frac{2 \Xi \ln\left(\frac{\alpha r_s}{4 \pi}\right)}{5r^{\frac{3}{4}}_s},
\label{er111}    
\end{equation}
\begin{eqnarray}
  \epsilon^1_{r12}(r_s,\zeta) &=& \frac{\delta_2}{r^{\frac{3}{4}}_s}\left[4+\frac{\delta_3}{2\Gamma(\frac{3}{4})} 
  - 5\ln\left(\frac{\alpha r_s}{3\pi}\right)\right],
\label{er112}    
\end{eqnarray}
where $\delta_2=\Xi/25$, $\delta_3=5 M_3 \Gamma(\frac{1}{4})/\sqrt{\pi}$ and $M_3$ is the following integral:
\begin{equation}
\int_0^{\infty} dy \frac{\ln\left(1+y^2\right)}{\left(1+y^2 \right)^{\frac{5}{4}}} = 1.02849.
\label{M3}    
\end{equation}

For the last term in Eq.~\ref{ring_11_subparts}, which we call $\epsilon^1_{r13}(r_s,\zeta)$, we did a change of variable to remove the $r_s$ and $y$ dependence in the limit of integration by using the transformation $Q = \overstar{k}(r_s,y) p$.  In such integrand, we can do an expansion of the Lindhard function in the large $r_s$ limit. After doing this transformation, we obtain the following:
\begin{equation}
g\left(\frac{2Q}{x_{\sigma}(\zeta)},i\frac{2Q^2y}{x_{\sigma}^2(\zeta)} \right) = \frac{{2 x_{\sigma}^2}(3\pi)^{\frac{1}{2}}}{3p^2(\alpha r_s)^{\frac{1}{2}}(1+y^2)^{\frac{1}{2}} }.
\label{g_largers}
\end{equation}

By replacing such expression in the last term inside of the bracket in Eq.~\ref{ring_11_subparts}, we obtain:
\begin{eqnarray}
  \epsilon^1_{r13}(r_s,\zeta) &\approx& \frac{\delta_2}{r^{\frac{3}{4}}_s} \left[4-\frac{\delta_3}{2\Gamma(\frac{3}{4})}-5\ln\left(\frac{3\alpha r_s}{16 \pi} \right)\right].
\label{er113_final}    
\end{eqnarray}
By summing the terms from Eqs.~\ref{er111},~\ref{er112}, and ~\ref{er113_final} we obtain the contribution to the coefficient of the $r_s^{-3/4}$:
\begin{equation}
\epsilon_{r11}(r_s,\zeta) = \frac{192 \Gamma(\frac{3}{4})}{25 \sqrt{\pi}\alpha^{\frac{3}{4}}(3 \pi)^{\frac{5}{4}}\Gamma(\frac{1}{4})r_s^{\frac{3}{4}}}.
\label{er11_final}    
\end{equation}

The calculation of $\epsilon_{r12}(r_s,\zeta)$ relies on using the same change of variable that we used to calculate $\epsilon^1_{r13}(r_s,\zeta)$ to make the limits of integration to be independent of $r_s$ and $y$. By expanding the logarithm and keeping track of all orders of expansion we obtain:
\begin{equation}
\epsilon_{r12}(r_s,\zeta) \approx \frac{48 \Gamma\left(\frac{3}{4}\right)M_4}{\sqrt{\pi}\alpha^{\frac{3}{4}}(3\pi)^{\frac{5}{4}}\Gamma(\frac{1}{4})r_s^{\frac{3}{4}}},
\label{er12_final}    
\end{equation}
where $M_4$ is a series that we calculated on Wolfram Mathematica, which yields:
\begin{equation}
M_4 = \sum_{n=1}^{\infty} \frac{(-1)^{n+1}}{n(4n+5)} \approx 0.08505.
\label{M4}    
\end{equation}

Similarly as done for $\epsilon_{r12}(r_s,\zeta)$, we can calculate the last integral in the first region of integration:
\begin{equation}
\epsilon_{r13}(r_s,\zeta) \approx -\frac{16 \Gamma(\frac{3}{4})}{\pi^{\frac{3}{2}}\alpha^{\frac{3}{4}}(3\pi)^{\frac{1}{4}} \Gamma(\frac{1}{4}) r_s^{\frac{3}{4}}}.
\label{er13_final}    
\end{equation}
By summing the three terms in Eq.~\ref{ring_1_step2} we obtain the leading term for the sum of ring diagrams in region 1:
\begin{eqnarray}
\epsilon_{r1} \approx -\frac{0.6823134}{r_s^{\frac{3}{4}}},
\label{er1_final}    
\end{eqnarray}
and by adding such a term to the result we have obtained in Region 2 given by Eq.~\ref{er2_final}, we obtain that the leading term of the integral is of the power form of $r_s^{-3/4}$ and its coefficient is given below:
\begin{equation}
\epsilon_{r}(r_s,\zeta) \approx -\frac{0.8031}{r_s^{\frac{3}{4}}},
\label{ring_largers}    
\end{equation}
which agrees with our numerical results from our fitted region for large values of $r_s$ within errors.

\section{Calculation of the kite diagram}
\label{Calculation-of-kite}

We avoid the issue of the ``sign'' problem, by doing a set of transformations to the wavevector and frequency variable which will also simplify the expression and reduce the dimension of the integral given in Eq.~\ref{E2b_correction}. This is easily done by isolating one of the frequency variables from the rest in the argument of the dielectric function at the Goldstone diagram given by Eq.~\ref{E2b_correction}. This allows the possibility of doing the other two frequency integrals by using Cauchy's residue theorem and thus avoiding taking into consideration the branch-cuts that arise from the logarithmic terms in the expression of the dielectric function. After doing the two frequency integrals variables $k^0$ and $q_2^0$, we obtain the sum of two terms $\Delta_1\left[n_{\uparrow},n_{\downarrow}\right]$ and $\Delta_2\left[n_{\uparrow},n_{\downarrow} \right]$ given by the following integral expressions:
\begin{equation}
    \Delta E^{RPA}_{2b}\left[n_{\uparrow},n_{\downarrow}\right] = \Delta_1\left[n_{\uparrow}, n_{\downarrow} \right] + \Delta_2\left[n_{\uparrow},n_{\downarrow} \right],
\label{E2b_correction_2}    
\end{equation}
where the two main contributions that we have so far for the correction of the kite diagram are given by the following:
\begin{widetext}
\begin{eqnarray}
\Delta_1 \left[n_{\uparrow},n_{\downarrow} \right] = -\frac{i V}{2 \hbar (2 \pi)^{10}}\sum_{\{\sigma \}=\pm}\int_0^1 d\lambda \int d^4q_1 \int d^3q_2 \int d^3k \frac{\lambda^2 V_0(\vec{k}-\vec{q_1}-\vec{q_2}) V_0^2(q_1)}{\epsilon_{{\lambda}}(q_1,q_1^0)(q_1^0+\omega_{\vec{q_2},\sigma}-\omega_{\vec{q_1}+\vec{q_2},\sigma}+i \eta)}\nonumber\\
\Pi^0(q_1,q_1^0)\Theta(k_{F \sigma}-q_2)\Theta(|\vec{q_1}+\vec{q_2}|-k_{F \sigma})\left[\frac{\Theta(k-k_{F \sigma}) \Theta(k_{F \sigma}-|\vec{k}-\vec{q_1}|)}{q_1^0+\omega_{\vec{k}-\vec{q_1},\sigma}-\omega_{\vec{k},\sigma}+i\eta} - \frac{\Theta(k_{F \sigma}-k) \Theta(|\vec{k}-\vec{q_1}|-k_{F \sigma})}{q_1^0+\omega_{\vec{k}-\vec{q_1},\sigma}-\omega_{\vec{k},\sigma}-i\eta}  \right],
\label{delta1}\\ 
\Delta_2 \left[n_{\uparrow},n_{\downarrow} \right] = \frac{i V}{2 \hbar (2 \pi)^{10}}\sum_{\{\sigma \}=\pm}\int_0^1 d\lambda \int d^4q_1 \int d^3q_2 \int d^3k \frac{\lambda^2 V_0(\vec{k}-\vec{q_1}-\vec{q_2}) V_0^2(q_1)}{\epsilon_{{\lambda}}(q_1,q_1^0)(q_1^0+\omega_{\vec{q_2},\sigma}-\omega_{\vec{q_1}+\vec{q_2},\sigma}-i \eta)}\nonumber\\
\Pi^0(q_1,q_1^0)\Theta(q_2-k_{F \sigma})\Theta(k_{F \sigma}-|\vec{q_1}+\vec{q_2}|)\left[\frac{\Theta(k-k_{F \sigma}) \Theta(k_{F \sigma}-|\vec{k}-\vec{q_1}|)}{q_1^0+\omega_{\vec{k}-\vec{q_1},\sigma}-\omega_{\vec{k},\sigma}+i\eta} - \frac{\Theta(k_{F \sigma}-k) \Theta(|\vec{k}-\vec{q_1}|-k_{F \sigma})}{q_1^0+\omega_{\vec{k}-\vec{q_1},\sigma}-\omega_{\vec{k},\sigma}-i\eta}  \right],
\label{delta2} 
\end{eqnarray}
\end{widetext}
where the frequencies $\omega_{\vec{k} \sigma}$ are related to the energy dispersion of the electrons as $\omega_{\vec{k} \sigma} = \epsilon_{\vec{k} \sigma}/\hbar$ and $\Pi^0(q_1,q_1^0)$ is the spin-trace of the polarization tensor.

In the integral expression of $\Delta_2[n_{\uparrow},n_{\downarrow}]$, we can do the transformations $\vec{q_2} \rightarrow \vec{q_2}-\vec{q_1}$ followed by $\vec{q_1}\rightarrow -\vec{q_1}$. These momentum transformations are what allow a common product of Heaviside functions for both $\Delta_1[n_{\uparrow},n_{\downarrow}]$ and $\Delta_2[n_{\uparrow},n_{\downarrow}]$ . We obtain:
\begin{widetext}
  \begin{multline}
\Delta_2 \left[n_{\uparrow},n_{\downarrow} \right] = \frac{i V}{2 \hbar (2 \pi)^{10}}\sum_{\{\sigma \}=\pm}\int_0^1 d\lambda \int d^4q_1 \int d^3q_2 \int d^3k \frac{\lambda^2 V_0(\vec{k}-\vec{q_2}) V_0^2(q_1)}{\epsilon_{{\lambda}}(q_1,q_1^0)(q_1^0+\omega_{\vec{q_1}+\vec{q_2},\sigma}-\omega_{\vec{q_2},\sigma}-i \eta)}\\
\Pi^0(q_1,q_1^0)\Theta(k_{F \sigma}-q_2)\Theta(|\vec{q_1}+\vec{q_2}|-k_{F \sigma})\left[\frac{\Theta(k-k_{F \sigma}) \Theta(k_{F \sigma}-|\vec{k}+\vec{q_1}|)}{q_1^0+\omega_{\vec{k}+\vec{q_1},\sigma}-\omega_{\vec{k},\sigma}+i\eta} - \frac{\Theta(k_{F \sigma}-k) \Theta(|\vec{k}+\vec{q_1}|-k_{F \sigma})}{q_1^0+\omega_{\vec{k}+\vec{q_1},\sigma}-\omega_{\vec{k},\sigma}-i\eta}  \right].
\label{delta2,2} 
\end{multline}
\end{widetext}

A sequence of wave-vector transformations has been done for both expressions given by Eq.~\ref{delta1} and Eq,~\ref{delta2,2} to simplify them and to find a common product of Heaviside functions since this makes it easier the computation these two 11 dimension integrals by the MC method. For $\Delta_1[n_{\uparrow},n_{\downarrow}]$ the sequence of wave-vector transformations goes as follows: First, on both terms of Eq.~\ref{delta1} flip the parity of wave-vector $\vec{k} \rightarrow - \vec{k}$. After this, on the resulting integral expression, transform the first of the two 11 dimension integral by the following translation on the wavevector: $\vec{k}\rightarrow\vec{k}-\vec{q_1}$. Lastly, on the first integrand do the following two parity transformations: $\vec{q_2} \rightarrow -\vec{q_2}$ and $\vec{q_1} \rightarrow -\vec{q_1}$. This yields the following simplified expression for $\Delta_1[n_{\uparrow},n_{\downarrow}]$:
\begin{widetext}
\begin{multline}
\Delta_1 \left[n_{\uparrow},n_{\downarrow} \right] = -\frac{i V}{2 \hbar (2 \pi)^{10}}\sum_{\{\sigma \}=\pm}\int_0^1 d\lambda \int d^4q_1 \int_{q_2 \leq k_{F \sigma}} d^3q_2 \int_{k< k_{F \sigma}} d^3k \frac{\lambda^2 V_0^2(q_1) \Pi^0(\vec{q_1},q_1^0)}{\epsilon_{{\lambda}}(q_1,q_1^0)(q_1^0 + \omega_{\vec{q_2},\sigma}-\omega_{\vec{q_1}+\vec{q_2},\sigma}+i\eta)} \\ 
\Theta(|\vec{q_1}+\vec{q_2}|-k_{F \sigma}) \Theta(|\vec{k}+\vec{q_1}|-k_{F \sigma}) \Bigl [ \frac{V_0(\vec{k}-\vec{q_2})}{(q_1^0+\omega_{\vec{k},\sigma}-\omega_{\vec{k}+\vec{q_1},\sigma}+i\eta)}
-\frac{V_0(\vec{k}+\vec{q_1}+\vec{q_2}) }{(q_1^0+\omega_{\vec{k}+\vec{q_1},\sigma}-\omega_{\vec{k},\sigma}-i\eta)}  \Bigr ].
\label{delta1,3}     
\end{multline}
\end{widetext}

The expression of $\Delta_2 \left[n_{\uparrow},n_{\downarrow} \right]$ can be simplified in a similar way as done for $\Delta_1 \left[n_{\uparrow},n_{\downarrow} \right]$ by a sequence of transformations of wave-vectors. For the first term of the integrand in Eq.~\ref{delta2,2}, we do the translation $\vec{k} \rightarrow \vec{k} - \vec{q_1}$. To the integrand resulting from the previous step, the following transformation has to be performed: $\vec{k} \rightarrow -\vec{k}$, which leads to the following expression:
\begin{widetext}
\begin{eqnarray}
\Delta_2 \left [n_{\uparrow},n_{\downarrow} \right] = \frac{i V}{2 \hbar (2 \pi)^{10}}\sum_{\{\sigma \}=\pm}\int_0^1 d\lambda \int d^4q_1 \int_{q_2 \leq k_{F \sigma}} d^3q_2 \int_{k< k_{F \sigma}} d^3k
  \frac{\lambda^2 V_0^2(q_1) \Pi^0(\vec{q_1},q_1^0)}{\epsilon_{{\lambda}}(q_1,q_1^0)(q_1^0 + \omega_{\vec{q_1}+\vec{q_2},\sigma}-\omega_{\vec{q_2},\sigma}-i\eta)}\nonumber \\ 
\Theta(|\vec{q_1}+\vec{q_2}|-k_{F \sigma}) \Theta(|\vec{k}+\vec{q_1}|-k_{F \sigma}) \Bigl [ \frac{V_0(\vec{k} + \vec{q_1} + \vec{q_2})}{(q_1^0+\omega_{\vec{k},\sigma}-\omega_{\vec{k}+\vec{q}_1,\sigma}+i\eta)} 
- \frac{V_0(\vec{k}-\vec{q_2}) }{(q_1^0+\omega_{\vec{k}+\vec{q}_1,\sigma}-\omega_{\vec{k},\sigma} - i\eta) } \Bigr ].
\label{delta2,3}     
\end{eqnarray}
\end{widetext}

In order to simplify the computation of the correction of the kite diagram in RPA, instead of only integrating $q_1^0$ over the real line, we can instead exploit Cauchy's theorem. We can choose to integrate along a contour path in the complex plane. The contour path $C$ can be separated into 4 path integrals: $C_1$, $C_2$, $C_3$, $C_4$. The path $C_1$ corresponds to the line integral over the real line, $C_2$ would correspond to the quarter of the circle on the first quadrant of the complex plane with counterclockwise orientation for a radius that tends to infinity, where its contribution to the complex path integral is zero. Path $C_3$ is the downward oriented line integral along the imaginary line, while $C_4$ corresponds to a quarter of a circle path on the third quadrant of the complex plane with clockwise orientation, where its contribution to the complex path integral is also zero when taking the radius of the circular path to $\infty$.  Due to the constraints given by the Heavisides, the complex poles in the expressions given by Eq.~\ref{delta1,3} and ~\ref{delta2,3}, can only be on the second or fourth quadrant of the complex plane. Since none of the complex poles are enclosed by the contour complex path, this means that the integration along the real line can be directly mapped into the integration along the imaginary axis. After doing this and exploiting the fact that the polarization tensor is an even function in frequency $q_1^0$, we obtain the following expression:
\begin{widetext}
\begin{multline}
    \Delta E^{RPA}_{2b} \left[n_{\uparrow}, n_{\downarrow} \right] = -\frac{V}{\hbar(2 \pi)^{10}}\sum_{\{\sigma \}=\pm}\int_0^1 d\lambda \int_{k<k_{F \sigma}} d^3k \int_{q_2\leq k_{F \sigma}} d^3q_2 \int d^3q_1 \int_{-\infty}^{\infty} d \nu \frac{{\lambda^2} (V_0(q_1))^2 }{\epsilon_{{\lambda}}(q_1,i\nu ))(\nu+(\omega_{\vec{k}+\vec{q_1},\sigma}-\omega_{k,\sigma})i)}\\
  \Pi^0_{\sigma}(q_1,i \nu)  \Theta(|\vec{k}+\vec{q_1}|-k_{F \sigma}) \Theta(|\vec{q_1}+\vec{q_2}|-  k_{F \sigma}) \left[ \frac{V_0(\vec{k}-\vec{q_1})}{(\nu+(\omega_{\vec{q_1}+\vec{q_2},\sigma}  - \omega_{\vec{q_2},\sigma})i))  }\right.
  \left. -\frac{V_0(\vec{k}+\vec{q_1}+\vec{q_2})}{(\nu+(\omega_{\vec{q_2},\sigma}  - \omega_{\vec{q_1}+ \vec{q_2},\sigma})i)   }
  \right].
\label{e2b_rpa_imag}
\end{multline}
\end{widetext}
The fact that the polarization tensor is completely real, makes it easier to compute using the stochastic integration technique. From Eq.~\ref{e2b_rpa_imag} and by doing the following variable transformation: $\vec{k}\rightarrow k_{F \sigma} \vec{k} $, $\vec{q_1} \rightarrow k_{F \sigma} \vec{q_1}$, $\vec{q_2} \rightarrow k_{F \sigma} \vec{q_2}$ and $\nu \rightarrow \hbar k_{F \sigma}^2 \nu/m$, we can extract the units from the expression of the integrand. Also, the correction to the exchange energy per particle $\Delta \epsilon_{2b}\left[n_{\uparrow}, n_{\downarrow} \right]$ from the kite-diagram series(in Ry) can be further simplified by a doing a simple variable transformation such as $\nu = a(\vec{q_1},\vec{q_2}) \tan(u)$ and $\nu = a(\vec{q_1},\vec{k}) \tan(u)$, where $a(\vec{q_1},\vec{q_2})$ is related to the difference between the energy dispersions given by the expression shown in Eq.~\ref{a_b}. The kite diagram correction term per particle (in Ry) given from the correction that arises from the renormalized interaction line yields two final results we have in Eq.~\ref{e2b_1} and Eq~\ref{e2b_2} that we discussed in the kite diagram section. 

\section{Ferromagnetism}
\label{ferromagnetism}

     \begin{figure*}[htp]
       \begin{center}
         \subfigure[]{
            \includegraphics[scale=0.3]{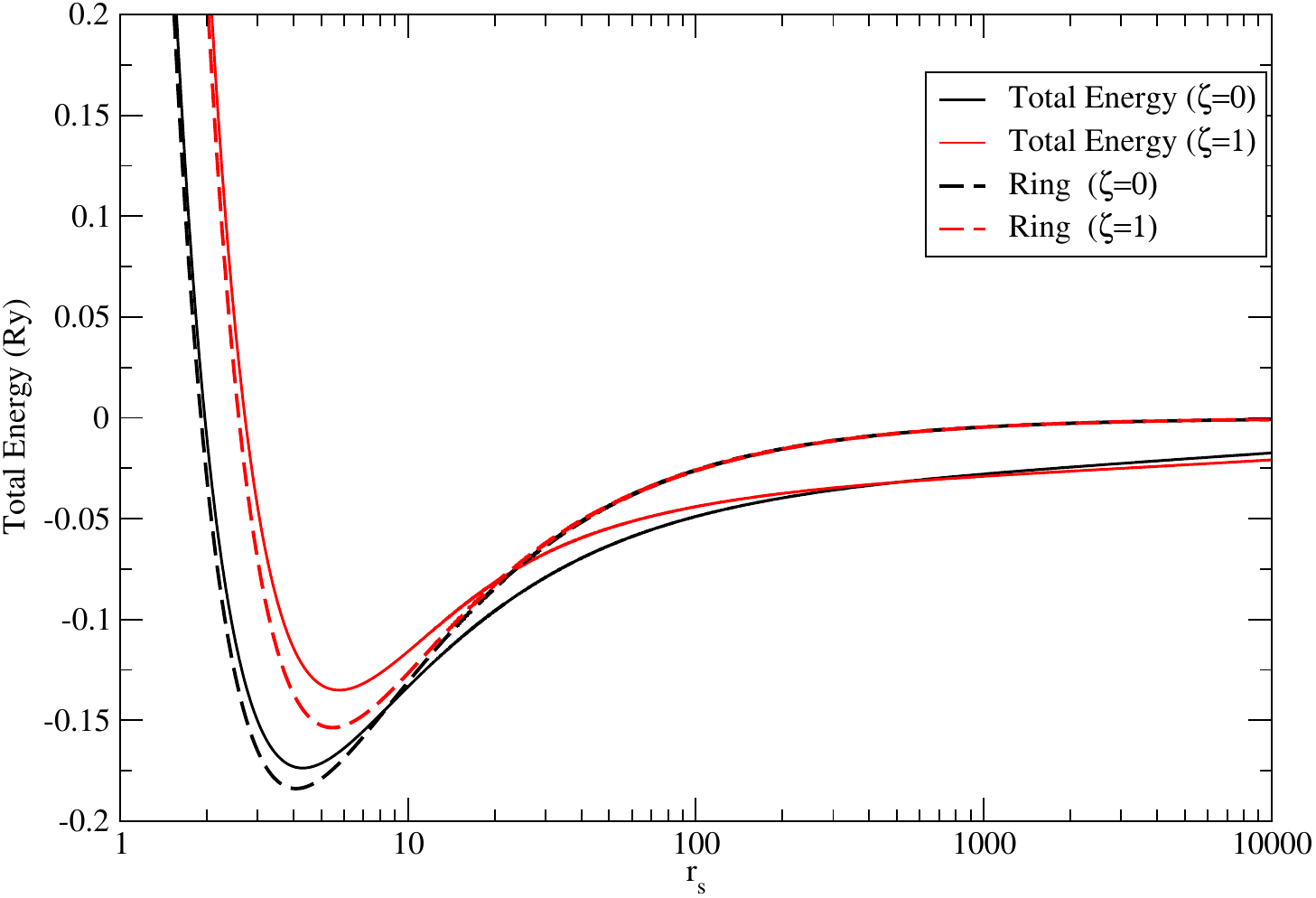} 
   \label{total-energy}
         }
         \subfigure[]{
               \includegraphics[scale=0.3]{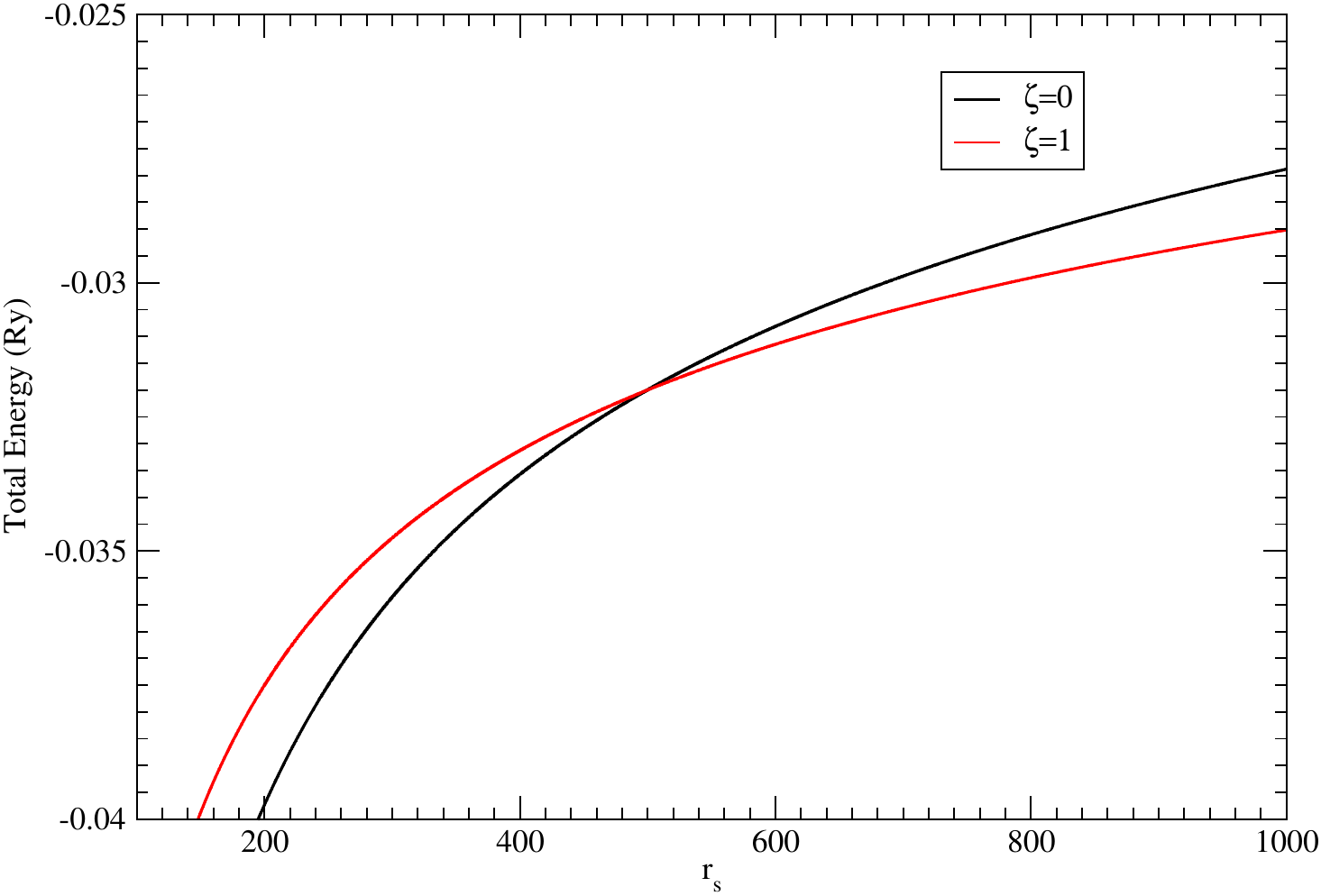} 
   \label{total-energy-zoom}
         }
   \end{center}
       \caption{(a) Comparison of the total energy for $\zeta=0$ and $\zeta=1$
         The solid lines is the total energy which includes the contribution of
         the kite-diagram series, while the dashed lines does not.
         (b) A zoom in the region where there is a crossing between
       the energy of the fully polarized with that of the unpolarized.}
       \end{figure*}
     The magnetic response and the magnetic phase diagram of the correlated electron fluid within RPA
     or otherwise has a long history\cite{PhysRevB.17.385,PhysRev.112.328,PhysRev.143.183} (see also Ref.~\onlinecite{PhysRevB.72.115317} and References
     therein). Here we wish to use our derived functional to ask the question
     of the stability  of the unpolarized phase against the
     fully spin-polarized one.
Sometime ago~\cite{PhysRev.140.A1645}, it was found
that, if one considers the contribution to the total energy, the
kinetic and the exchange term, and only the ring-series term in the correlation
energy, there is a value of
$r_s$ close to the ``physical'' region where the fully polarized
electron-gas becomes energetically more stable than the unpolarized one.
However, this turns out to be caused by the approximations used to
obtain an estimate for the contribution of the ring-series.
As shown in Fig.~\ref{total-energy} the dashed-black and dashed-red
lines correspond to the total energy when only the ring-series is
included in the correlation energy for $\zeta=0$ and $\zeta=1$ respectively. The
polarized fluid becomes marginally energetically favorable against the
unpolarized for $r_s > 70$ (as also found in Ref.~\onlinecite{PhysRevB.72.115317}, although the energy difference even at
$r_s=100$ is of the order of $10^{-4}$ Ry which is not
comparable with the exchange energy in Fe for example). However, when we include the contribution of the
kite-diagram series, we obtain the solid-black ($\zeta=0$) and
solid-red ($\zeta=1$) lines. There is a very large value of $r_s$
of the order of 500, where as illustrated in the
zoomed-in Fig.~\ref{total-energy-zoom} the polarized becomes more
stable. The energy difference between the fully-polarized and
the unpolarized electron fluid is sizable in the physical region
and it is about the same with or without the contribution of the kite-diagram
series.

This is in qualitative agreement with other functionals
derived from QMC\cite{PhysRevLett.45.566,PhysRevLett.82.5317} that the fully polarized electron gas becomes
favorable for values of $r_s$ well outside the ``physical'' region.

\section{Tables of numerical results}
\label{tables}
In this section of the Appendix we give tables of various results
discussed in the main part of the manuscript. The captions explain
the content of each table.

\begin{table*}{\bf Results of the DFT calculations}
  \vskip 0.2 in
  \begin{tabular}{ |c|c|c|c|c|c|c||c|c|c|c|c|c|c|}
    \hline
    & \multicolumn{3}{c|}{$a_{0}$} & \multicolumn{3}{c||}{$B_{0}$} & & \multicolumn{3}{c}{$a_{0}$} & \multicolumn{3}{|c|}{$B_{0}$} \\ \cline{1-14}
Crystal & PW & RPAF & Expt. & PW & RPAF & Expt. & Crystal & PW & RPAF & Expt. & PW & RPAF & Expt. \\
\hline
Li (A2) & 3.366 & 3.389 & 3.477 & 15.0 & 14.3 & 13.0 & NaF (B1) & 4.509 & 4.521 & 4.609 & 60.9 & 59.4 & 51.4 \\
\hline
Na (A2) & 4.060 & 4.090 & 4.225 & 9.0 & 8.5 & 7.5 & NaCl (B1) & 5.470 & 5.489 & 5.595 & 31.8 & 31.1 & 26.6 \\
\hline
K (A2) & 5.045 & 5.096 & 5.225 & 4.5 & 4.2 & 3.7 & MgO (B1) & 4.166 & 4.174 & 4.207 & 171.9 & 169.9 & 165 \\
\hline
Rb (A2) & 5.380 & 5.439 & 5.585 & 3.5 & 3.2 & 3.1 & MgS (B1) & 5.134 & 5.148 & 5.202 & 82.0 & 85.0 & 78.9 \\
\hline
Ca (A1) & 5.341 & 5.374 & 5.565 & 19.5 & 18.9 & 15 & CaO (B1) & 4.709 & 4.719 & 4.803 & 128.8 & 126.0 & 110 \\
\hline
Sr (A1) & 5.789 & 5.831 & 6.048 & 13.7 & 13.3 & 12 & TiC (B1) & 4.268 & 4.276 & 4.23 & 281.6 & 276.5 & 233 \\
\hline
Ba (A2) & 4.772 & 4.815 & 5.007 & 11.0 & 10.3 & 10 & TiN (B1) & 4.179 & 4.187 & 4.239 & 319.0 & 314.3 & 288 \\
\hline
V (A2) & 2.937 & 2.944 & 3.028 & 202.7 & 200.2 & 162 & ZrC (B1) & 4.643 & 4.652 & 4.696 & 245.3 & 243.7 & 265 \\
\hline
Nb (A2) & 3.245 & 3.253 & 3.296 & 176.5 & 173.0 & 170 & ZrN (B1) & 4.528 & 4.536 & 4.585 & 283.2 & 277.6 & 215 \\
\hline
Ta (A2) & 3.262 & 3.269 & 3.301 & 221.7 & 219.5 & 194 & HfC (B1) & 4.572 & 4.580 & 4.638 & 261.9 & 259.0 & 270 \\
\hline
Mo (A2) & 3.110 & 3.116 & 3.144 & 295.5 & 285.3 & 272 & HfN (B1) & 4.468 & 4.476 & 4.52 & 302.0 & 295.6 & 306 \\
\hline
W (A2) & 3.136 & 3.141 & 3.162 & 339.4 & 335.5 & 296 & VC (B1) & 4.094 & 4.102 & 4.16 & 344.4 & 337.3 & 303 \\
\hline
Fe (A2) & 2.700 & 2.705 & 2.861 & 323.6 & 317.5 & 167 & VN (B1) & 4.050 & 4.057 & 4.141 & 365.4 & 360.2 & 233 \\
\hline
Rh (A1) & 3.755 & 3.761 & 3.798 & 313.4 & 315.8 & 269 & NbC (B1) & 4.429 & 4.436 & 4.47 & 333.3 & 328.1 & 315 \\
\hline
Ir (A1) & 3.812 & 3.817 & 3.835 & 405.9 & 400.3 & 355 & NbN (B1) & 4.361 & 4.368 & 4.392 & 352.7 & 345.9 & 292 \\
\hline
Ni (A1) & 3.421 & 3.428 & 3.516 & 258.9 & 252.4 & 184 & FeAl (B2) & 2.814 & 2.819 & 2.889 & 207.2 & 204.0 & 136 \\
\hline
Pd (A1) & 3.839 & 3.846 & 3.881 & 229.1 & 225.6 & 195 & CoAl (B2) & 2.795 & 2.801 & 2.861 & 207.8 & 203.7 & 162 \\
\hline
Pt (A1) & 3.893 & 3.900 & 3.916 & 307.7 & 302.2 & 277 & NiAl (B2) & 2.833 & 2.840 & 2.887 & 186.9 & 183.4 & 156 \\
\hline
Cu (A1) & 3.522 & 3.530 & 3.603 & 179.6 & 178.6 & 133 & BN (B3) & 3.586 & 3.592 & 3.607 & 402.2 & 397.4 & 369 \\
\hline
Ag (A1) & 4.007 & 4.015 & 4.069 & 136.7 & 142.1 & 109 & BP (B3) & 4.492 & 4.502 & 4.538 & 175.8 & 172.6 & 173 \\
\hline
Au (A1) & 4.050 & 4.058 & 4.065 & 188.8 & 186.5 & 167 & BAs (B3) & 4.736 & 4.747 & 4.777 & 146.7 & 144.1 & 148 \\
\hline
Al (A1) & 3.979 & 3.994 & 4.032 & 78.2 & 74.4 & 73 & AlP (B3) & 5.434 & 5.447 & 5.46 & 90.3 & 88.6 & 86 \\
\hline
C (A4) & 3.534 & 3.539 & 3.567 & 465.2 & 462.1 & 443 & AlAs (B3) & 5.629 & 5.644 & 5.658 & 75.6 & 73.9 & 82 \\
\hline
Si (A4) & 5.401 & 5.414 & 5.43 & 97.0 & 94.8 & 99.2 & GaN (B3) & 4.460 & 4.468 & 4.531 & 202.2 & 200.1 & 190 \\
\hline
Ge (A4) & 5.623 & 5.639 & 5.652 & 68.5 & 70.5 & 75.8 & GaP (B3) & 5.392 & 5.406 & 5.448 & 89.8 & 88.5 & 88 \\
\hline
Sn (A4) & 6.476 & 6.498 & 6.482 & 44.3 & 42.8 & 53 & GaAs (B3) & 5.605 & 5.620 & 5.648 & 74.3 & 72.5 & 75.6 \\
\hline
Pb (A1) & 4.874 & 4.891 & 4.916 & 52.9 & 51.8 & 45.8 & InP (B3) & 5.832 & 5.847 & 5.866 & 69.9 & 68.7 & 72 \\
\hline
Th (A1) & 4.899 & 4.925 & 5.074 & 70.8 & 50.7 & 58 & InAs (B3) & 6.031 & 6.048 & 6.054 & 59.8 & 58.5 & 58 \\
\hline
LiF (B1) & 3.909 & 3.921 & 4.01 & 87.1 & 85.2 & 69.8 & SiC (B3) & 4.330 & 4.338 & 4.358 & 230.5 & 226.5 & 225 \\
\hline
LiCl (B1) & 4.965 & 4.984 & 5.106 & 40.9 & 39.7 & 35.4 & CeO2 (C1) & 5.364 & 5.370 & 5.411 & 214.0 & 202.5 & 220 \\
\hline

\multicolumn{2}{|c}{} & \multicolumn{6}{|c}{$a_{0}$} & \multicolumn{6}{|c|}{$B_{0}$} \\ \cline{1-14}
\multicolumn{2}{|c}{} & \multicolumn{3}{|c}{PW} & \multicolumn{3}{|c}{RPAF} & \multicolumn{3}{|c}{PW} & \multicolumn{3}{|c|}{RPAF} \\ \cline{1-14}

\multicolumn{2}{|c}{me} & \multicolumn{3}{|c}{-0.072} & \multicolumn{3}{|c}{-0.059} & \multicolumn{3}{|c}{21.6} & \multicolumn{3}{|c|}{18.6} \\

\multicolumn{2}{|c}{mae} & \multicolumn{3}{|c}{0.074} & \multicolumn{3}{|c}{0.061} & \multicolumn{3}{|c}{23.9} & \multicolumn{3}{|c|}{22.1} \\

\multicolumn{2}{|c}{mre (\%)} & \multicolumn{3}{|c}{-1.654} & \multicolumn{3}{|c}{-1.363} & \multicolumn{3}{|c}{14.1} & \multicolumn{3}{|c|}{11.3} \\

\multicolumn{2}{|c}{mare (\%)} & \multicolumn{3}{|c}{1.684} & \multicolumn{3}{|c}{1.407} & \multicolumn{3}{|c}{16.0} & \multicolumn{3}{|c|}{14.2} \\ \cline{1-14}

\multicolumn{2}{|c}{variance} & \multicolumn{3}{|c}{0.0088} & \multicolumn{3}{|c}{0.0061} & \multicolumn{3}{|c}{1468.4} & \multicolumn{3}{|c|}{1298.3} \\

\multicolumn{2}{|c}{relative variance} & \multicolumn{3}{|c}{0.0004} & \multicolumn{3}{|c}{0.0003} & \multicolumn{3}{|c}{0.050} & \multicolumn{3}{|c|}{0.043} \\ \cline{1-14}

\hline
  \end{tabular}
  \caption{Equilibrium lattice constants $a_{0}$ ({\AA}) and bulk modulus $B_{0}$ (GPa) of the 60 crystals listed in Ref.~\onlinecite{PhysRevB.79.085104}. The solid Strukturbericht symbols are in parenthesis and used for the crystals as follows: fcc (A1), bcc (A2), diamond (A4), rocksalt (B1), CsCl (B2), zinc blended (B3), and fluorite (C1). The statistics reported at the end of the table are: mean error (me), mean absolute error (mae), mean relative error (mre), mean absolute relative error (mare), variance, and relative variance. The experimental bulk moduli for most crystals were obtained from Ref.~\onlinecite{PhysRevB.75.115131}, with Pb~\cite{brandes2013smithells}, HfC~\cite{liang2020insights}, BAs~\cite{10.1063/1.5116025}, and CeO\textsubscript{2}~\cite{gerward2005bulk}. }
		\label{table:4}
\end{table*}

\begin{table*}
  \begin{center}
\begin{tabular}{|l|c|c|c|c|c|c|c|c|c|c|c|c|c|c|c|c|} 
    \hline
$r_{s}\backslash \zeta$&0.00&0.20&0.40&0.60&0.80&0.90&0.91&0.92&0.93&0.94&0.95&0.96&0.97&0.98&0.99&1.00\\
\hline
0.01&-429.0&-424.4&-410.2&-384.4&-341.6&-308.3&-304.2&-299.8&-295.2&-290.3&-285.0&-279.2&-272.9&-265.7&-256.9&-243.1\\
\hline
0.09&-294.5&-291.4&-281.8&-264.5&-236.1&-214.5&-211.8&-209.0&-206.1&-203.0&-199.6&-196.0&-192.1&-187.7&-182.6&-175.3\\
\hline
0.1&-288.1&-285.1&-275.7&-258.9&-231.2&-210.1&-207.5&-204.8&-201.9&-198.9&-195.6&-192.1&-188.3&-184.1&-179.1&-172.1\\
\hline
0.2&-247.0&-244.4&-236.5&-222.2&-199.0&-181.4&-179.3&-177.1&-174.7&-172.2&-169.6&-166.7&-163.6&-160.2&-156.3&-151.0\\
\hline
0.3&-223.5&-221.2&-214.1&-201.3&-180.6&-165.1&-163.2&-161.2&-159.2&-157.0&-154.7&-152.2&-149.5&-146.6&-143.2&-138.8\\
\hline
0.4&-207.2&-205.1&-198.5&-186.8&-167.9&-153.7&-152.0&-150.2&-148.4&-146.4&-144.3&-142.1&-139.7&-137.1&-134.1&-130.3\\
\hline
0.5&-194.7&-192.7&-186.6&-175.7&-158.1&-145.1&-143.5&-141.9&-140.1&-138.3&-136.4&-134.4&-132.2&-129.8&-127.1&-123.7\\
\hline
0.6&-184.7&-182.8&-177.1&-166.9&-150.4&-138.1&-136.6&-135.1&-133.5&-131.8&-130.1&-128.2&-126.2&-124.0&-121.5&-118.4\\
\hline
0.7&-176.4&-174.7&-169.2&-159.5&-143.9&-132.3&-130.9&-129.5&-128.0&-126.4&-124.8&-123.0&-121.1&-119.1&-116.8&-113.9\\
\hline
0.8&-169.3&-167.6&-162.4&-153.2&-138.3&-127.4&-126.1&-124.7&-123.3&-121.8&-120.3&-118.6&-116.8&-114.9&-112.7&-110.1\\
\hline
0.9&-163.1&-161.5&-156.5&-147.7&-133.5&-123.1&-121.8&-120.5&-119.2&-117.8&-116.3&-114.7&-113.1&-111.2&-109.2&-106.8\\
\hline
1&-157.6&-156.1&-151.3&-142.8&-129.2&-119.2&-118.1&-116.8&-115.6&-114.2&-112.8&-111.3&-109.7&-108.0&-106.1&-103.8\\
\hline
2&-123.9&-122.7&-119.2&-112.8&-102.9&-95.4&-94.6&-93.7&-92.8&-91.9&-90.9&-89.9&-88.8&-87.6&-86.3&-84.8\\
\hline
3&-105.7&-104.7&-101.8&-96.6&-88.5&-82.6&-82.0&-81.3&-80.6&-79.8&-79.1&-78.3&-77.4&-76.5&-75.5&-74.4\\
\hline
4&-93.7&-92.9&-90.4&-86.0&-79.1&-74.1&-73.6&-73.0&-72.4&-71.8&-71.1&-70.5&-69.8&-69.0&-68.2&-67.3\\
\hline
5&-85.0&-84.3&-82.1&-78.2&-72.2&-67.9&-67.4&-66.9&-66.4&-65.9&-65.3&-64.7&-64.1&-63.5&-62.8&-62.0\\
\hline
6&-78.3&-77.6&-75.7&-72.2&-66.8&-63.0&-62.6&-62.2&-61.7&-61.2&-60.7&-60.2&-59.7&-59.1&-58.5&-57.8\\
\hline
7&-72.9&-72.3&-70.5&-67.3&-62.5&-59.1&-58.7&-58.3&-57.9&-57.5&-57.0&-56.6&-56.1&-55.6&-55.0&-54.4\\
\hline
8&-68.4&-67.9&-66.2&-63.3&-58.9&-55.8&-55.4&-55.1&-54.7&-54.3&-53.9&-53.5&-53.1&-52.6&-52.1&-51.6\\
\hline
9&-64.6&-64.1&-62.6&-59.9&-55.8&-53.0&-52.6&-52.3&-52.0&-51.6&-51.2&-50.9&-50.5&-50.1&-49.6&-49.1\\
\hline
10&-61.3&-60.9&-59.5&-57.0&-53.2&-50.5&-50.2&-49.9&-49.6&-49.3&-48.9&-48.6&-48.2&-47.8&-47.4&-47.0\\
\hline
20&-42.8&-42.5&-41.6&-40.1&-37.9&-36.4&-36.2&-36.0&-35.8&-35.6&-35.4&-35.2&-35.0&-34.8&-34.6&-34.3\\
\hline
30&-34.1&-33.9&-33.3&-32.2&-30.6&-29.5&-29.4&-29.3&-29.2&-29.0&-28.9&-28.7&-28.6&-28.4&-28.3&-28.1\\
\hline
40&-28.9&-28.8&-28.3&-27.4&-26.2&-25.3&-25.2&-25.1&-25.0&-24.9&-24.8&-24.7&-24.6&-24.5&-24.3&-24.2\\
\hline
50&-25.4&-25.2&-24.8&-24.1&-23.1&-22.4&-22.3&-22.2&-22.2&-22.1&-22.0&-21.9&-21.8&-21.7&-21.6&-21.5\\
\hline
60&-22.7&-22.6&-22.3&-21.7&-20.8&-20.2&-20.1&-20.1&-20.0&-19.9&-19.9&-19.8&-19.7&-19.6&-19.5&-19.4\\
\hline
70&-20.7&-20.6&-20.3&-19.8&-19.0&-18.5&-18.4&-18.4&-18.3&-18.3&-18.2&-18.1&-18.1&-18.0&-17.9&-17.8\\
\hline
80&-19.1&-19.0&-18.7&-18.3&-17.6&-17.1&-17.1&-17.0&-17.0&-16.9&-16.9&-16.8&-16.7&-16.7&-16.6&-16.5\\
\hline
90&-17.7&-17.7&-17.4&-17.0&-16.4&-16.0&-15.9&-15.9&-15.8&-15.8&-15.7&-15.7&-15.6&-15.6&-15.5&-15.4\\
\hline
100&-16.6&-16.5&-16.3&-15.9&-15.4&-15.0&-15.0&-14.9&-14.9&-14.8&-14.8&-14.7&-14.7&-14.6&-14.6&-14.5\\
\hline
200&-10.7&-10.6&-10.5&-10.3&-10.0&-9.8&-9.8&-9.8&-9.8&-9.7&-9.7&-9.7&-9.7&-9.6&-9.6&-9.6\\
\hline
300&-8.2&-8.1&-8.1&-7.9&-7.7&-7.6&-7.6&-7.6&-7.6&-7.5&-7.5&-7.5&-7.5&-7.5&-7.5&-7.4\\
\hline
500&-5.8&-5.8&-5.7&-5.7&-5.5&-5.5&-5.5&-5.4&-5.4&-5.4&-5.4&-5.4&-5.4&-5.4&-5.4&-5.4\\
\hline
1000&-3.6&-3.6&-3.6&-3.5&-3.5&-3.4&-3.4&-3.4&-3.4&-3.4&-3.4&-3.4&-3.4&-3.4&-3.4&-3.4\\
\hline
2000&-2.2&-2.2&-2.2&-2.2&-2.2&-2.1&-2.1&-2.1&-2.1&-2.1&-2.1&-2.1&-2.1&-2.1&-2.1&-2.1\\
\hline
3000&-1.7&-1.7&-1.7&-1.7&-1.6&-1.6&-1.6&-1.6&-1.6&-1.6&-1.6&-1.6&-1.6&-1.6&-1.6&-1.6\\
\hline
4000&-1.4&-1.4&-1.4&-1.4&-1.3&-1.3&-1.3&-1.3&-1.3&-1.3&-1.3&-1.3&-1.3&-1.3&-1.3&-1.3\\
\hline
5000&-1.2&-1.2&-1.2&-1.2&-1.1&-1.1&-1.1&-1.1&-1.1&-1.1&-1.1&-1.1&-1.1&-1.1&-1.1&-1.1\\
\hline
20000&-0.43&-0.43&-0.43&-0.43&-0.43&-0.42&-0.42&-0.42&-0.42&-0.42&-0.42&-0.42&-0.42&-0.42&-0.42&-0.42\\
\hline
200000&-0.08&-0.08&-0.08&-0.08&-0.08&-0.079&-0.079&-0.079&-0.079&-0.079&-0.079&-0.079&-0.079&-0.079&-0.079&-0.079\\
\hline
1000000&-0.024&-0.024&-0.024&-0.024&-0.024&-0.024&-0.024&-0.024&-0.024&-0.024&-0.024&-0.024&-0.024&-0.024&-0.024&-0.024\\
    \hline
\end{tabular}
  \end{center}
\caption{Numerical results for the series of the ring-diagrams data (in mRy) generated using adaptive quadrature integration method for many values of $\zeta$ and $r_{s}$ covering the small $r_{s}$ region, the large $r_{s}$ region, and the physical region where materials are realized. }
  \label{table_ring_bare}
  \end{table*}

\begin{table*}
\begin{center}
\begin{tabular}{|c|c||c|c|c|c|c|c|c|} 
\hline
$r_{s}$&$\zeta=0$&$r_{s}\backslash\zeta$&0.2&0.4&0.6&0.8&0.9&1.0\\
\hline
0.01 & 47.60 (10) & 0.2 & 40.75 (19) & 40.85 (44) & 41.66 (32) & 42.83 (12) & 43.3 (56) & 44.44 (14) \\
\hline
0.09 & 44.16 (16) & 0.4 & 36.23 (37) & 37.37 (20) & 37.77 (25) & 39.24 (16) & 39.5 (51) & 42.30 (9) \\
\hline
0.1 & 43.53 (18) & 0.6 & 33.79 (20) & 34.31 (16) & 35.02 (20) & 36.85 (45) & 37.6 (49) & 39.89 (18) \\
\hline
0.2 & 40.85 (25) & 0.8 & 31.2 (19) & 31.75 (27) & 32.68 (22) & 34.46 (35) & 36.0 (47) & 37.76 (32) \\
\hline
0.3 & 38.45 (26) & 1 & 27.77 (61) & 29.42 (28) & 30.09 (32) & 32.46 (47) & 33.6 (44) & 36.43 (27) \\
\hline
0.4 & 36.00 (37) & 1.2 & 26.41 (35) & 27.79 (28) & 27.59 (55) & 30.34 (45) & 32.2 (42) & 34.83 (32) \\
\hline
0.5 & 34.26 (36) & 1.4 & 23.90 (62) & 25.51 (59) & 27.08 (40) & 29.68 (25) & 31.0 (40) & 34.05 (27) \\
\hline
0.6 & 33.90 (19) & 1.6 & 23.10 (23) & 24.02 (31) & 25.37 (35) & 27.10 (75) & 29.7 (39) & 32.65 (28) \\
\hline
0.7 & 31.89 (60) & 1.8 & 21.62 (29) & 21.38 (92) & 24.49 (24) & 26.79 (40) & 28.9 (38) & 31.71 (20) \\
\hline
0.8 & 30.2 (17) & 2 & 20.55 (29) & 21.63 (40) & 23.23 (21) & 25.73 (27) & 27.5 (36) & 30.46 (40) \\
\hline
0.9 & 28.79 (54) & 2.2 & 19.29 (23) & 19.67 (40) & 21.95 (28) & 24.23 (44) & 26.9 (35) & 29.48 (49) \\
\hline
1 & 27.62 (84) & 2.4 & 17.04 (73) & 18.66 (38) & 21.12 (19) & 24.26 (26) & 25.5 (33) & 28.80 (19) \\
\hline
2 & 20.31 (33) & 2.6 & 17.08 (26) & 17.44 (74) & 19.69 (28) & 21.97 (74) & 24.6 (32) & 27.82 (23) \\
\hline
3 & 14.85 (68) & 2.8 & 16.16 (19) & 17.28 (22) & 18.96 (35) & 22.02 (29) & 24.5 (32) & 26.90 (38) \\
\hline
4 & 10.56 (44) & 3 & 14.58 (40) & 15.97 (29) & 17.01 (60) & 21.18 (29) & 23.3 (31) & 25.93 (57) \\
\hline
5 & 8.04 (34) & 3.2 & 14.02 (25) & 14.17 (67) & 16.63 (42) & 20.41 (31) & 22.18 (62) & 24.71 (63) \\
\hline
6 & 5.07 (39) & 3.4 & 13.27 (55) & 14.41 (31) & 16.44 (29) & 19.70 (28) & 21.6 (28) & 24.81 (43) \\
\hline
7 & 2.23 (41) & 3.6 & 12.13 (33) & 13.17 (53) & 15.82 (32) & 18.91 (25) & 21.0 (28) & 23.81 (31) \\
\hline
8 & 0.78 (69) & 3.8 & 12.14 (33) & 12.61 (36) & 14.97 (66) & 18.34 (27) & 20.3 (27) & 23.47 (26) \\
\hline
9 & -1.70 (72) & 4 & 10.59 (49) & 11.72 (51) & 14.9 (22) & 17.79 (22) & 19.69 (23) & 22.32 (44) \\
\hline
10 & -2.11 (41) & 4.2 & 9.81 (35) & 11.35 (26) & 13.53 (33) & 17.14 (24) & 18.8 (25) & 22.21 (54) \\
\hline
20 & -11.13 (90) & 4.4 & 9.08 (67) & 10.42 (29) & 12.89 (28) & 15.74 (38) & 18.8 (25) & 21.4 (13) \\
\hline
30 & -14.85 (28) & 4.6 & 9.01 (29) & 9.50 (41) & 11.95 (28) & 15.47 (54) & 18.26 (19) & 20.87 (28) \\
\hline
40 & -17.91 (46) & 4.8 & 8.10 (56) & 9.49 (27) & 11.91 (18) & 14.79 (58) & 17.3 (23) & 20.14 (71) \\
\hline
50 & -19.45 (65) & 5 & 7.49 (65) & 8.59 (80) & 11.58 (39) & 14.21 (35) & 16.1 (11) & 20.04 (33) \\
\hline
60 & -20.27 (43) & 5.2 & 6.74 (51) & 7.94 (39) & 10.98 (30) & 14.50 (29) & 16.7 (22) & 19.01 (35) \\
\hline
70 & -21.67 (58) & 5.4 & 6.17 (36) & 7.68 (32) & 10.47 (31) & 12.99 (69) & 16.1 (21) & 19.13 (23) \\
\hline
80 & -22.14 (28) & 5.6 & 5.91 (43) & 7.36 (22) & 9.62 (26) & 13.08 (37) & 15.9 (21) & 18.79 (24) \\
\hline
90 & -22.62 (48) & 5.8 & 4.85 (43) & 6.17 (42) & 8.54 (61) & 12.78 (27) & 14.8 (20) & 17.18 (70) \\
\hline
100 & -23.68 (80) & 6 & 4.66 (59) & 6.43 (27) & 8.93 (30) & 12.59 (21) & 14.8 (20) & 17.78 (24) \\
\hline
200 & -25.09 (62) & 10 & -1.76 (57) & -1.00 (68) & 2.01 (75) & 3.9 (11) & 8.20 (44) & 8.8 (12) \\
\hline
300 & -24.57 (25) & 20 & -13.0 (27) & -9.57 (67) & -6.71 (65) & -4.01 (48) & -2.5 (12) & 1.35 (62) \\
\hline
500 & -24.71 (49) & 30 & -15.18 (66) & -12.76 (31) & -11.0 (15) & -8.91 (50) & -7.59 (51) & -4.60 (49) \\
\hline
1000 & -23.55 (36) & 40 & -17.05 (47) & -15.81 (47) & -14.76 (86) & -11.00 (42) & -9.56 (77) & -9.6 (22) \\
\hline
2000 & -20.77 (48) & 50 & -19.25 (61) & -17.42 (38) & -16.01 (41) & -13.29 (65) & -12.02 (40) & -11.57 (82) \\
\hline
3000 & -20.87 (73) & 60 & -20.47 (76) & -19.34 (44) & -18.42 (98) & -15.53 (52) & -13.72 (34) & -12.24 (47) \\
\hline
4000 & -19.49 (22) & 70 & -22.0 (11) & -19.29 (31) & -17.92 (33) & -16.08 (45) & -15.10 (40) & -13.32 (33) \\
\hline
5000 & -18.44 (18) & 80 & -21.38 (49) & -21.2 (32) & -19.7 (47) & -17.92 (53) & -17.15 (74) & -15.39 (39) \\
\hline
20000 & -14.35 (22) & 90 & -22.46 (54) & -20.81 (31) & -20.28 (45) & -18.38 (55) & -17.31 (61) & -16.07 (33) \\
\hline
200000 & -7.13 (15) & 100 & -23.15 (72) & -21.6 (15) & -21.01 (44) & -18.63 (43) & -18.60 (50) & -17.22 (58) \\
\hline
1000000 & -3.89 (16) & 200 & -24.43 (49) & -23.43 (33) & -23.56 (79) & -22.46 (34) & -22.30 (47) & -21.97 (61) \\
\hline
 &  & 300 & -24.24 (48) & -24.88 (50) & -24.06 (40) & -23.92 (57) & -24.04 (60) & -23.12 (48) \\
 \hline
 &  & 400 & -25.44 (76) & -24.10 (36) & -24.34 (29) & -23.52 (25) & -23.36 (30) & -24.33 (53) \\
 \hline
 &  & 500 & -24.28 (33) & -24.50 (43) & -24.89 (70) & -24.86 (70) & -23.49 (32) & -26.9 (22) \\
 \hline
 &  & 1000 & -23.56 (39) & -22.88 (28) & -23.11 (32) & -23.99 (48) & -23.94 (33) & -24.59 (62) \\
 \hline
 &  & 2000 & -21.16 (25) & -21.95 (70) & -22.18 (56) & -22.10 (21) & -22.88 (49) & -23.71 (36) \\
 \hline
 &  & 3000 & -20.16 (38) & -20.44 (22) & -20.46 (30) & -21.79 (50) & -21.69 (26) & -22.92 (36) \\
 \hline
 &  & 4000 & -19.45 (37) & -19.53 (24) & -20.06 (33) & -20.85 (29) & -20.76 (21) & -21.68 (23) \\
 \hline
 &  & 5000 & -19.76 (69) & -18.96 (27) & -19.86 (55) & -20.32 (77) & -20.15 (23) & -21.79 (45) \\
 \hline
 &  & 20000 & -13.98 (24) & -14.31 (21) & -14.58 (17) & -16.34 (56) & -15.93 (26) & -17.77 (47) \\
 \hline
 &  & 200000 & -7.00 (20) & -7.55 (19) & -7.46 (14) & -8.80 (30) & -9.06 (22) & -9.88 (30) \\
 \hline
 &  & 1000000 & -4.26 (39) & -3.78 (19) & -4.72 (25) & -4.41 (19) & -5.52 (32) & -5.56 (19) \\
\hline
\end{tabular}
\end{center}
\caption{Numerical results for the kite-diagram series
(in mRy) generated using Monte Carlo integration method for many values of $\zeta$ and $r_{s}$ covering the small $r_{s}$ region, the large $r_{s}$ region, and the physical region where materials are realized. (The standard deviations are reported in parenthesis).}
\label{table_kite_bare}
\end{table*}

\end{document}